\documentclass[12pt]{article}

\usepackage{amstext,amsmath,amssymb,amsfonts,bbm,slashed}
\usepackage[latin1]{inputenc}
\usepackage{epsfig}
\usepackage{hyperref}
\usepackage{amsthm}
\usepackage{subfigure}
\usepackage{color}
\usepackage{multirow}
\usepackage{calc}

\usepackage{stmaryrd}

\usepackage{cancel}

\newcommand{\kin}{{\rm{kin\,}}}
\newcommand{\ext}{{\rm{ext\,}}}
\newcommand{\inter}{{\rm{int\,}}}

\newcommand{\od}{{\rm{od}}}
\newcommand{\ev}{{\rm{ev}}}

\newcommand{\sset}{\subset}

\def\C{{\mathbbm C}}
\def\N{{\mathbbm N}}
\def\Z{{\mathbbm Z}}

\newcommand{\cG}{{\mathcal G}}
\newcommand{\bG}{{\partial\mathcal G}}
\newcommand{\tJ}{{\widetilde{J}}}

\newcommand{\cexG}{\mathcal G_{\text{color}}}
\newcommand{\bJ}{ J_{\partial} }

\newcommand{\cO}{{\mathcal O}}
\newcommand{\cR}{{\mathcal R}}
\newcommand{\cL}{{\mathcal L}}
\newcommand{\cV}{{\mathcal V}}
\newcommand{\cF}{{\mathcal F}}
\newcommand{\bcF}{{\overline{\mathcal F}}}
\newcommand{\cP}{{\mathcal P}}

\newcommand{\cU}{{\mathcal U}}

 \newcommand{\cE}{\mathcal{E}}

 \newcommand{\HU}{  {\rm{HU}} }

\newcommand{\cH}{\mathcal{HR}}
\newcommand{\cHE}{\mathcal{HE}}

\newcommand{\cFL}{\mathcal{FL}}

\newcommand{\bdel}{ {\boldsymbol{\delta}} }

\newcommand{\Om}{ {\boldsymbol{\Omega}} }

\newcommand{\eps}{\epsilon}
\newcommand{\beps}{\bar\epsilon}

\newtheorem{lemma}{Lemma}

\newtheorem{remark}{Remark}
\newtheorem{definition}{Definition}
\newtheorem{theorem}{Theorem}
\newtheorem{proposition}{Proposition}

\newcommand{\bea}{\begin{eqnarray}}
\newcommand{\eea}{\end{eqnarray}}
\newcommand{\beq}{\begin{equation}}
\newcommand{\eeq}{\end{equation}}

\makeatletter

\begin{document}

\begin{titlepage}
\begin{flushright}
ICMPA-MPA/2014/21
\end{flushright}

\vspace{20pt}

\begin{center}

{\Large\bf Parametric Representation \\
 of Rank $d$ Tensorial Group Field Theory: \\
\medskip 
Abelian Models with Kinetic Term $\sum_{s}|p_s| + \mu$}
\vspace{15pt}

{\large Joseph Ben Geloun$^{a,c,\dag}$ and Reiko Toriumi$^{b,\ddag} $}

\vspace{15pt}

$^{a}${\sl Max Planck Institute for Gravitational Physics, Albert Einstein Institute}\\
{\sl Am M\"uhlenberg 1, 14476, Potsdam, Germany }\\
\vspace{5pt}

$^{b}${\sl Centre de Physique Th\'eorique, CNRS-Luminy, Aix-Marseille Universit\'e}\\
{\sl Case 907, 13288 Marseille, France}

\vspace{5pt}

$^{c}${\sl International Chair in Mathematical Physics 
and Applications\\ (ICMPA-UNESCO Chair), University of Abomey-Calavi,\\
072B.P.50, Cotonou, Rep. of Benin}\\

\vspace{5pt}

E-mails:  {\sl $^{\dag}$jbengeloun@aei.mpg.de, 
$^\ddag$Reiko.Toriumi@cpt.univ-mrs.fr}

\vspace{10pt}

\begin{abstract}
We consider the parametric representation of the amplitudes of
Abelian models in the so-called framework of  
rank $d$ Tensorial Group Field Theory. 
These models are called Abelian because
their  fields live on $U(1)^D$. 
We concentrate on the case
when these models are endowed with particular
kinetic terms involving a linear power in momenta.  
New dimensional regularization and renormalization 
schemes are introduced for particular models in this
class: a rank 3 tensor model, an infinite tower of matrix 
models $\phi^{2n}$ over $U(1)$, and 
a matrix model over $U(1)^2$. 
For all divergent amplitudes, we identify a domain of meromorphicity in a strip
determined by the real part of the group dimension 
$D$.  From this point, the ordinary subtraction program
 is applied and leads to convergent and analytic
renormalized integrals. Furthermore, we identify and
study in depth the Symanzik polynomials
provided by the parametric amplitudes of generic
rank $d$ Abelian models. We find that these
polynomials do not satisfy the ordinary Tutte's rules (contraction/deletion).
By scrutinizing the ``face''-structure of these polynomials, 
we find a generalized polynomial which turns out to be stable only under contraction. 
\end{abstract}

\today

\end{center}

\noindent  Pacs numbers:  11.10.Gh, 04.60.-m, 02.10.Ox
\\

\bigskip

\setcounter{footnote}{0}

\end{titlepage}


\tableofcontents

\section{Introduction}
\label{intro}

Tensorial group field theories (TGFTs) provide a background independent framework to quantum gravity which is intimately based on the idea that the fundamental building blocks (quanta) of space-time are discrete \cite{oriti}--\cite{Rivasseau:2014ima}. 
Within this approach, the fields are rank $d$ tensors labeled by abstract group representations. From such a discrete structure, one dually associates tensor fields with basic  $d-1$ dimensional simplexes and their possible
interactions with $d$ dimensional simplicial building blocks. At the level of the partition function, the Feynman diagrams generated by the theory represent discretizations of a manifold in $d$ dimensions.
Thus, in essence, TGFTs which randomly generate topologies 
and geometries in covariant and algebraic ways,
 can be rightfully called quantum field theories of spacetime. 
One of the main efforts in this research program is to seek what types of phases the theory exhibits.
More to the point, one may ask if any of these phases give our geometric universe described by General Relativity from the pre-geometric cellular-complex picture that the bare theory gives \cite{Konopka:2006hu}.
This question is accompanied by a further suggestion that the relevant phase corresponds to a condensate of the microscopic degrees of freedom \cite{oriti,Rivasseau:2011hm}.
Note that this question has found a partial answer in \cite{gielen,gielen2}.

Because they are field theories, TGFTs can  certainly be
scrutinized using several different lenses. In particular, one
of the main successes of quantum field theory which 
is a Renormalization Group analysis turns out to have a counterpart
in TGFTs. We recall that renormalizability of any quantum field theory is a desirable feature since  it ensures that the theory survives after several energy scales.
In fact, so far, all known interactions of the standard model are renormalizable.
Quantum field theory predictions rely on the fact that, from the Wilsonian renormalization group point of view, the infinities that appear in the theory should locally reflect a change in the form of the theory \cite{Rivasseau:1991ub}.
In particular, if TGFTs are to describe any physical reality like our spacetime at a low energy scale, one is certainly interested in probing the flow of this theory.
The renormalization program 
suitably provides a mechanism to study the flow of a theory with respect to scales and also might lead to predictions. 
Within TGFTs, this renormalization program can be addressed in several ways and, indeed, has known important recent developments \cite{BenGeloun:2011rc}--\cite{Carrozza:2014rba}. The simplest setting in which one can think within TGFTs is given in purely combinatorial terms as tensor models. 

Tensor models, originally introduced in \cite{ambj3dqg}--\cite{oog}, especially enjoy the knowledge of their lower dimensional cousins: matrix models \cite{Di Francesco:1993nw}. These latter models are nowadays 
well developed and understood through rich statistical tools \cite{'tHooft:1973jz}--\cite{conf1}. 
 Specifically, the Feynman integral of matrix models generates ribbon graphs organized in a $1/N$ (or genus) expansion \cite{'tHooft:1973jz}. In short, this statistical sum is analytically well controlled. 
It is only recent that the notion of large $N$ expansion 
was extended to tensor models \cite{Gur3}--\cite{Gur4}
(for however the class of colored models \cite{color}--\cite{Gurau:2010nd}). 
From this point, important progresses have been unlocked 
\cite{Gurau:2011xp}--\cite{Bonzom:2014oua}
 and a renormalization program for tensor models
uncovered (for a review, see \cite{Carrozza:2013mna,Geloun:2013saa}).

Back to renormalization and its applicability to TGTFs, one
notes that, in anterior works, thriving efforts were developed 
on the so-called multi-scale renormalization \cite{Rivasseau:1991ub}. 
It is also worth and advantageous to understand how 
other known tools in renormalization
(like the Polchinski equation or Functional  Renormalization Group
methods \cite{thomas,Eichhorn:2013isa,Rivasseau:2014ima}) 
can shed light and even convey more insights in 
the present class of models and thereby enrich their Physics. 
Among these well-known renormalization procedures,
there is the celebrated dimensional regularization.   

In ordinary quantum field theory, dimensional regularization is an important scheme as it delivers finite (regularized) amplitudes and
respects, at the same time, the symmetries of gauge theories (preserves field equations and Ward identities) \cite{Bergere:1980sm,Bergere:1977ft}.
Of course, in our present class of non-local models, there exists a notion of 
invariance but it is an open issue to show that 
their associated Ward identities \cite{Samary:2014tja} will be preserved or not after the dimensional regularization and its subtraction program introduced
in the present work.
 Nevertheless, a dimensional regularization is a very interesting tool
that one may want to have in TGFTs. It allows 
one to understand the fine structure of the amplitudes:
it makes easy to locate the divergences in any amplitude as it picks out the divergences in the form of poles and exhibit meromorphic structure of
these integrals. 

$\bullet$ As a first upshot of the present paper, and for a particular class of 
TGFT models defined over Abelian groups $U(1)^{\bdel}$, we show that
a dimensional regularization procedure can be defined and yields
finite renormalized amplitudes. Under their parametric form
and by complexifying the group dimension $D\in \C$, 
amplitudes are proved to be meromorphic functions in a
extended strip  $0<\Re(D)< \bdel + \varepsilon$,
where $\varepsilon>0$ is a small parameter. 
A subtraction operator can be defined at this
stage and will provide finite amplitudes. 
Theorems \ref{theo:extan} and \ref{theo:merom}
contain our main results on this part. 
During the analysis, it appears possible 
to introduce another complex parameter
 associated with the rank $d$ of the theory.
Although, we did not address this issue here,
it is a new and interesting fact that 
another type of regularization (that
one can call  a ``rank regularization'' scheme) 
could be introduced
using the parametric amplitudes in tensor models.  This will require further investigations  elsewhere.

The parametric representation of Feynman amplitudes 
has several other interesting properties. 
For instance, it allows one to read off the so-called Symanzik polynomials. In standard quantum field theory \cite{riv} and
even extended to noncommutative field theory \cite{Krajewski:2010pt},
these polynomials satisfy particular contraction/deletion
rules like the Tutte polynomial, an important invariant
in graph theory. 

$\bullet$ As a second set of results, we 
sort the structure of the ``Symanzik polynomials''
associated with the parametric amplitudes of
any rank $d$ Abelian models (not only the ones
assumed to be renormalizable). We show that
these polynomials fail to satisfy a contraction/deletion
rule. Under specific assumptions, the first Symanzik polynomial
that we found can be mapped
onto the invariant by Krajewski and co-workers
\cite{Krajewski:2010pt}. As an interesting
feature of these polynomials, we will observe
 that they respect a peculiar ``face''-structure 
of the tensor graph. A way to stabilize the polynomials
under some recurrence rules is to fully consider
this structure and to enlarge the space 
under which one must consider the recurrence.
Given a graph $\cG$, we will consider its so-called
set of internal faces $\cF_{\inter}$ (these are closed loops). 
The new invariant that we construct is defined 
over $\cG \times \cP(\cF_{\inter})^{\times 2}\times \{\od,\ev\}^{\times 2}$, where $ \cP(\cF_{\inter})$ is
the power set of $\cF_{\inter}$ and $\{\od,\ev\}$
is a parity set. The new invariant is stable only under
 contraction operations and this result is  new to
the best  of our knowledge. Theorems \ref{theo:gencdel} and \ref{theo:tensorrecurrence}
embody our key results on this part.

This paper is organized as follows.
Section \ref{sect:sgraphs} covers definitions and terminologies associated with important graph concepts used throughout the text, for the rank $d \ge 3$ colored tensor graphs and the rank $d = 2$ ribbon graphs. 
For notations closer to our discussion, we refer to 
the survey given in Section II in \cite{Geloun:2013saa}. 
Section \ref{sect:param} presents the models including rank $d = 2$ matrix and $d \ge 3$ tensor models that we shall study. The parametric representation of the amplitudes and the new Symanzik polynomials $U^{\od/\ev}$ and $\tilde{W}$ are presented.
In Section \ref{sect:facto}, we develop dimensional regularization and renormalization of particular models presented in Section \ref{sect:param}. 
The proof of the amplitude factorization  (which is necessary for showing 
that the pole extraction is equivalent  to adding counterterms of the form of the initial theory) and  the exploration of meromorphic structure of the amplitudes in the complex dimension parameter $D$ are undertaken. 
Then, a paragraph on the subtraction operator and the
procedure leading to renormalized amplitudes is  sketched. 
Section \ref{sect:poly} explores the properties of the newly found Symanzik polynomials $U^{\od/\ev}$ and $\tilde{W}$. Then, we identify a polynomial $\cU^{\eps, \beps}$ which is an extended version of $U^{\od/\ev}$ with a stable recurrence relation based only on contraction operations on a graph.
Section \ref{concl} is devoted to a summary and perspectives
of the present work.

\section{Stranded graphs}
\label{sect:sgraphs}

Before starting the study of the parametric amplitudes
associated with Feynman graphs in tensor models, 
it is worth fixing the basic definitions of the type 
of graphs we will be analyzing in these models.

In the following, we shall give a survey of the main ingredients of two types of graphs: 

- the so-called rank $d>2$ colored tensor graphs
which we will describe only from the field
theoretical point of view 
(for a mathematical definition, we will refer to \cite{avohou});

- ribbon graphs with half-ribbons also called  
rank 2 graphs in this paper. These graphs are quite well understood
 and still  intensively investigated. 
For a complete definition of ribbon graphs, we will refer to 
one of the following standard references
\cite{bollo,bollo2, bollo3, schmu, joanna} (the last
reference offers an up-to-date survey). 
The case of ribbon graphs with half-edges or half-ribbons 
and their relation to Physics, the work by Krajewski
and co-workers \cite{Krajewski:2010pt} is seminal.
However, our notations are closer to \cite{avohou2}.

\subsection{Rank $d>2$ colored stranded graphs}
\label{subsect:graphs}

Colored tensor models \cite{color} expand in perturbation theory as colored Feynman graphs
endowed with a rich structure. 
From these colored tensor graphs, one builds another type of graphs called uncolored \cite{Gurau:2011tj}--\cite{Bonzom:2012hw}. 
These graphs will form the useful category of graphs we will be dealing with
at the quantum field theoretical level. 
In this section, we provide  a lightning  review of 
the basic definitions of objects in the above references.
Most of our illustrations focus on the rank 3 situation which
is already a nontrivial case mostly discussed in 
our following sections; we invite the reader to more illustrations in \cite{Geloun:2013saa}.

\medskip 
\noindent{\bf Colored tensor graphs.}
In a rank $d$ colored tensor model, a graph is a collection of edges or lines and vertices with an incidence relation enforced by quantum field theory rules.
In such a theory, we call graph a (rank $d$) colored tensor graph.
This graph has a stranded structure described by the following properties \cite{avohou}:

\begin{figure}[h]
 \centering
     \begin{minipage}[t]{.8\textwidth}
      \centering
\includegraphics[angle=0, width=3cm, height=1.0cm]{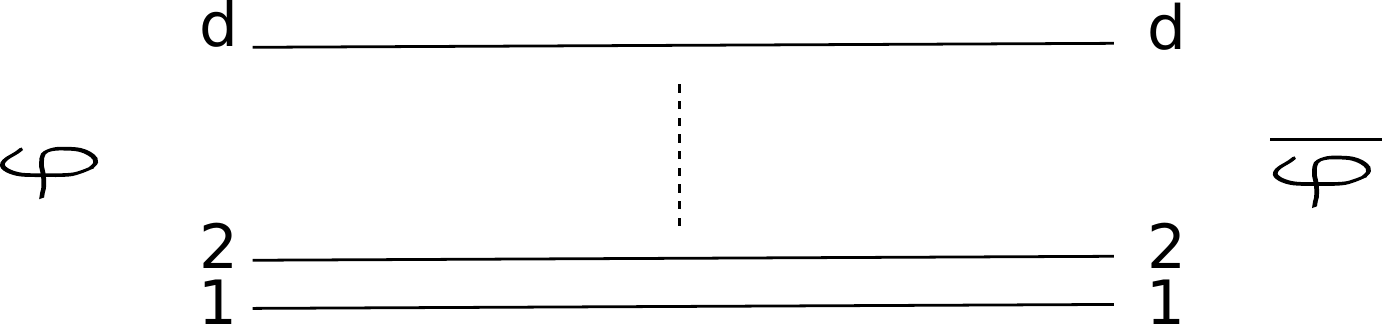}
\caption{ {\small A stranded propagator or line in a rank $d$ tensor model. }}
\label{fig:propa}
\end{minipage}
\end{figure}

- each edge corresponds to a propagator and is represented by a line with $d$ strands (see Fig.\ref{fig:propa}). Fields $\varphi$ are half-lines with the same structure;

- there exists a $(d+1)$ edge or line coloring;

- each vertex has coordination or valence $d+1$ with each leg connecting all half-lines hooked to the vertex. Due to the stranded structure at the vertex and the existence of an edge coloring, one defines a strand bi-coloring: each strand leaves a leg of color $a$ and joins a leg of color $b$, $a \ne b$, in the vertex;

\begin{figure}[h]
 \centering
     \begin{minipage}[t]{0.8\textwidth}
      \centering
\includegraphics[angle=0, width=7cm, height=2.5cm]{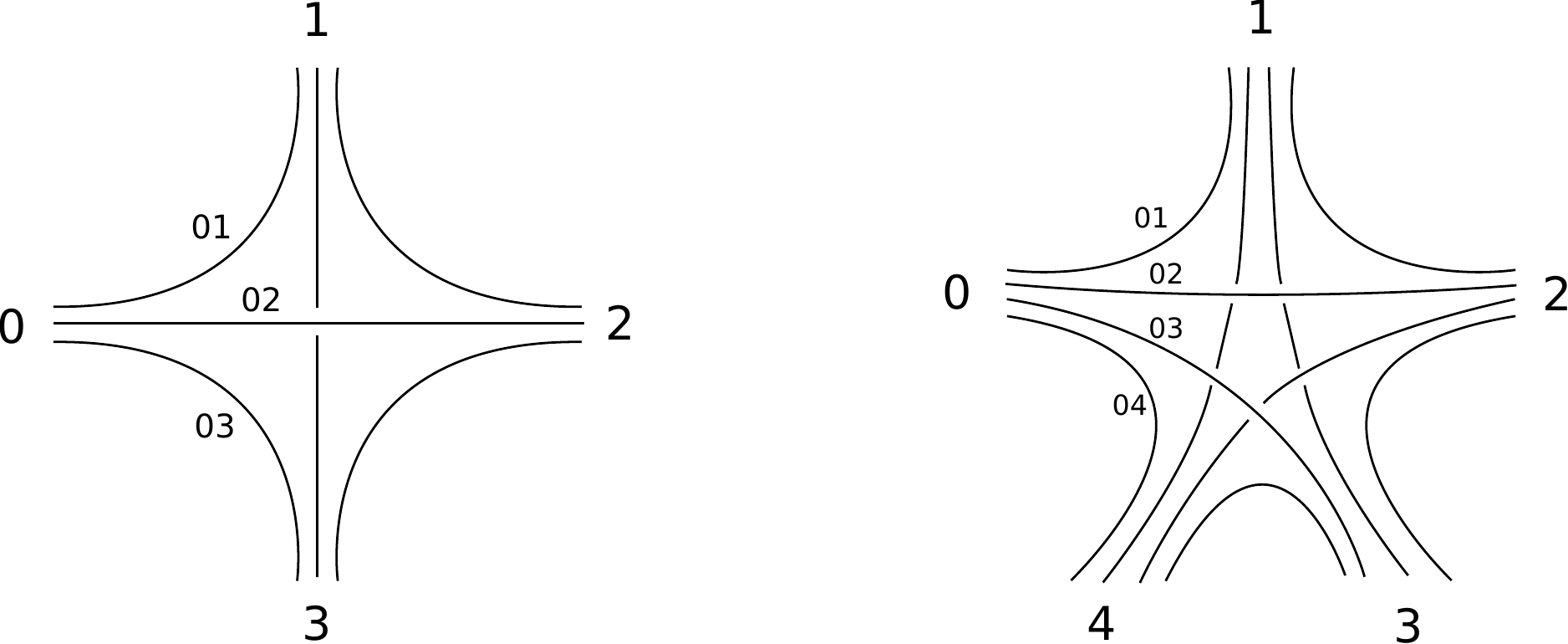}
\caption{ {\small Two vertices in rank $d=3$ (left) and $d=4$ (right) colored models.
Strands have a bi-color label. }}
\label{fig:kd1}
\end{minipage}
\end{figure}
- there are two types of vertices, black and white,
and we require the graph to be bi-partite.  
Illustrations on rank $d =3, 4$ white vertices are depicted in Fig.\ref{fig:kd1}.
Black vertices, on the other hand, are associated with barred labels 
and drawn with counterclockwise orientations. 

\begin{figure}[h]
 \centering
     \begin{minipage}[t]{0.8\textwidth}
      \centering
\includegraphics[angle=0, width=6cm, height=2.4cm]{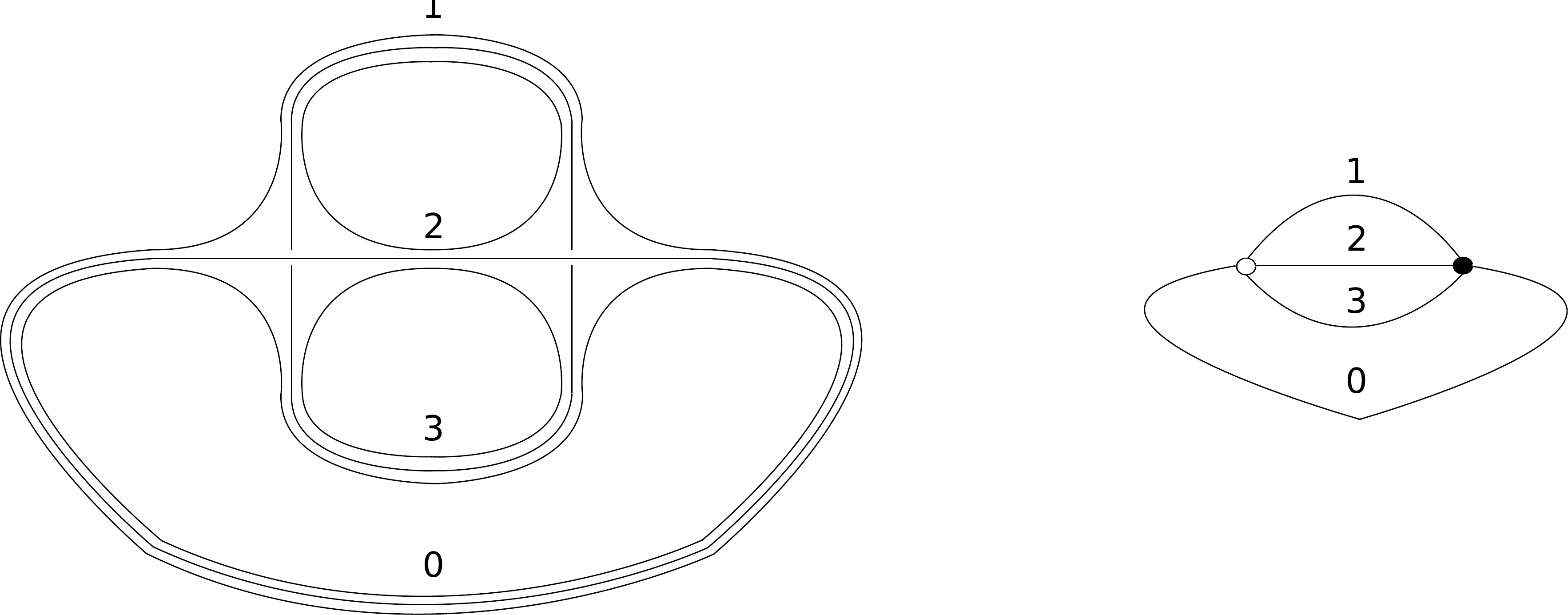}%
\caption{ {\small A rank $3$ colored tensor graph and its compact colored bi-partite representation (right).}}
\label{fig:repcomp}
\end{minipage}
\end{figure}

We may use a simplified diagram which collapses all the stranded structure into a simple colored graph. The resulting graph still captures all the information of the former. Fig.\ref{fig:repcomp} illustrates an example
of such a collapsed graph.

All rank $d$ tensor graphs (without color) have
a nice dual geometrical interpretation. The rank $d$ 
vertex determines a $d$ simplex and the fields
represent $(d-1)$ simplexes. A generic graph is
therefore a $d$ dimensional simplicial complex
obtained from the gluing of $d$ simplexes along 
their boundaries.  
The key role of colors in  tensor graphs
was put forward in \cite{Gurau:2010nd}. These colored graphs
are dual to simplicial pseudo-manifold in any dimension
$d$. 

\medskip

\begin{figure}[h]
 \centering
     \begin{minipage}[t]{0.8\textwidth}
      \centering
\includegraphics[angle=0, width=6cm, height=2.4cm]{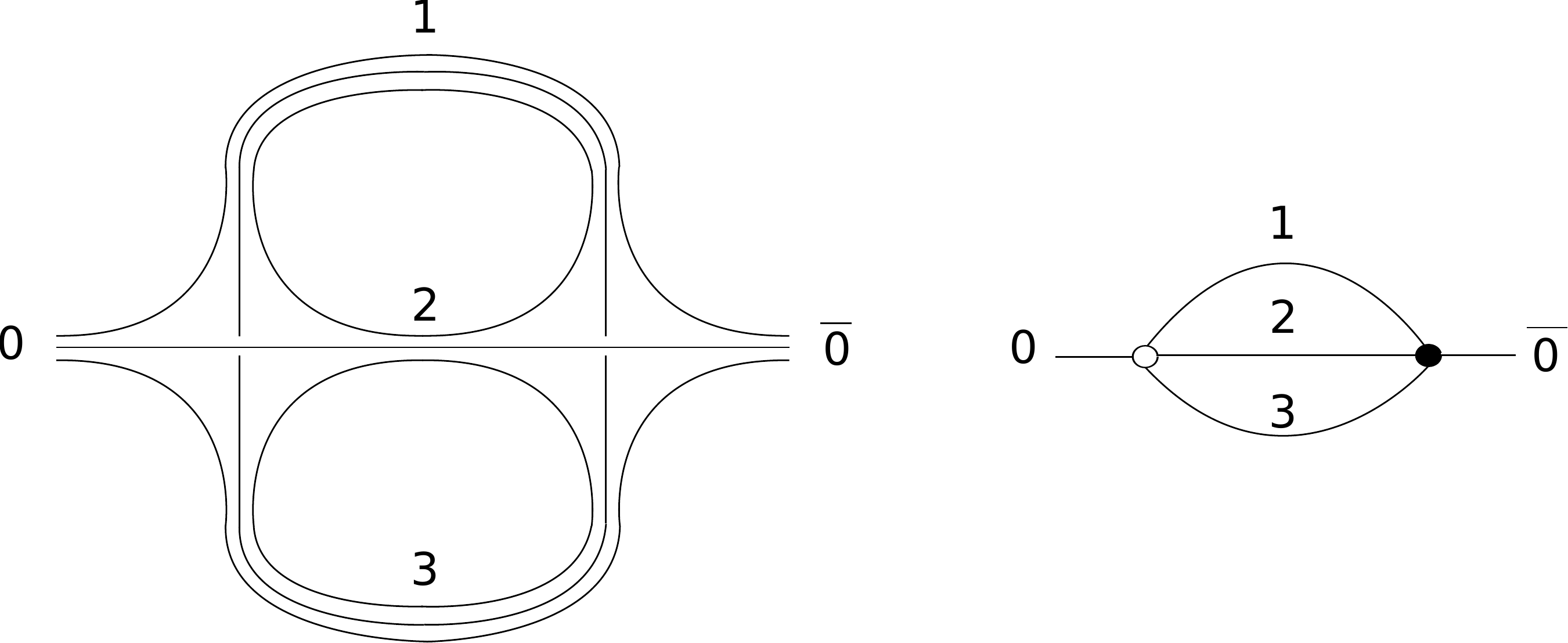}
\caption{ {\small A rank 3 open colored tensor graph and its compact representation with half-edges.}}
\label{fig:open}
\end{minipage}
\end{figure}

\noindent{\bf Open and closed graphs.}
A graph is said to be closed if it does not contain half-lines
(also called half-edges).
It is open otherwise. One refers such half-lines to external legs representing external fields in usual field theory.
We give an example of a rank 3 open graph in Fig.\ref{fig:open}.

\begin{figure}[h]
 \centering
     \begin{minipage}[t]{0.8\textwidth}
      \centering
\includegraphics[angle=0, width=6cm, height=2.4cm]{compact.pdf}%
\quad\quad
\includegraphics[angle=0, width=6cm, height=2.4cm]{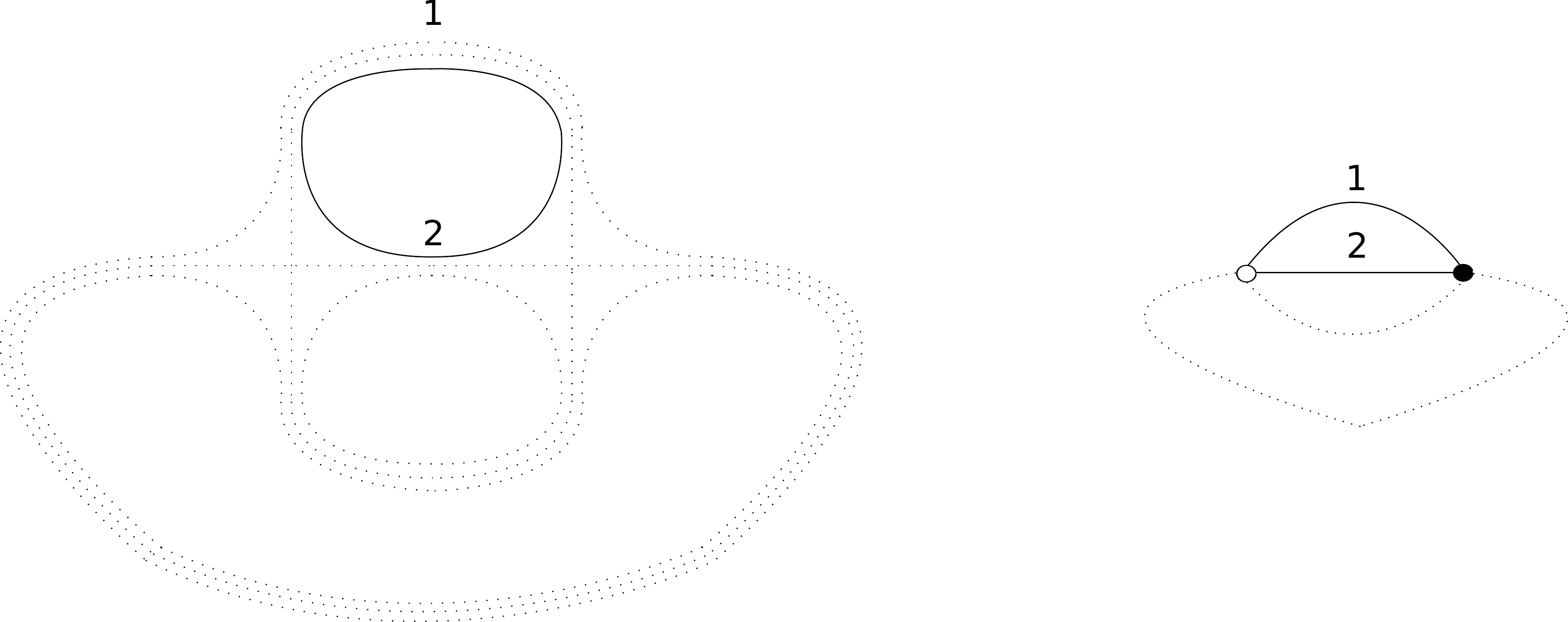}
\caption{ {\small Deleting colors $0$ and $3$ in the graph on
the left, one obtains a $2$-bubble, the face $f_{12}$ (right).}}
\label{fig:face}
\end{minipage}
\end{figure}

\medskip
\noindent{\bf $p$-bubbles and faces.}
Appearing as one of their most striking features, 
colored tensor graphs in any rank $d$ have a homological cellular structure \cite{color}.
A $p$-bubble is a maximally connected component subgraph of the collapsed colored graph associated with a rank $d$ colored tensor graph, 
with $p$ the number of colors of the edges of that subgraph.
For example, a $0$-bubble is a vertex, a $1$-bubble is a line.
A $2$-bubble is called a face.
Faces can be viewed in the simplified colored graph as cycles of edges with alternate colors, see Fig.\ref{fig:face}. They will play a major role in all of our next developments.

We have few remarks: 

- In the full expansion of the colored graph using strands, a face is nothing but a connected component made with one strand. The color of strands alternates when passing through the edges which define the face.

- A $p$-bubble is open if it contains an external half-line, otherwise it is closed. For instance, there exist three open $3$-bubbles (${\bf b}_{012}$, ${\bf b}_{013}$ and ${\bf b}_{023}$) and one closed bubble ${\bf b}_{123}$ in the graph $\cG$ in Fig.\ref{fig:open}.

\medskip

\begin{figure}[h]
 \centering
     \begin{minipage}[t]{0.8\textwidth}
      \centering
\includegraphics[angle=0, width=12cm, height=2.5cm]{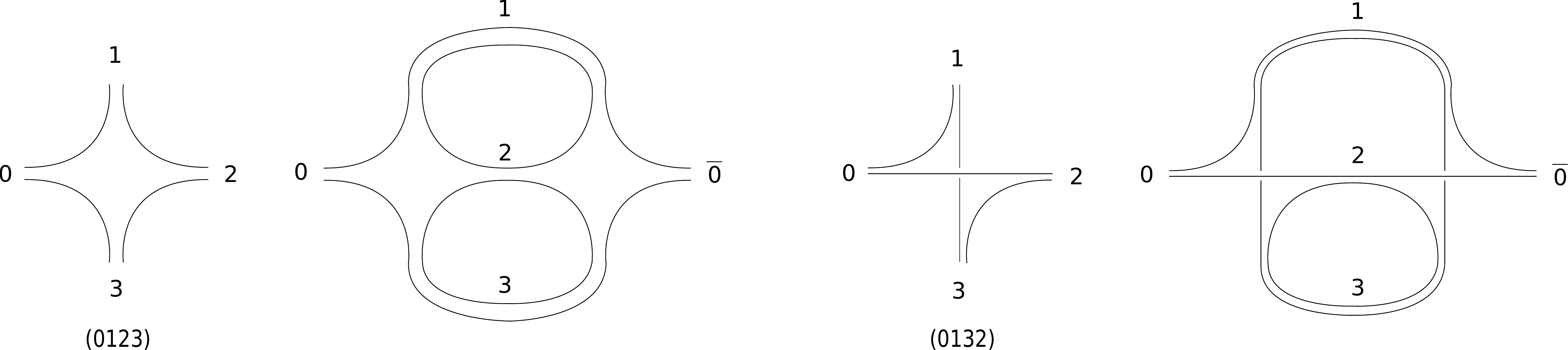}
\caption{ {\small Two open jackets, $J_{0123}$ (left) and $J_{0132}$ 
(right) of the graph in Fig.\ref{fig:open}. The subscripts stand for 
a given color cyclic permutation used to decompose
the colored tensor vertex in another-ribbon like vertex.}}
\label{fig:jacketg}
\end{minipage}
\end{figure}

\noindent{\bf Jackets.}
Jackets are ribbon graphs coming from a decomposition of a colored tensor graph. Following \cite{Gur4,Ryan:2011qm}, a jacket in rank $d$ colored tensor graph is defined by a permutation of $\{1, \cdots, d\}$ namely $(0, a_1, \cdots, a_d)$,  $a_i \in \llbracket 1, d\rrbracket$, up to orientation.  One divides the $(d+1)$ valent vertex into cycles of colors using the strands with color pairs $(0 a_1)$, $(a_1 a_2), \cdots, (a_{d-1} a_d)$  and proceed in the same way with rank $d$ edges. 
Some jackets are illustrated in Fig.\ref{fig:jacketg}. 
Open and closed jackets follow the standard definition of
having or not having external legs, respectively. 

\medskip

\noindent{\bf Boundary graphs.}
Tensor graphs with external legs are dual to simplicial complexes with boundaries. 
This boundary itself inherits a simplicial (even homological) complex structure in the context of  colored models \cite{Gurau:2009tz}. From the field theoretical 
perspective, we are interested in graphs with external
legs,\footnote{External legs allow us to probe events happening at a much higher scale as compared to the scale of their own.} therefore in the present context, in simplicial complexes with boundaries. 

One can map the boundary complex of a rank $d$ colored graph to
a tensor graph with lower rank $d-1$ endowed with a vertex-edge coloring \cite{avohou}.
The procedure for achieving this mapping is known as ``pinching'' (or 
closing of open tensor graphs):
one inserts a $d$-valent vertex at each external leg of a rank $d$ open  tensor graph. 
The boundary $\partial \cG$ of rank $d$ colored tensor graph $\cG$
is then a graph 

- the vertex set of which is one-to-one with the set of external legs of $\cG$ and is the set of $d$-valent vertices inserted;

-  the edge set of which is one-to-one with the set of open faces of $\cG$.

\begin{figure}[h]
 \centering
     \begin{minipage}[t]{0.8\textwidth}
      \centering
\includegraphics[angle=0, width=7cm, height=2.5cm]{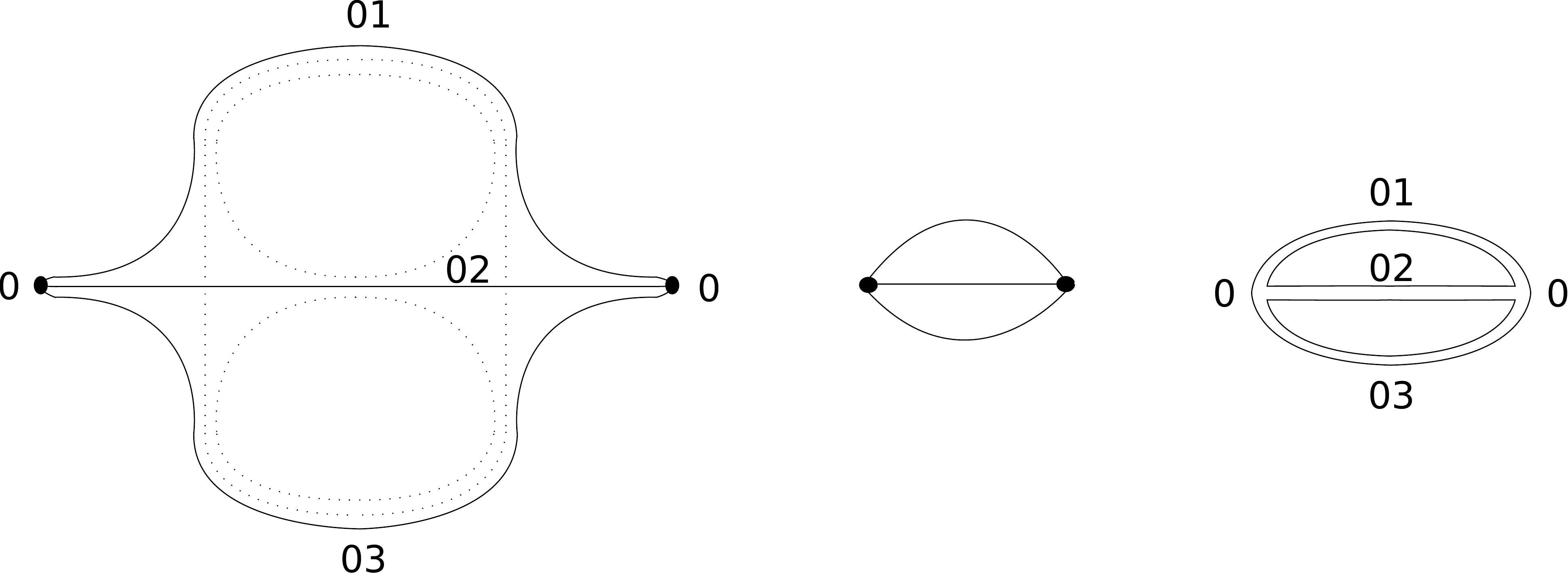}
\caption{ {\small The boundary graph $\bG$
of the graph $\cG$ of Fig.\ref{fig:open}. 
$\bG$ (graph in the middle) is obtained by inserting a $d=3$ valent vertex 
at each external leg in $\cG$ and erasing the closed internal faces.
$\bG$ has a rank $d-1=2$ structure (most right).}}
\label{fig:boundary}
\end{minipage}
\end{figure}

As a direct consequence, the boundary graph has a vertex coloring inherited from the edge coloring and has an edge bi-coloring coming from the bi-coloring of the faces of the initial graph. See Fig.\ref{fig:boundary}  as an illustration in a rank 3 colored
tensor graph.
Note, for example, that in rank $d=3$, the boundary of a rank $3$ colored tensor graph is a ribbon graph.

\medskip

\noindent{\bf Degree of a colored tensor graph.}
Organizing the divergences occurring in the perturbation series of rank $d$ colored tensor graphs, one introduces the following quantity called degree of the colored tensor graph $\cG$ \cite{Gur3,GurRiv,Gur4}
\beq
\omega(\cG) = \sum_J g_J\,,
\eeq
where $g_J$ is the genus of the jacket $J$ and the sum is performed over all jackets in the colored tensor graph $\cG$. 
For an open graph, one might use instead pinched jackets $\tJ$ 
for defining the degree. A graph for which $\omega(\cG)=0$
is called a ``melon'' or ``melonic'' graph \cite{Bonzom:2011zz}.
This quantity is at the core of the extension of the notion of genus expansion (t'Hooft large $N$ expansion in matrix models) 
now for colored tensor models. It is 
at the basis of the success of finding a way to 
analytically resum the perturbation series in colored tensor models 
at leading order and even beyond
\cite{Bonzom:2011zz}--\cite{Bonzom:2014oua}.

\begin{figure}[h]
 \centering
     \begin{minipage}[t]{0.8\textwidth}
      \centering
\includegraphics[angle=0, width=5cm, height=1cm]{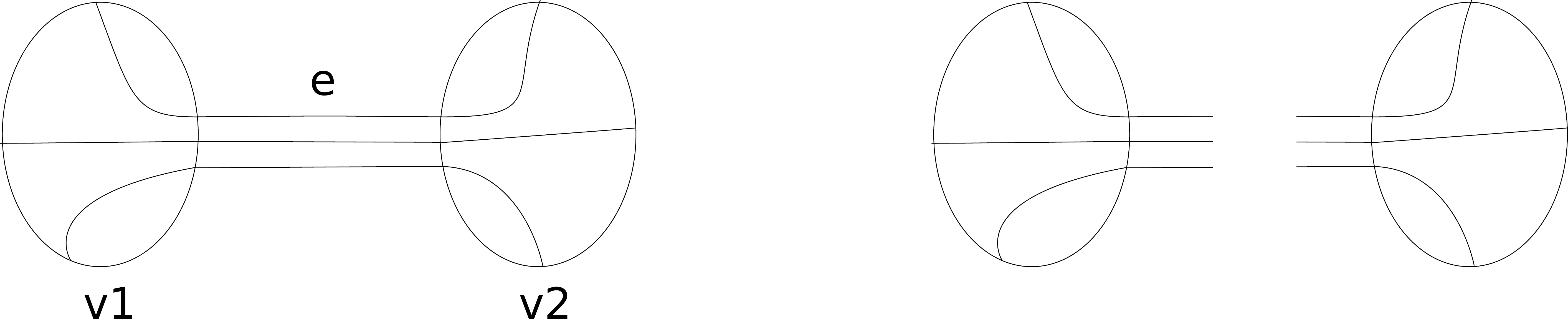}
\vspace{0.2cm}
\caption{ {\small Cut of a stranded line $e$.}}
\label{fig:cutstr}
\end{minipage}
\put(-236,-15){\tiny $\cG$}
\put(-160,-15){\tiny $\cG\vee e$}
\end{figure}

\noindent{\bf Contraction and cut of
a stranded edge.} As in ordinary graph theory, 
an edge can be regular or special (bridge and
loop). We will consider the following operations on
a tensor graph: 

- The {\bf cut operation} on an edge is intuitive:
we replace a stranded line by two  stranded half-lines
on the vertex or vertices where the edge was incident (see Fig.\ref{fig:cutstr}).
Importantly, we respect the bi-coloring of strands during the process. 
We denote $\cG \vee e$ the resulting graph after
cutting $e$ in $\cG$. We realize immediately that
cutting edges has a strong effect on the 
boundary graph. 

\begin{figure}[h]
 \centering
     \begin{minipage}[t]{0.8\textwidth}
      \centering
\includegraphics[angle=0, width=5cm, height=1.5cm]{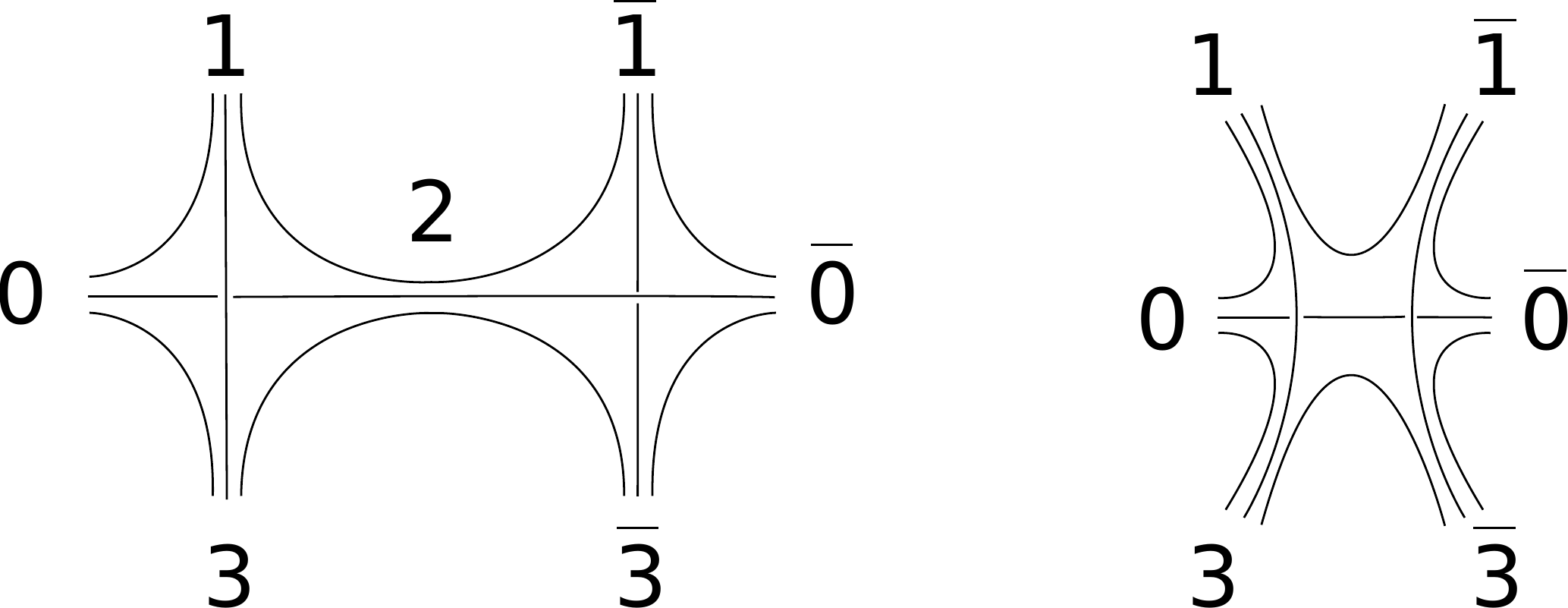}
\vspace{0.2cm}
\caption{ {\small Contraction of a stranded line 2.}}
\label{fig:strcont}
\end{minipage}
\put(-226,-15){\tiny $\cG$}
\put(-146,-15){\tiny $\cG/e$}
\end{figure}

- The {\bf contraction of a non-loop } rank $d$ stranded edge
is similar to ordinary contraction in graph theory.
The important point is, once again, to respect the stranded structure. 
The contraction of an edge $e$ incident to $v_1$
and $v_2$ is performed by removing $e$ and its end vertices and
 introduce another vertex containing all the remaining edges 
incident to $v_1$ and $v_2$ in such a way to conserve their stranded
structure and incidence relations (see Fig.\ref{fig:strcont}). Starting from a colored graph,
such an operation immediately leads  to a non colored graph.
However, the stranded structure and stranded bi-coloring
are preserved. These are the important ingredients that
we need in our next developments.

A colored graph does not have loop edges 
(a loop edge is incident to the same vertex). Thus 
our initial class of rank $d$ colored tensor graphs does not generate
any loops. However, after contractions of regular edges, it is 
easy to imagine that one might end up with configurations with loops
from a generic graph. Since
we will be interested in situations where such configurations arise and  
where we must further perform contractions,
a definition of loop contraction is required.  
In \cite{avohou}, such a contraction has been defined in
the case of a trivial loop.\footnote{After the contraction of a tree of regular edges, 
we always end up  with a generalized stranded rosette graph. 
In ribbon graphs, a loop on a
rosette is called trivial if it does not interlace with any other loops.
In stranded graphs, one might impose further conditions 
categorized by possible consequences of the contraction of these
 loops before calling them trivial.} 
We provide here a straightforward generalization
of this definition which turns out to be useful for our following study. 
For simplicity, we restrict to the rank $3$ colored case, and
the general situation can easily be recovered from this point.

\begin{figure}[h]
 \centering
     \begin{minipage}[t]{0.8\textwidth}
      \centering
\includegraphics[angle=0, width=6cm, height=3cm]{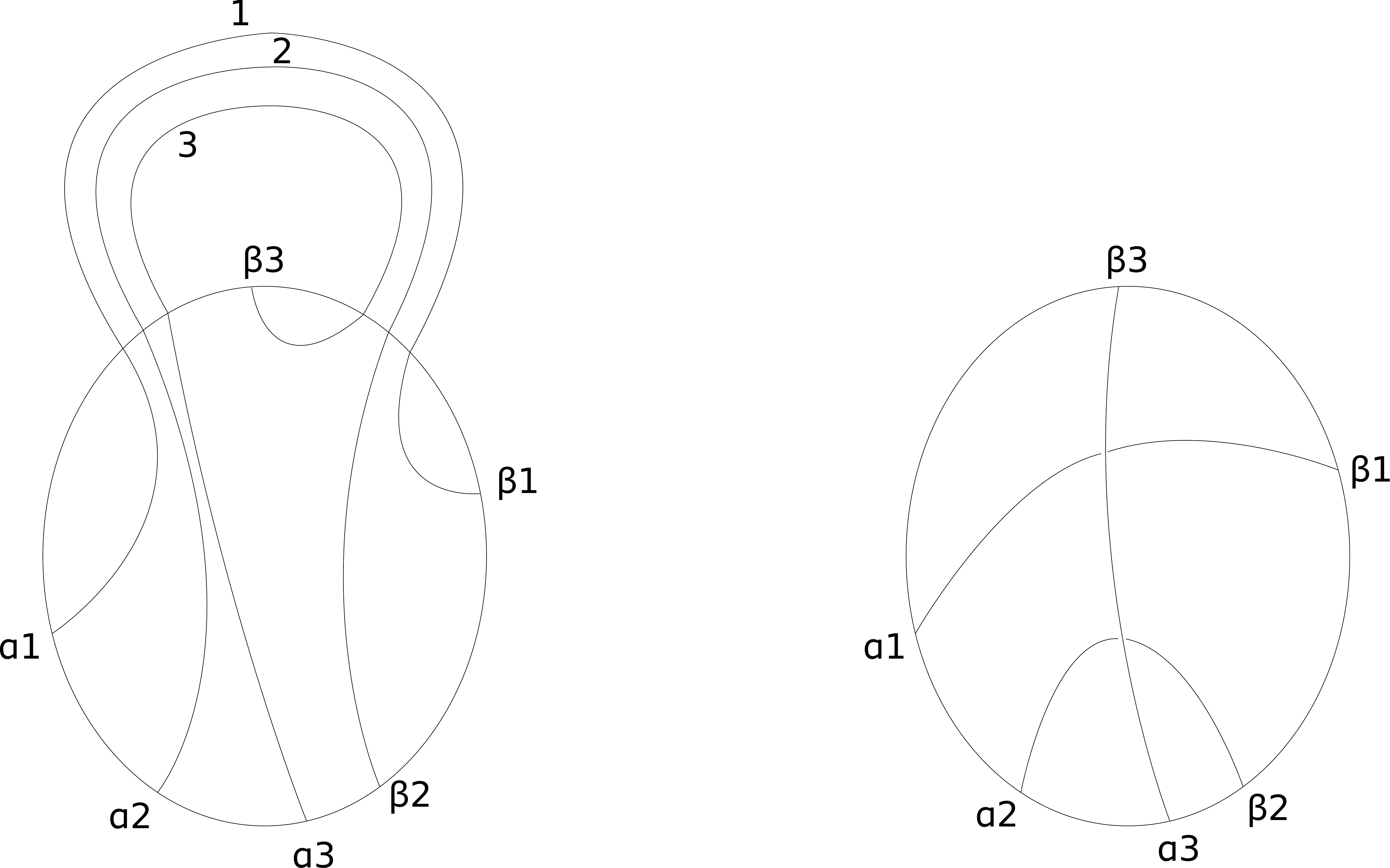}
\vspace{0.2cm}
\caption{ {\small A loop in a graph $\cG$ and its bi-colored strands $i=1,2,3$.
After contraction, in the graph $\cG/e$, the sectors $\alpha_i$ are joined with the $\beta_i$'s in the resulting vertex. }}
\label{fig:cstrloop}
\end{minipage}
\put(-246,-15){\tiny $\cG$}
\put(-142,-15){\tiny $\cG/e$}
\end{figure}

\begin{figure}[h]
 \centering
     \begin{minipage}[t]{0.4\textwidth}
      \centering
\includegraphics[angle=0, width=5.5cm, height=3cm]{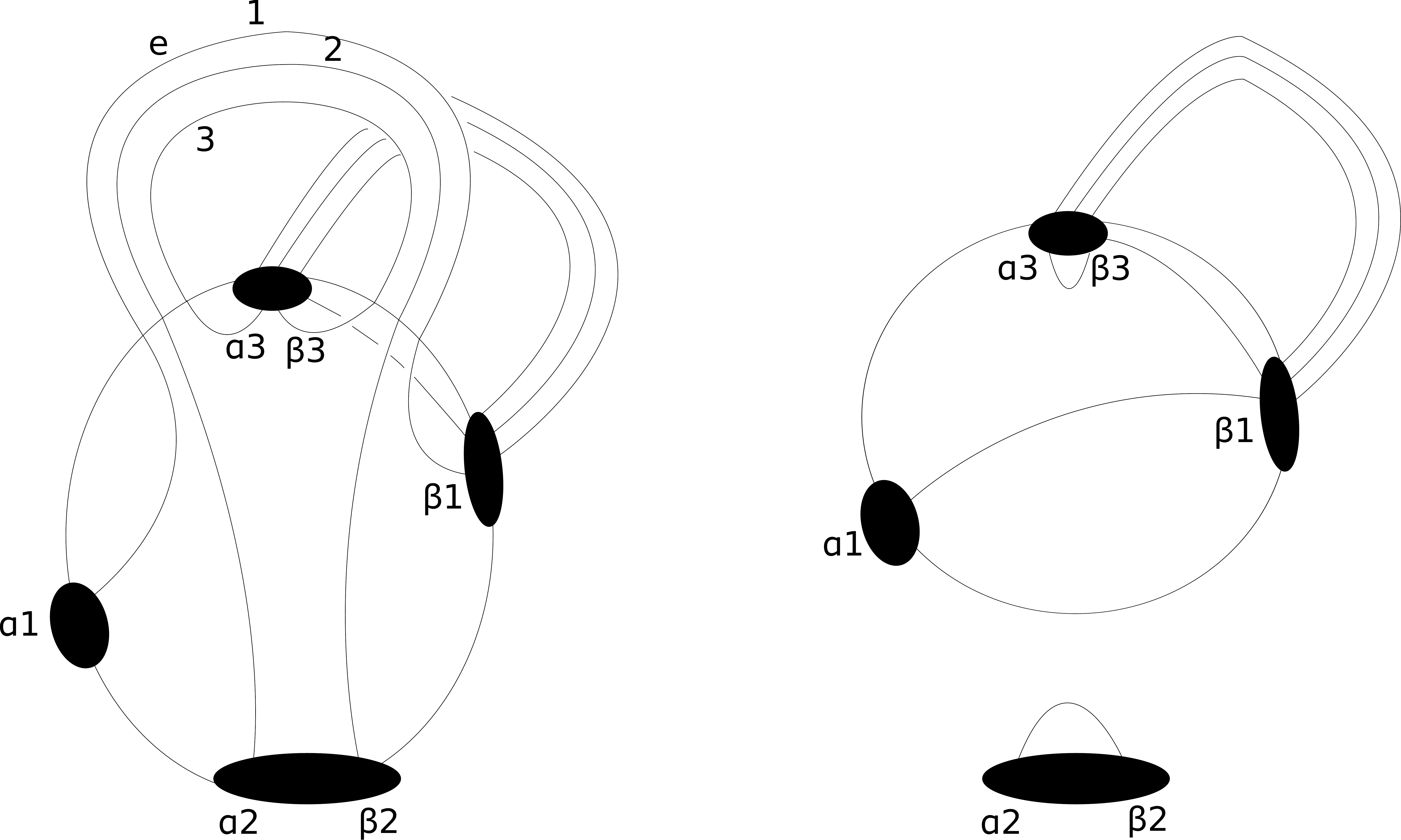}
\vspace{0.2cm}
\caption{ {\small A loop contraction: black sectors represent some parts
of the graph where the $\alpha_i$ and $\beta_i$ are connected. After
contraction, the vertex splits. }}
\label{fig:cstrloop2}
\end{minipage} \hspace{1.2cm}
\begin{minipage}[t]{0.4\textwidth}
      \centering
\includegraphics[angle=0, width=5.5cm, height=3cm]{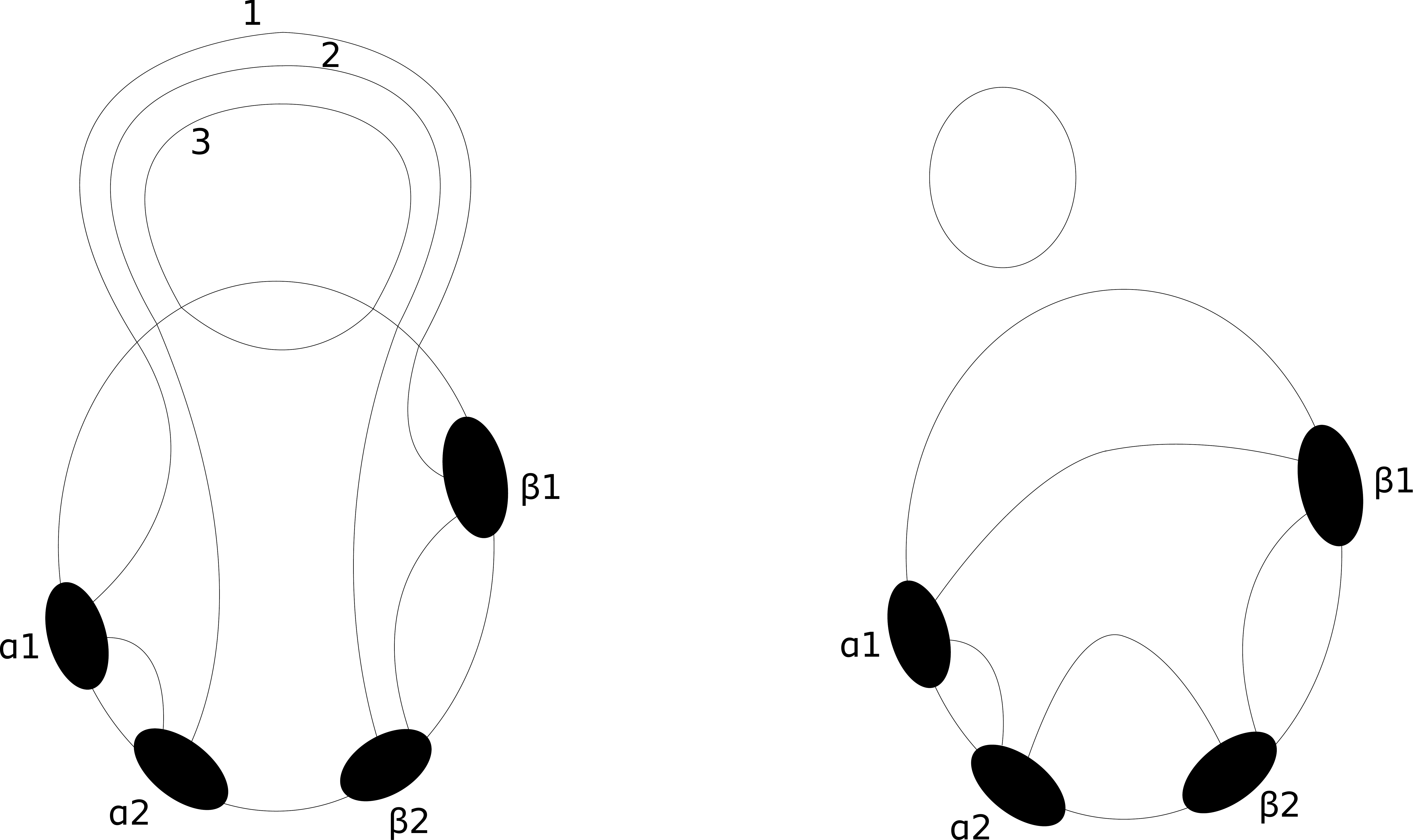}
\vspace{0.2cm}
\caption{ {\small A loop contraction: the strand 3 is 
closed and does not pass by 
any other edges. After contraction, $\cG$ splits and $\cG/e$ contains a disc. }}
\label{fig:cstrloop3}
\end{minipage}
\put(-376,-15){\tiny $\cG$}
\put(-292,-15){\tiny $\cG/e$}
\put(-150,-15){\tiny $\cG$}
\put(-54,-15){\tiny $\cG/e$}
\end{figure}

- The {\bf contraction of a loop} stranded edge:
Consider a loop edge $e$, and its bi-colored strands called $i=1,2,3$. 
Call $\alpha_i$ and $\beta_i$'s, $1\leq i\leq 3$, 
the points where the strands connect other  half-lines (or legs) of the vertex 
(see Fig.\ref{fig:cstrloop}). We write $1\leq i\leq 3$, because
it may happen that the strand $i$ does not 
exit at
another leg of the vertex but directly becomes a loop. 
Note that the $\alpha_i$'s (and $\beta_i$'s) are all pairwise
distinct by definition of a bi-colored stranded vertex. 
The contraction of $e$ incident to a vertex $v$ in $\cG$ 
is performed by removing $e$ and  directly connecting all $\alpha_i$
to $\beta_i$ with the same color index. 
Several situations may occur.  The graph might split if the resulting
parts of the vertex form themselves vertices with their incident
edges (see Fig.\ref{fig:cstrloop2}). If there is a closed strand passing through  
$e$ and $v$ only, the resulting graph, by convention, contains 
a disc issued from this closed strand (see Fig.\ref{fig:cstrloop3}).  
We will see that this procedure will extend the similar contraction in 
the case of ribbon graphs. 

The above contraction has been called ``soft'' in \cite{avohou}
as opposed to the so-called ``hard'' contraction. 
The hard contraction follows the same rules of
the soft contraction but whenever a disc graph
(without any edges) is generated during the procedure 
we remove it from the resulting graph. 
Note that hard contraction cannot be distinguished
from soft contraction on non-loop edges and
even on specific loops which do not contain these
particular closed strands. 
Hard contraction is useful in the quantum field theory setting.
However, during the study of invariant polynomials
on graph structures, considering soft contraction which preserves
the number of faces becomes capital to achieve all main results
and recurrence relations.

\subsection{Ribbon graphs}
\label{subsect:ribbon}

Let us define the type of graphs for the rank $d=2$
case that will retain our attention.

\begin{definition}[Ribbon graphs \cite{bollo}\cite{Krajewski:2010pt}]\label{def:ribbongraph}
	A ribbon graph $\cG$ is  a (not necessarily orientable) surface with boundary represented as the union of two 
sets of closed topological discs called vertices $\cV$ and edges $\cE.$ These sets satisfy the following:

$\bullet$ Vertices and edges intersect by  disjoint line segment,

$\bullet$ each such line segment lies on the boundary of precisely one vertex and one edge,

$\bullet$ every edge contains exactly two such line segments.
\end{definition}
In the following, when no ambiguity can occur, we might simply call ribbon graphs  as graphs. 

Ribbon edges can be  twisted or not and this induces 
consequences on the orientability and genus of the ribbon graph as a surface. 

Defining the class of ribbon graphs, we  take 
the point of view of Bollob\`as and Riordan \cite{bollo}. 
Arbitrary cyclic orientation (+ or -) signs on vertices are fixed, and then one assigns to each ribbon edge an orientation, + or -, according to
the fact that the orientation of its end-vertices across the edge 
are consistent or not, respectively. 
Note that flipping a vertex (or reversing its cyclic ordering)
has the effect of changing the orientation of all its incident edges except its ``loops'' (ribbon edges
incident to the same vertex).
Two ribbon graphs are isomorphic if there exist a
series of vertex flips composed with isomorphisms of cyclic
graphs \cite{bollo2} which transform one into the other. 
Now, according to the class
of ribbon graphs, only  
the parity of the number of twists matters.

The notions of regular ribbon edges and bridges are direct 
(these can be also called non-loop edges). 
The notion of loop in ribbon graphs must be clarified. 
A loop is a ribbon edge incident to the same vertex.
In particular \cite{bollo}, we say that a loop $e$ at a vertex $v$ of a ribbon graph $\cG$ is twisted if $v \cup e$ forms a M\"obius band
as opposed to an annulus for an untwisted loop. A loop $e$ is called trivial if there is no cycle in $\cG$ which can be contracted to form a loop $e'$ interlaced with $e$.

An edge is called special if it is either a bridge or a loop.  
A ribbon graph is called a terminal form when it contains only special edges.

\medskip 
\noindent{\bf Ribbon graph operations.}
Let us first address the notion of contraction and
deletion for ribbon edges: Let $\cG$ be a ribbon graph and $e$ one of its edges. 

$\bullet$ We call $\cG-e$ the ribbon graph obtained from 
$\cG$ by deleting $e$. 

\begin{figure}[h]
\centering
\begin{minipage}[t]{0.8\textwidth}
\def\svgwidth{0.9\columnwidth}
\tiny{
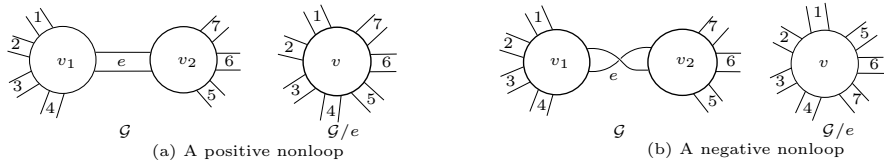
}
\caption{\small Non-loop edge contractions. 
}\label{fig:notaloop}  
\end{minipage}
\end{figure}

$\bullet$ If $e$ is not a loop and is positive, consider its
end-vertices $v_1$ and $v_2$. The graph $\cG/e$  obtained by contracting $e$ is defined from $\cG$ by replacing $e$, 
$v_1$ and $v_2$ by a single vertex disc $e\cup v_1 \cup v_2$
\cite{joanna}.
If $e$ is a negative non-loop, then untwist it (by flipping
one of its incident vertex) and  contract. 
Both contractions are illustrated in Fig.\ref{fig:notaloop}.

\begin{figure}[h]
\centering
\begin{minipage}[t]{0.8\textwidth}
\centering
\def\svgwidth{0.5\columnwidth}
\tiny{
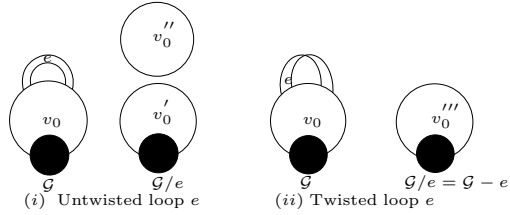
}
\caption{\small 
(i) The contraction of the untwisted trivial loop $e$ generates two separate graphs one of which is a vertex.
(ii) The contraction of the trivial twisted loop $e$ in $\cG$
is the same as its deletion. 
}
\label{fig:terminalform} 
\end{minipage}
\end{figure}

$\bullet$ If $e$ is a trivial twisted loop, 
contraction is deletion: $\cG-e = \cG/e$. 
The contraction of a trivial untwisted loop  $e$
is the deletion of $e$ and the addition of a new 
connected component vertex $v_0$ to 
the graph  $\cG-e$. We write $\cG/e = (\cG-e)\sqcup \{v_0\}$
(see Fig.\ref{fig:terminalform}). 

$\bullet$ If $e$ is general loop (not necessarily trivial), the definition of a
contraction becomes a little bit more involved.
One way to address this can be done within the framework of arrow presentations
\cite{joanna}. In the end, the result can be simply 
described as follows: 

\begin{figure}[h]
\centering
\begin{minipage}[t]{0.8\textwidth}
\centering
\def\svgwidth{0.9\columnwidth}
\tiny{
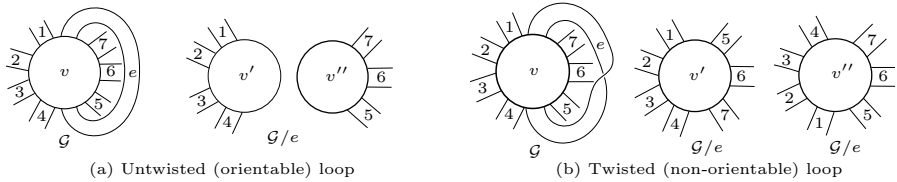
}
\caption{\small General loop contractions. 
}\label{fig:loop} 
\end{minipage}
\end{figure}
- if the loop is positive (orientable), the vertex splits into two  parts
which were previously separated by the edge $e$ in the
vertex.  
Each new vertex has the same ribbons in the same 
cyclic order that they appeared before (see Fig.\ref{fig:loop}a);

- if the loop is negative (non-orientable), then the vertex does not
split. Consider the part $\alpha$ and $\beta$ on
the vertex which 
are separated by the edge (see Fig.\ref{fig:loop}b,
$\alpha=\{1,2,3,4\}$ and $\beta=\{5,6,7\}$). 
The result of the contraction is given by the graph obtained after removing $e$ and drawing on a new vertex $v'$
the part $\alpha$ in the same cyclic order
and the part $\beta$ drawn in opposite cyclic
order. Note that using a vertex flip on $v'$, one could
achieve the equivalent vertex configuration $v''$
obtained by reversing the role of $\alpha$ and $\beta$.

\begin{figure}[h]
\centering
\begin{minipage}[t]{0.8\textwidth}
\centering
\def\svgwidth{0.52\columnwidth}
\tiny{
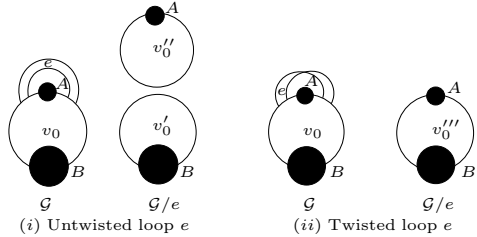
}
\caption{\small 
(i) Contraction of the untwisted $e$ in $\cG$ generates two separate graphs. 
(ii) Contraction of the twisted $e$ in $\cG$ generates one graph.
}\label{fig:terminalformAB} 
 \end{minipage}
\end{figure}

In practice, we will be interested in generic situations listed
in Fig.\ref{fig:terminalformAB}.

In this context of loop contraction, one can also introduce
the concept of hard contraction removing  extra discs generated. 
There exist other types of operations that
are useful in ordinary graph theory and extends to 
ribbon graphs. In our developments, we will only need 
the disjoint union of graphs $\cG_1 \sqcup \cG_2$
which needs no comment.

\begin{definition}[Faces \cite{bollo}]
A face is a component of a boundary of $\cG$ considered as a geometric ribbon graph, and hence as a surface with boundary. 
\end{definition}
Note that vertex graph made with one disc has one face. 

The notion of ribbon graphs being properly introduced,
we can proceed further and define an extended class of 
ribbon graphs. The class in question is called the class of ribbon graphs with half-ribbons.
In the work  by Krajewski et al. \cite{Krajewski:2010pt},   the authors called these graphs ribbon graphs with flags. 

\begin{figure}[h]
 \centering
     \begin{minipage}[t]{.8\textwidth}
      \centering
\includegraphics[angle=0, width=1.8cm, height=1cm]{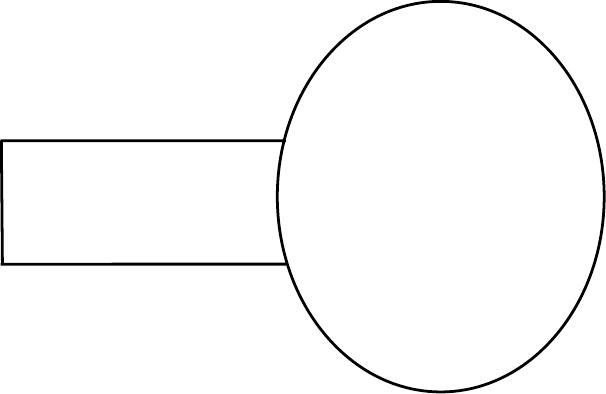}
\caption{ {\small A half-ribbon $h$ incident to one vertex disc. }}
\label{fig:halfribbon}
\end{minipage}
\put(-202,20){$h$}
\end{figure}

\begin{definition}[Half-ribbons and half-edges]
\label{ribflag}
A  half-ribbon is a ribbon incident to a unique vertex by a unique segment and without forming a loop.
(An illustration is given in Figure \ref{fig:halfribbon}.)
\end{definition}

As opposed to ribbon edges, we do not 
assign any orientation to half-ribbons. 

\begin{figure}[h]
 \centering
     \begin{minipage}[t]{.8\textwidth}
      \centering
\includegraphics[angle=0, width=7cm, height=1cm]{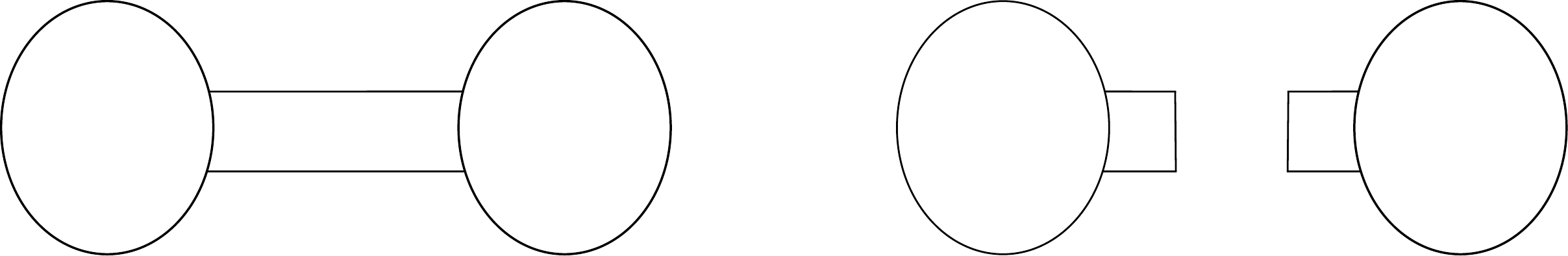}
\caption{ {\small Cutting a ribbon edge. }}
\label{fig:cut}
\end{minipage}
\end{figure}

\begin{definition}[Cut of a ribbon edge \cite{Krajewski:2010pt}]
 Let $\cG$ be a ribbon graph and let $e$ be one of its ribbon edge.
The cut graph $\cG \vee e$, is the graph obtained by 
removing $e$ and let two half-ribbons attached at the end vertices of $e$ (see Fig.\ref{fig:cut}). If $e$ is a loop, the two half-ribbons  are on the same vertex.
\end{definition}

- A half-ribbon generated by the cut of a ribbon edge
is called a half-ribbon edge, but sometimes it will
be simply referred to as half-edge. 

- A ribbon graph with half-ribbons 
is a ribbon graph together with a set of half-ribbons attached
to its discs. 

- The set of half-ribbons  is denoted by $\cH$ 
(with cardinal $HR$) and it includes the set of half-edges by $\cHE$ (with cardinal $HE$). The rest
of the half-ribbons will be called flags  and denoted by
$\cFL$ (with cardinal $FL$). Thus $\cH= \cHE \cup \cFL$. 

Precisions must be  now given on the equivalence
relation of ribbon graphs we will be working on.
First, one must extend the notion of cyclic graphs
to cyclic graphs with half-edges (the notion of
``half-edge''  in simple graph theory exists). Then  
two ribbon graphs with half-ribbons are isomorphic if there exist a series of vertex flips composed with isomorphisms of cyclic graphs with half-edges which transform one into the other. 

The cut of a ribbon edge modifies the boundary faces of 
the ribbon graph.  After the procedure,
the new boundary faces follow the contour of the half-ribbons. It is always possible to introduce
a distinction between this type of new faces and the initial
ones.  We will give a precision on this below. 

\begin{figure}[h]
 \centering
     \begin{minipage}[t]{.8\textwidth}
      \centering
\includegraphics[angle=0, width=1.3cm, height=1cm]{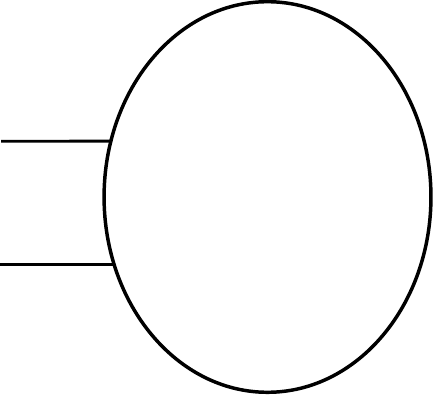}
\caption{ {\small Stranded structure of a half-ribbon. }}
\label{fig:halfribbon2}
\end{minipage}
\end{figure}

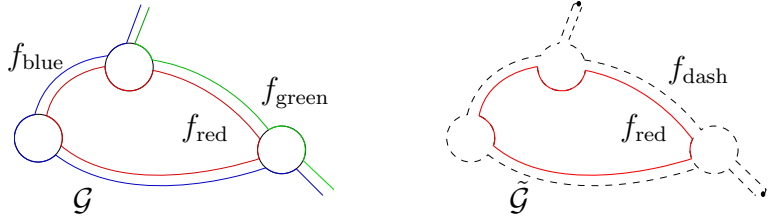
\begin{figure}[h]
 \centering
     \begin{minipage}[t]{.8\textwidth}
      \centering
\includegraphics[angle=0, width=10cm, height=2.6cm]{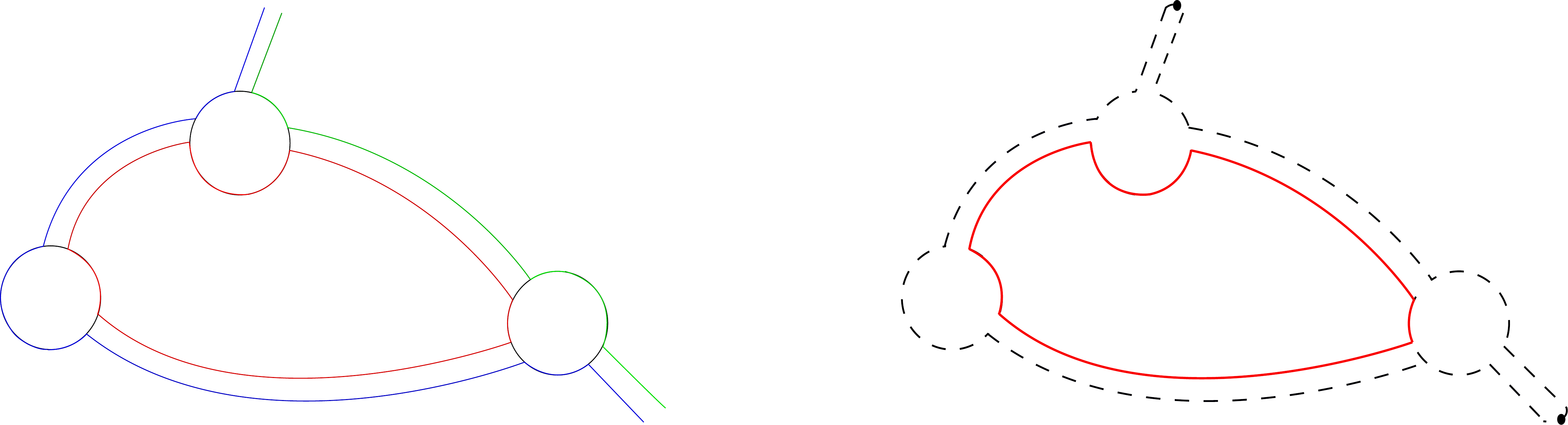}
\vspace{0.1cm}
\caption{ {\small An open ribbon graph $\cG$ with a closed face $f_{\rm red}$ and open faces $f_{\rm green,blue}$ (with suggestive labels).
The pinched graph $\tilde{\cG}$ and 
the boundary $\bG$ of $\cG$ represented in dashed lines.}}
\label{fig:ribbongraph}
\end{minipage}
\put(-269,23){$f_{\rm red}$}
\put(-239,37){$f_{\rm green}$}
\put(-335,50){$f_{\rm blue}$}
\put(-310,-5){$\cG$}
\put(-85,45){$f_{\rm dash}$}
\put(-103,23){$f_{\rm red}$}
\put(-145,-5){$\tilde{\cG}$}
\end{figure}

As defined in  Section \ref{subsect:graphs},
the notion of open and closed graphs and their constituents (forgetting the coloring) can be also addressed here. A closed
ribbon graph does not have half-ribbons, otherwise it
is called open. To harmonize our notations with Section \ref{subsect:graphs} and make transparent the link with the above
tensor models, we will explicitly draw half ribbons as two parallel strands,
see Fig.\ref{fig:halfribbon2}.
We can now introduce a definition for closed or open face as simply closed or open strand, respectively. The notions of pinched
and boundary graphs find equivalent notions
in ribbon graphs. We will refrain to
introduce  more definitions at this point (Fig.\ref{fig:ribbongraph}
 illustrates an open ribbon graph, with open and
closed faces, its pinched and boundary graph). 

\section{Parametric representation of amplitudes}
\label{sect:param}

We start by reviewing our notations for tensor models. 
From the following subsection, we present
new results on the parametric form of the amplitudes
of these models. 

\subsection{Abelian rank $d$ models}
\label{subsect:models}

Consider a rank $d\geq 2$ complex field $\varphi$ over the Lie
group $G_D= U(1)^D$, $D\in \N\setminus \{0\}$, $\varphi: (G_D)^d \to \C$, decomposed in Fourier 
components as 
\beq
\varphi(h_1,h_2,\dots,h_d)
= \sum_{P_{I_s}} \tilde\varphi_{P_{I_1},P_{I_2},\dots,P_{I_d}} D^{P_{I_1}}(h_1) D^{P_{I_2}}(h_2) \dots D^{P_{I_d}}(h_d) \,,
\label{field}
\eeq
where  $h_s \in G_D$. The sum is performed over all values of momenta $P_{I_s}$. $P_{I_s}$ are labeled by multi-indices $I_s$, with $s=1,2,\dots,d,$ where $I_s$ defines the representation indices of the group element $h_s$ 
in the momentum space. $D^{P_{I_s}}(h_s)$ plays the role of the plane wave in that representation. More specifically, one has 
\bea
&&
h_s=(h_{s,1},\dots, h_{s,D}) \in G_D\,,\;\,
h_{s,l} = e^{i \theta_{s,l}}\in U(1)\,, \;\,
 D^{P_{I_s}}(h_s) = \prod_{l=1}^D e^{i p_{s,l} \theta_{s,l}}\,,\;\, p_{s,l}\in \Z\crcr
&&
P_{I_s} = \{p_{s,1},\dots,p_{s,D}\}\,,\;\,
I_s = \{(s,1),\dots,(s,D)\}\,.
\label{repu1}
\eea
Concerning the tensor $\tilde \varphi$, we will simply use the notation 
$\varphi_{[I]}:=\tilde\varphi_{P_{I_1},P_{I_2},\dots,P_{I_d}}$, where
the super index $[I]$ collects all momentum labels, i.e. $[I]=\{I_1,I_2,\dots,I_d\}$.  Note that no symmetry under permutation of  the arguments is assumed for  $\varphi_{[I]}$. We rewrite \eqref{field} 
in these shorthand notations as
\beq
\varphi(h_1,h_2,\dots,h_d) = \sum_{P_{[I]}} \varphi_{[I]}
D^{I_1}(h_1) D^{I_2}(h_2) \dots D^{I_d}(h_d) \,,\;\, 
D^{I_s}(h_s):= D^{P_{I_s}}(h_s)\,.
\eeq
Restricting to $d=2$, $\varphi_{I_1,I_2}$ will be referred to a matrix.  

\medskip 

\noindent{\bf Kinetic term.}
Upon writing an action, we  must define a kinetic term and, in the present
higher rank models, several interactions. 
In the momentum space, we define as kinetic term for our model
\beq
S^{\kin} =
 \sum_{P_{[I]}}
\bar\varphi_{P_{[I]}}
\Big(\sum_{s=1}^d |P_{I_s}| + \mu\Big)\varphi_{P_{[I]}}\,,\qquad
 |P_{I_s}| := \sum_{l=1}^D |p_{s,l}|\,, 
\label{skin}
\eeq
where the sum is performed over all values
of the momenta $p_{s,l}\in \Z$ and $\mu\geq 0$
is a mass coupling constant. 

In direct space formulation, 
the term \eqref{skin} corresponds to a kinetic term  defined by $\sum_s |\Delta_s|^{\frac12} + \mu$ and acts 
on the field $\varphi$. 
The non-integer power of the Laplacian can be motivated
from several points of view: 
\begin{enumerate}

\item[(i)] With the exact power of momentum in the propagator,
there exist  rank $d$ models that are renormalizable among
which we have a rank 3 tensor model and several matrix models
\cite{Geloun:2013saa}. They will be the prototype
models on which our following dimensional regularization
procedure will be applied. 

\item[(ii)] From axiomatic quantum field theory, models 
with $\Delta^{a}$, where $a \in (0,1] $ are susceptible 
to be Osterwalder-Schrader positive \cite{Rivasseau:2013uca,
Rivasseau:2014ima}. 

\item[(iii)] To the above significant features, we add
the fact that,  with this power of the momenta, the parametric
amplitudes of the models find a summable and tractable formula
with interesting properties worth to be investigated in 
greater details. 

\end{enumerate}

Passing to the quantum realm, we introduce a Gaussian measure on the tensor fields as $d\nu_C(\varphi,\bar\varphi)$ with a covariance given by
\beq
C[\{P_{I_s}\},\{\tilde P_{I_{s}}\}] = \Big[\prod_{s=1}^d \delta_{P_{I_s},\tilde P_{I_{s}}} \Big] \left(\sum_{s=1}^d  |P_{I_s}| + \mu\right)^{-1}\,,
\label{chew0}
\eeq
such that,  $\delta_{P_{I_s},\tilde P_{I_{s}}} := \prod_{l=1}^D \delta_{p_{s,l},\tilde p_{s,l} }$. 
Using the Schwinger trick, the covariance can be 
recast as
\beq
C[\{P_{I_s}\},\{\tilde P_{I_{s}}\}] =\Big[\prod_{s=1}^d \delta_{P_{I_s},\tilde P_{I_{s}}} \Big]\int_0^{\infty} d\alpha\,
 e^{-\alpha(\sum_{s=1}^d  |P_{I_s}|^a + \mu^2)}\,.
\label{chew}
\eeq
The propagator is represented by a line made as a collection of  $d$
strands, see Fig.\ref{fig:propa}. 

\medskip 

\noindent{\bf Interactions.} Depending on the rank $d$,
two types of interactions dictated by the possible notions
of invariance will be discussed.

\begin{figure}[h]
 \centering
     \begin{minipage}[t]{.8\textwidth}
      \centering
\includegraphics[angle=0, width=10cm, height=5cm]{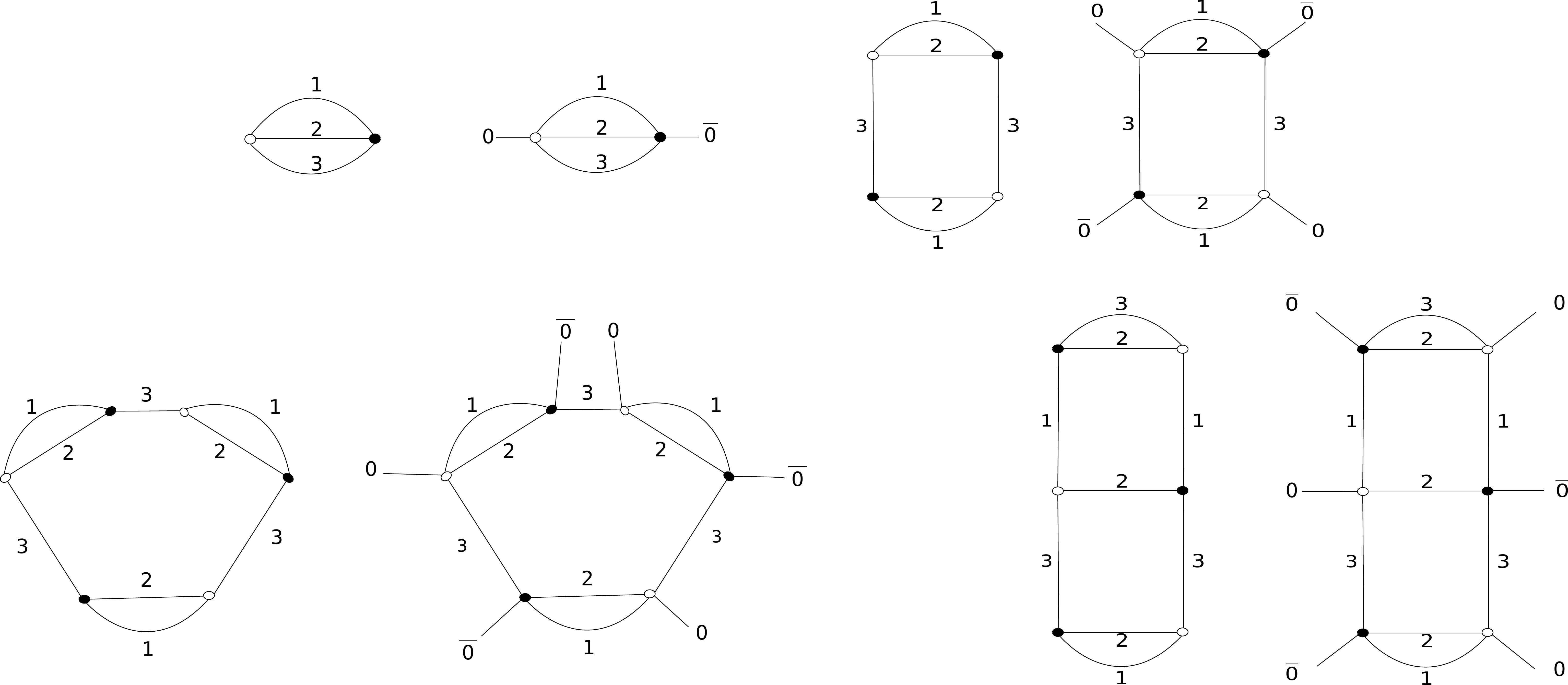}
\vspace{0.1cm}
\caption{ {\small Colored $3$-bubbles and their corresponding tensor invariants (in compact representation):
The tensor fields are $0$ and $\bar 0$ and are contracted
according to the pattern of the 3-bubble they are
associated with.}}
\label{fig:tensinv}
\end{minipage}
\end{figure}

- In rank $d\geq 3$: the interactions of the models considered are effective interactions obtained after integrating $d$ colors in the rank $d+1$ colored tensor model \cite{Gurau:2011tj} as discussed in 
Section \ref{subsect:graphs} (for a complete discussion, we
refer to \cite{Geloun:2013saa}). The above field $\varphi$
is nothing but the remaining field $\varphi^0=\varphi$.  
An interaction term is defined from unsymmetrized tensors as 
unitary tensor invariant objects and built from the particular convolution of arguments of some set of tensors $\varphi_{[I]}$ and $\bar\varphi_{[I']}$. 
Such a contraction is performed only between the $s^{th}$ label
of some $\varphi_{[I]}$ to another $s^{th}$ label of some $\bar\varphi_{[I']}$. It turns out that the total contraction 
of these tensors follows the pattern of a 
connected $d$-colored graphs called $d$-bubbles denoted ${\mathbf{b}}$ (we recall 
that $p$-bubble were introduced in Section \ref{subsect:graphs};
see Fig.\ref{fig:tensinv}). 

 In rank $d\geq 3$, a  general interaction can be written: 
\beq
S^{\inter}(\varphi,\bar\varphi) = \sum_{{\mathbf{b}} \in {\mathcal B}} \lambda_{{\mathbf{b}}} I_{{\mathbf{b}}}(\varphi,\bar\varphi)\,,
\label{intergen}
\eeq
where the sum is over a finite set ${\mathcal B}$ of rank 
$d$ colored tensor bubble graphs and  $\lambda_{{\mathbf{b}}}$ is a coupling constant associated with that interaction. 
To each $I_{{\mathbf{b}}}(\varphi,\bar\varphi)$ corresponds
a vertex operator identifying incoming and outgoing momenta
and is of the form of a product of delta functions. 
In Fig.\ref{fig:tensinv}, we have illustrated some of these tensor
invariants in rank $3$ models. 

\begin{figure}[h]
 \centering
     \begin{minipage}[t]{.8\textwidth}
      \centering
\includegraphics[angle=0, width=5cm, height=2.5cm]{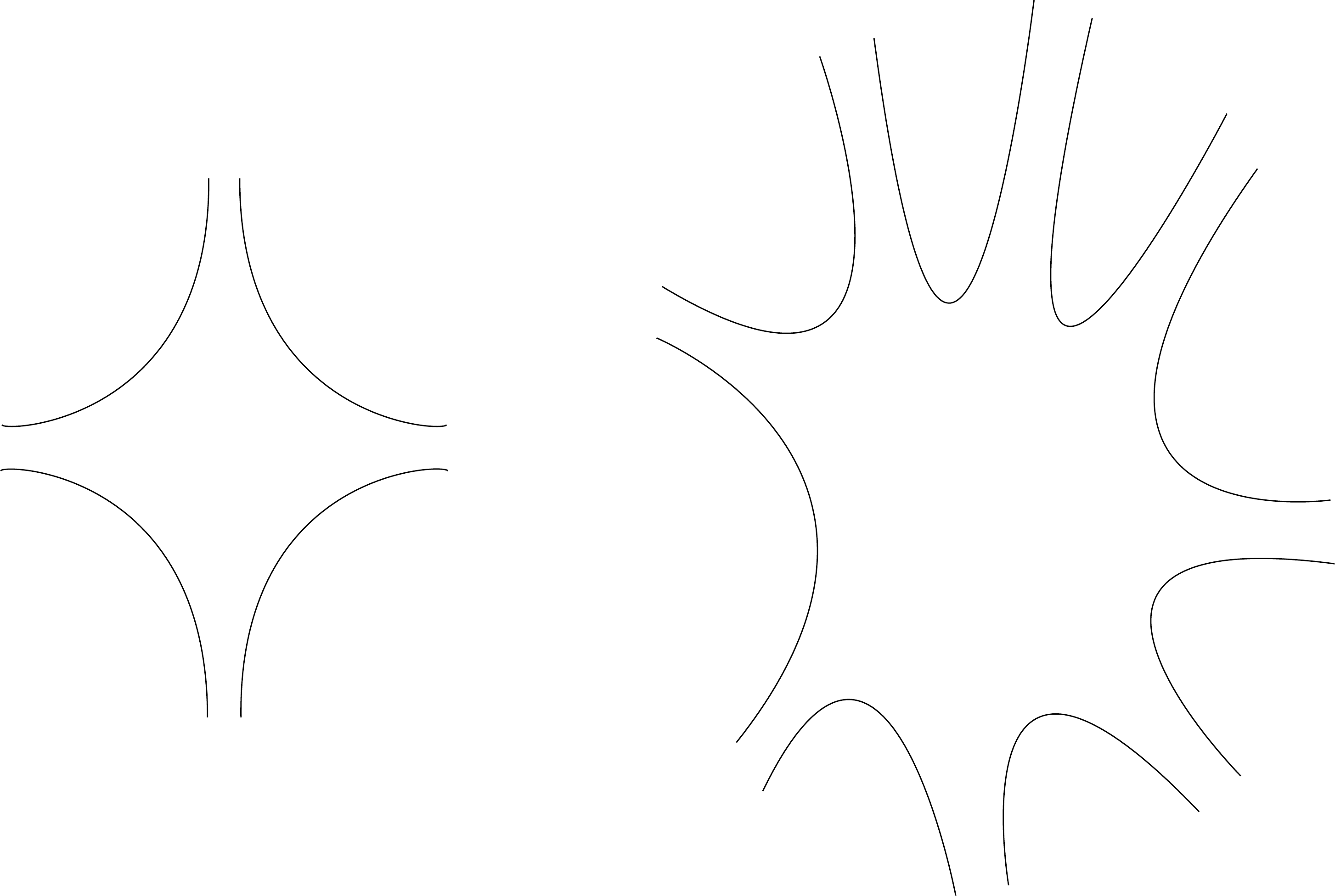}
\vspace{0.1cm}
\caption{ {\small Examples of matrix model cyclic invariants: $ {\text{tr}}(\varphi^4)$
(left) and $ {\text{tr}}(\varphi^8)$ (right). }}
\label{fig:matrixInv}
\end{minipage}
\end{figure}

\begin{figure}[h]
 \centering
     \begin{minipage}[t]{.8\textwidth}
      \centering
\includegraphics[angle=0, width=5cm, height=2.5cm]{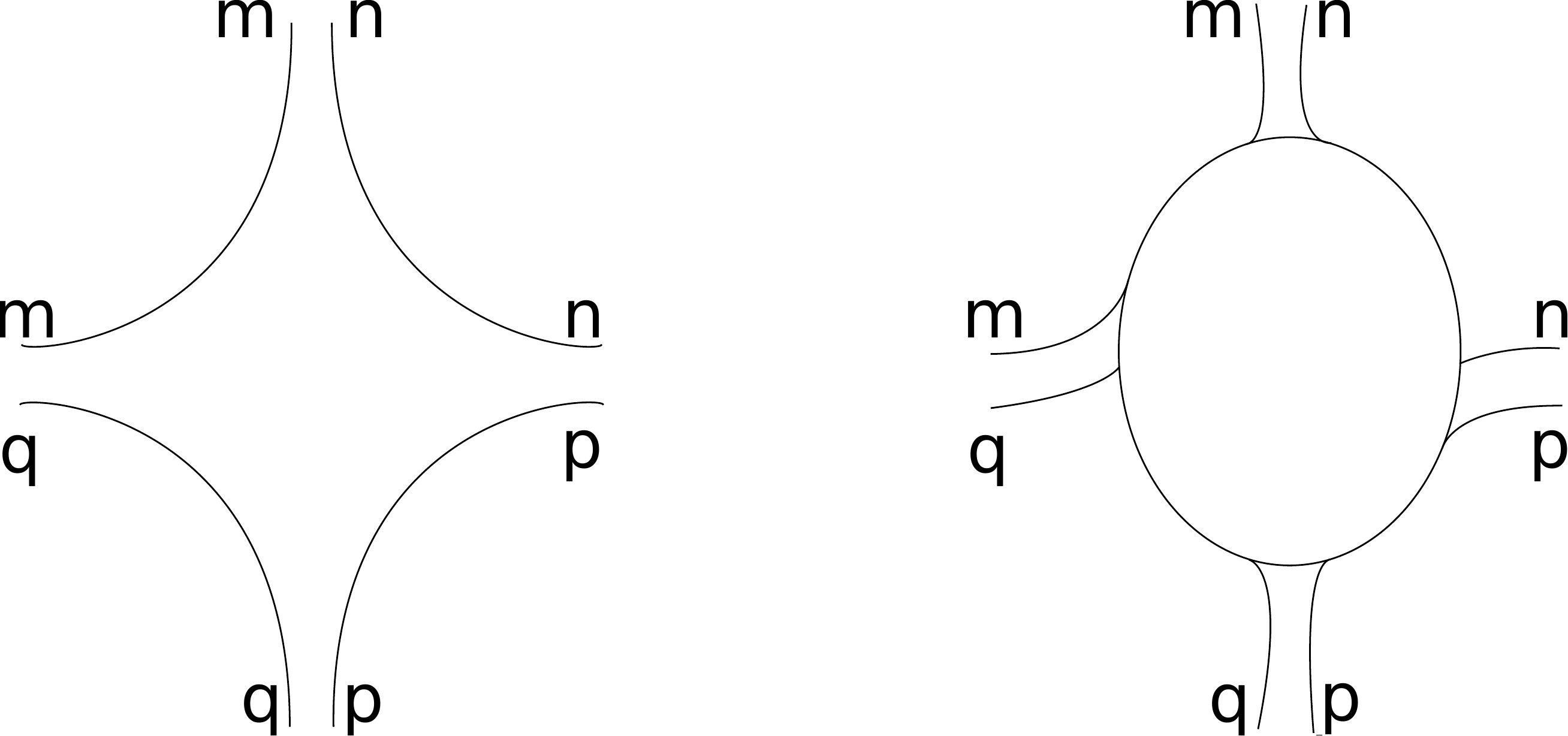}
\vspace{0.1cm}
\caption{ {\small A $ {\text{tr}}(\varphi^4)$ vertex as a disc with half-ribbons and the set of data. }}
\label{fig:quad}
\end{minipage}
\end{figure}

- In rank $d=2$ or matrix  models, the interactions are
simply trace invariants in the ordinary  sense:
\beq
S^{\inter}(\varphi,\bar\varphi)=\sum_{p=2}^{p_{\max}}\lambda_p S^{\inter}_p(\varphi,\bar\varphi)\,,
\qquad 
 S^{\inter}_p(\varphi,\bar\varphi) = {\text{tr}} [(\bar\varphi\varphi)^{p}]\,,
\label{matrixinter}
\eeq
where $\lambda_p$ stands for a coupling constant. 
Graphically, each term in \eqref{matrixinter} is represented by a cyclic graph with $p$ external legs, see Fig.\ref{fig:matrixInv}.
One might wonder how the graphs obtained in matrix models
relate to the ribbon graphs with flags explained earlier
in Section \ref{subsect:ribbon}. The answer to this is simple 
since one maps the vertices of matrix models to discs with half-ribbons
(see Fig.\ref{fig:quad}) whereas propagators are viewed as ribbon lines.
 In order to achieve the mapping, one must attach the 
 vertex/propagator data to the abstract discs with half ribbons
and ribbon lines. 

\subsection{Parametric amplitudes}
\label{subsect:param2}

 The partition function of any models described 
above  is of the form
\bea
Z = \int d\nu_C(\varphi,\bar\varphi) \,e^{-S^{\inter}(\varphi,\bar\varphi)}\,,
\label{part}
\eea
where $C$ is given by \eqref{chew} and $S^{\inter}$ given 
either by \eqref{intergen} for rank $d\geq 3$ or by \eqref{matrixinter} in the case $d=2$.

As it is in the ordinary case, Feynman amplitudes are obtained  from  Wick's theorem. 
We compute for any connected graph 
$\cG$ made with the set $\cL$ of lines and the set $\cV$ of vertices,  the amplitude
\beq
A_{\cG} = \lambda_{\cG}\sum_{P_{[I](v)}} \,\prod_{\ell \in \cL}
C_{\ell}[\{P_{I_s(\ell);\, v(\ell)}\},\{\tilde P_{I_{s}(\ell);\, v'(\ell)}\}] \prod_{v\in \cV;s}
\delta_{P_{I_s;\,v};P'_{I_s;\, v}}\,,
\label{ampli}
\eeq
where $\lambda_{\cG}$ incorporates all coupling constants
and the symmetry factors, and where
 the sum is performed over all values of the momenta $P_{[I](v)}$ associated
with vertices $v$ on which the propagator lines are incident. The propagators $C_\ell$ possess line labels $\ell \in \cL$. 

Due to the fact that vertex operators and propagators
are product of delta's enforcing conservation of momenta
along a strand, the amplitude \eqref{ampli} factorizes in terms of 
connected strand components (faces) of the graph.
There exist two types of faces: 
 open faces  the set of which will be denoted by $\cF_{\ext}$ 
(with cardinal $F_{\ext}=|\cF_{\ext}|$) and closed faces (or closed strands) the set of which will be denoted by $\cF_{\inter}$ (with cardinal $F_{\inter}=|\cF_{\inter}|$). 
Evaluating  \eqref{ampli} using \eqref{chew}, one gets
\beq
A_{\cG} = \lambda_{\cG} \sum_{P_{I_f}} \int \Big[\prod_{\ell \in \cL} d\alpha_\ell\Big]\Big\{
\prod_{f \in \cF_{\ext}}\Big[
e^{-(\sum_{\ell \in f}\alpha_\ell) |P^{\ext}_{I_f}|} \Big]
\prod_{f \in \cF_{\inter}}
\Big[ 
e^{-(\sum_{\ell \in f}\alpha_\ell) |P_{I_f}|} \Big] 
\Big\} \,,
\label{amfa}
\eeq
where $P^{\ext}_{I_f}$ are external momenta (not summed and labeled
by external faces) and the sum is over all values of internal momenta $P_{I_f}$ (indexed by internal faces). 

 It turns out that, from the linear dependency in 
momenta of the propagator, all momentum dependency
in the amplitude can be summed. The following proposition holds:
\begin{proposition}
\label{prop:propamp}
Let $\cG$ be graph, $\cL_{\cG}$ 
its set of lines, $\cF_{\inter;\,\cG}$ its set of internal faces, $\cF_{\ext;\,\cG}$ its set of external faces,
we denote  the cardinal $L_\cG = \vert \cL_\cG \vert$. 
Then, the amplitude $A_\cG$ of $\cG$ is given by
\beq
A_\cG = c\, \lambda_{\cG} \int_{[0,1]^{L_{\cG}}} \; \Big[\prod_{l \in \cL_\cG} d t_l  {\left( 1- t_l \right)^{\mu -1} \over \left( 1 + t_l\right)^{\mu +1} } \Big] \; { W_\cG ( \{m_f\}; \{ t_l \} ) \over \left[ U^{\od}_\cG ( \{ t_l\} ) \right]^{D}} \,, 
\label{eq:defa}
\eeq
where $c=2^{L_{\cG}}$ is an inessential factor, and
\bea
&&
W_\cG ( \{m_f\}; \{ t_l \} ) 
= 
\tilde{W}_\cG ( \{m_f\}; \{ t_l \} )
  \Big[U_{\cG}^{\ev}\left(\{t_l\} \right) \Big]^{D},
\label{eq:defv} \\\cr
&&
\tilde{W}_\cG ( \{m_f\}; \{ t_l \} ) = 
\prod_{f \in \cF_{\ext;\,\cG} } \Big(A_f^{\ext}\Big) ^{|m_f|} \,,\qquad
A_f^{\ext} \equiv \prod_{l \in f} \frac{ 1 - t_l}{1 + t_l} \,,\crcr
&&
U_{\cG}^{\ev/\od}\left(\{t_l\} \right)
 = \prod_{f \in \cF_{\inter;\,\cG}}  \; A^{\ev/\od}_f \,,
\qquad 
A^{\ev/\od}_f
\equiv
\sum_{\shortstack{ $_{A \subset f}$ \\ $_{|A| \; {\rm even/odd} }$} } \prod_{l \in A}  t_l \,, \label{eq:defu}  
\eea
where $m_f$ is the external momentum associated with an open face $f$. 
\end{proposition}

\proof  For any connected graph $\cG$, we re-write 
the amplitude \eqref{amfa} as 
\beq
A_\cG 
= \lambda_\cG   \int  [\prod_{l \in {\mathcal {L}}_\cG} d \alpha_l e^{ - \, \alpha_l \mu}] 
\Big[\prod_{f \in \cF_{\ext;\,\cG}}  e^{-\left( \sum_{l \in f} \alpha_l \right)   |m_f|}\Big]
 \prod_{f \in \cF_{\inter;\,\cG}} \Big[  
\frac{ 1+ e^{-\left( \sum_{l \in f} \alpha_l  \right) }}{ 1 - e^{-\left( \sum_{l \in f} \alpha_l \right) }} 
 \Big]^{D}.
\label{eq:inacg}
\eeq
 Now, we change variable as 
\beq
t_l = \tanh { \alpha_l \over 2} \,,
\eeq
and obtain 
\bea
&&
A_\cG = 2^{L_{\cG}}
\int  \Big[\prod_{l \in \cL_\cG} dt_l   {\left( 1- t_l \right)^{\mu -1} \over \left( 1 + t_l\right)^{\mu +1} } \Big]\, 
\tilde{W}_\cG ( \{m_f\}; \{ t_l \} )
\frac{ \left(\prod_{f \in \cF_{\inter;\,\cG}} \; { \prod_{l \in f}  \left(1 + t_l\right) + \prod_{l \in f} \left( { 1 - t_l } \right) \over 2}\right)^{D} }{
\left(\prod_{f \in \cF_{\inter;\,\cG}} \; { \prod_{l \in f}  \left(1 + t_l\right) - \prod_{l \in f} \left( { 1 - t_l } \right) \over 2}\right)^{D} } \,, 
\cr\cr
&&\tilde{W}_\cG ( \{m_f\}; \{ t_l \} ) = 
\prod_{f \in \cF_{\ext;\,\cG} } \prod_{l \in f} \left({ 1 - t_l \over 1 + t_l }\right)^{|m_f|}.
 \label{eq:ampsteps}
\eea
Using now the definitions \eqref{eq:defu}, we can 
infer that the numerator in the amplitude is given by 
$\tilde{W}_\cG \, (U^{\ev}_{\cG})^D$, in other words \eqref{eq:defv}, and 
the denominator can be further expanded and yields
$U^{\od}_{\cG}$  \eqref{eq:defu}.

 \qed

The formulas \eqref{eq:defa} and \eqref{eq:inacg} provide, for any rank $d$ model over $G_D$ with a propagator linear in momentum, the parametric amplitude for a graph $\cG$. 
The parametric form \eqref{eq:defa} 
appears more adapted to our following developments. 
For the reduced rank $d=2$, the same
parametric amplitudes do not fully coincide with the analog amplitudes
of the GW model 
\cite{Grosse:2012uv,Grosse:2004yu,Grosse:2003nw} in the matrix basis neither in 2D nor in 4D \cite{Krajewski:2010pt}. The reason this occurs comes from the fact that the GW model in the matrix basis is described in terms of matrices $M_{m,n}$ with indices $n$ and $m$ having values only in positive integers $\N$ (2D) or $\N^2$ (4D). 
In order to recover the  amplitudes for the GW models from
\eqref{eq:defa}, one must  replace in $W_{\cG}$, $U^{\ev}_{\cG}(\{t_l\})$
by $c'\prod_{f \in \cF_{\inter;\,\cG}} \prod_{l \in f} (1+t_l)$ with $c'$
a inessential factor $2^{-D \cF_{\inter;\,\cG}}$ which should 
be combined with $c= 2^{L_{\cG}}$.

The polynomials $U^{\od/\ev}$ appear as a product over 
 faces of some other polynomials. The following analysis
rests strongly on this face structure. 
\begin{definition}[Odd, even and external face polynomial]
\label{def:oeface}
Let $f$ be an internal face in a tensor graph of the above models. 
We call $A^{\od/\ev}_f$ \eqref{eq:defu} the odd/even 
face polynomial in the variables $\{t_l\}_{l \in f}$ associated 
with $f$. If $f$ is external, then we call $A^{\ext}_f$
the external face polynomial associated with $f$  in the
variable $T_l=(1-t_l)/(1+t_l)$. 
\end{definition}

Some conventions must be set at this stage.
For the empty 
graph $\cG=\emptyset$ (no vertex), we set $U^{\od}_{\cG} = 1$
 and $U^{\ev}_{\cG} =\tilde{W}_{\cG} = W_{\cG}=1$.
Consider the vertex as a simple disc. As a graph we will denote
it by $\cG=o$. It has one closed face $f$
and, for such a graph, we set: 
\beq
\label{convface}
A^{\od}_{ f } = 0 \,, \qquad
A^{\ev}_{f } =1 \,.
\eeq
As a result, for the  vertex graph  $\cG=o$, 
we set $U^{\od}_{\cG} = 0$, $U^{\ev}_{\cG} =1$,
and $\tilde{W}_{\cG} = 1= W_{\cG}$. 
Furthermore, there exist open faces which do not have any lines. 
For these types of faces, we set 
\bea
A^{\ext}_f = 1\,.
\label{convfaext}
\eea
Now, for a graph $\cG$ without any lines but external faces,
we have  $U^{\od}_{\cG} = 1= U^{\ev}_{\cG} $
and $\tilde{W}_{\cG} = 1 = W_{\cG}$. 

Consider two distinct graphs $\cG_1$ and $\cG_2$,
we have
\bea
U^{\od/\ev}_{\cG_1\sqcup \cG_2} = U^{\od/\ev}_{\cG_1}U^{\od/\ev}_{\cG_2}\,,
\qquad 
\tilde{W}_{\cG_1\sqcup \cG_2} = \tilde{W}_{\cG_1} \tilde{W}_{ \cG_2} \,.
\eea
From this rule, a drastic consequence follows: for any 
graph $\cG$, $U^\od_{\cG \sqcup o} =U^\od_{\cG} U^\od_{ o}=0.$
This means that to (soft) contract arbitrary edges in a graph might
lead to vanishing polynomials on the resulting graphs. Thus, one
can have severe implications on the amplitudes of contracted
graphs that we will aim at studying in the following section.
Nevertheless, this present convention makes transparent 
the analysis of polynomials undertaken in Section \ref{sect:poly}. In any case,  there should
exist a set of conventions (for e.g. setting $U^\od_{ o}=1$),
under which the following amplitude analysis should be valid
and the analysis of polynomials should be slightly re-adjusted. 
In the next section, we will use hard contractions on rank $d$ graphs
and these, by definition, do not generate discs to avoid any issues.

\section{Dimensional regularization and renormalization}
\label{sect:facto}

In this section, we start the investigation of the parametric
amplitudes in view of a dimensional regularization and
its associated renormalization procedure.

  The idea of the subsequent  
procedure can be considered as a ``classic'' in the field
\cite{Bergere:1977ft, Bergere:1977wz,Bergere:1980sm}. 
It also proves to be powerful enough for nonlocal theories \cite{gurau3,gurau4} and can even lead to further applications
in noncommutative field theory \cite{Tanasa:2008bt,Tanasa:2007ai}.  
Let us review quickly this method in the ordinary field theoretical
formalism.

Using a parametric form of the quantum field amplitudes in a
$\mathbf{d}$ dimensional spacetime, the dimension $\mathbf{d}$
appears  as an explicit parameter in these amplitudes
and, as such, can be complexified. 
First, one must show that there exists a complex domain in $\mathbf{d}$
(which can be small) which guarantees the convergence
of all amplitudes and their analytic structure. Then, 
one extends the domain and show that the only possible
divergences occurring in the amplitudes are located
at distinct values of $\mathbf{d}$ involving only isolated poles.
As functions of $\mathbf{d}$ on this extended domain,
amplitudes are therefore meromorphic.  
From this point, the so-called amplitude regularization can be undertaken
by removing the problematic infinite contributions
using a neat subtraction operator. This operator acts on the amplitudes
and leads to  finite and analytic integrals on the whole meromorphicity domain. 
The new amplitudes are called renormalized.

To be complete, it is noteworthy to signal that, in order to prove the meromorphic structure of the Feynman amplitudes, there are
at least two known ways. One of the methods uses the so-called complete 
Mellin representation of the parametric amplitudes 
\cite{deCalan:1979ii,de Calan:1980qj, De Calan:1983qy} 
(which can be applied to the context
of noncommutative field theory \cite{Gurau:2007az}) and
the other introduces the method of Hepp sectors \cite{Bergere:1977ft, Bergere:1977wz} and factorization techniques. The first approach in 
the present context leads to peculiarities which
need to be understood. Using the second path, one
discovers that the method is well defined and 
finds a non-trivial counterpart for, at least, some 
just-renormalizable tensor models. 
We, thereafter, focus on this second alternative. 

\subsection{Regularization using Hepp sectors}
\label{sub:Hepp}

We now proceed with the dimensional regularization scheme. 
Using Hepp sectors (or a meaningful subgraphs' decomposition) of the amplitude, one can identify the singular part of any
diverging amplitude.
The singular part is expressed in terms of the complexified
dimension $D$. It is important to first study the factorization properties of the amplitudes in terms of divergent subgraphs. 

Our main concern is the regularization of the integral \eqref{eq:defa}
when $t_l \to  0$ corresponding to the UV (ultraviolet) limit of the model. One notices that when $t_l \to 1$, the integral is divergent when the mass $\mu$ is bounded as $0\leq \mu < 1 $ 
and if all external momenta $|m_f|$ are equally put to 0. 
For a massive field theory, one can assume the mass to be strictly larger than 1 with no loss of generality, and 
for a massless field theory, one can define fields without 
0-momentum modes. 
In the direct space formalism \cite{BenGeloun:2011rc},
 the same limit $t_l \to 1$ corresponds to an IR (infrared) limit, and the amplitude turns out to be bounded simply because of the compactness of $U(1)^{D}$. Given these reasons and since we discuss UV divergences,
 we will only investigate $t_l \to 0$. 

\begin{figure}[h]
 \centering
     \begin{minipage}[t]{.8\textwidth}
      \centering
\includegraphics[angle=0, width=3.5cm, height=3cm]{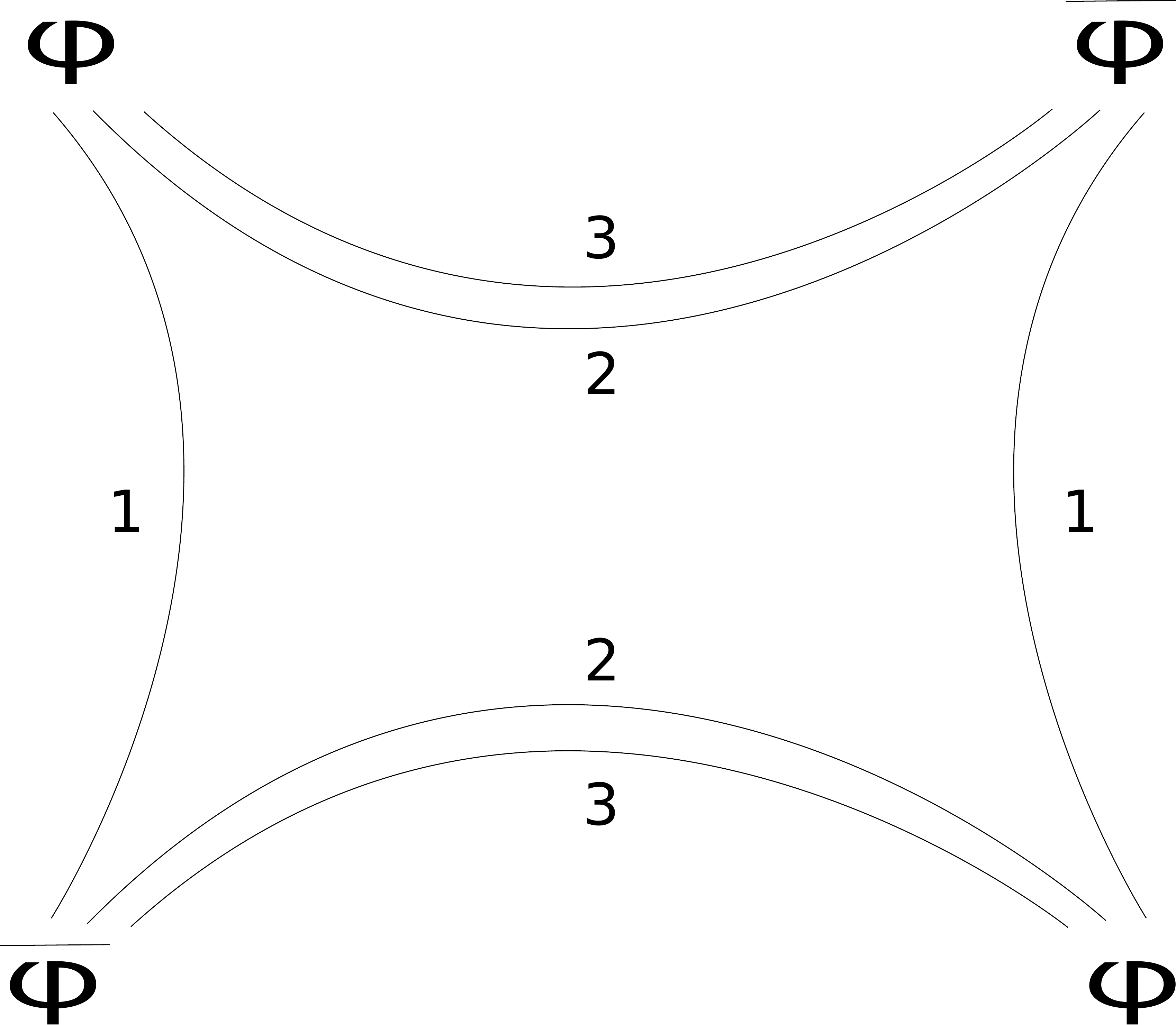}
\vspace{0.1cm}
\caption{ {\small A tensor invariant $\varphi^4$.  }}
\label{fig:tensphi4}
\end{minipage}
\end{figure}

In the following, we are interested in Abelian models
(i.e. $G_D = U(1)^D$) with a kinetic term of the form
$\sum_s |P_s| + \mu$. A generic model will be written as $_D\Phi^{k_{\max}}_{d}$
where $D$ refers to the dimension of the group $G_D$,
$k_{\max}$ to the maximal valence of the vertices, 
and $d$ to the theory rank.   
According to the analysis 
\cite{Geloun:2013saa}, only the following models 
respect these conditions and are perturbatively renormalizable (at all orders):
\bea
&&
  _1\Phi^{4}_{3}\,, \quad G_D= U(1)\;\;\; \text{(Just-renormalizable\; \cite{BenGeloun:2012pu})} \,; 
\crcr
&&  _2\Phi^{4}_{2}\,, \quad G_D= U(1)^2  \;\;\; \text{(Just-renormalizable)}\,;
\crcr
&&
\forall n \geq 2, \qquad  _1\Phi^{2n}_{2}\,, \quad G_D= U(1) \;\;\; \text{(Super-renormalizable)}\,.
\label{renmod}
\eea
We refer the last family of models $_1\Phi^{2n}_{2}$ 
to a tower of models parametrized by the maximal 
valence of its vertices $k_{\max}=2n$. 
The matrix  interactions are, as discussed in the previous
section,  single trace invariants. 
For the model $ _1\Phi^{4}_{3}$, the type 
of tensor invariant interactions that one considers 
are constructed with 4 tensors contracted according
to the pattern of a 3-bubble colored graph made with 
4 vertices (2 white and 2 black, see Fig.\ref{fig:tensinv}).
There are 3 colored symmetric connected invariants 
of this type. Fully expanded, one of these invariants
is drawn in Fig.\ref{fig:tensphi4}. The rest of
the invariants participating to the interaction of
$ _1\Phi^{4}_{3}$ can be obtained by color symmetry.

The graph amplitudes in rank $d\geq 2$
TGFTs were studied using
multi-scale analysis in \cite{Geloun:2013saa}. 
In this work, we provide a new and independent way of 
regularizing these divergent graphs using now the particular form 
of their parametric amplitude representation and their
 underlying meromorphic structure.

\subsubsection{Factorization of the amplitudes}
\label{subsub:facto}

A particular factorization property
of the parametric amplitudes is now investigated. Such a factorization is  necessary for undertaking the subsequent renormalization procedure of the models \eqref{renmod}.

The key idea is the following: 
we assign scales to propagators in a graph $\cG$ and
define the corresponding Hepp sectors.
Choose a subgraph $S$ in $\cG$
and contract all its lines to give $\cG/S$. The interesting
case is when $S$ is primitively divergent (determined by 
a set of conditions on the graph $S$). Roughly speaking, one must 
prove that the amplitude $A_{\cG}$ factorizes in two contributions:
one determined by $A_S$ and the other $A_{\cG /S}$ such 
that by replacing $S$ by a (counter) term of the Lagrangian,
the  integral $A_{\cG /S}$ becomes finite. 
This factorization  plays a crucial role in the definition
of a co-product for the Connes-Kreimer Hopf algebra structure  
intimately associated with the renormalization of the model
(see \cite{Connes:1999yr,Connes:2000fe} for
seminal works). How this applies
to tensor models can be found in \cite{Raasakka:2013kaa}.
For recent approaches in the framework of noncommutative field theory,
one can consult \cite{Krajewski:2012is,Tanasa:2007xa}).

\medskip

We shall need some information about the 
scaling of the polynomials $U^{\od/\ev}$. A specific terminology
and more notations are now introduced: 

- We strengthen the notations $\cL_{\cG}=\cL(\cG)$
and $\cF_{\inter;\cG} = \cF_{\inter}(\cG)$ making
explicit the dependence on the graph $\cG$.

- A subgraph $S$ of $\cG$ is defined by a subset $\cL(S)$ of lines
of $\cG$ and their incident vertices and cutting all remaining
lines incident to these vertices. Thus, from the field theory point of
view, we will always consider a
``subgraph'' as a ``cutting subgraph''.

- We call a divergent subgraph $S$ of $\cG$ 
a subgraph of $\cG$, such that $A_S$ is divergent. There
is a set on conditions under which it occurs.
We will come back on these in a subsequent section.

- We recall the following operations on subgraphs:
Consider a subgraph $S$ of $\cG$. 

$\bullet$ ``Contraction'' always refers in this section to hard contraction
unless otherwise explicitly stated. 

$\bullet$ Let $\cG$ be a graph 
and $e$ be one of its edges (lines). 
The graph $\cG/e$ is  defined as in Section \ref{sect:sgraphs} 
and is called the graph obtained after contraction of $e$.

\begin{figure}[h]
\centering
\begin{minipage}[t]{0.8\textwidth}
\centering
\def\svgwidth{0.5\columnwidth}
\tiny{
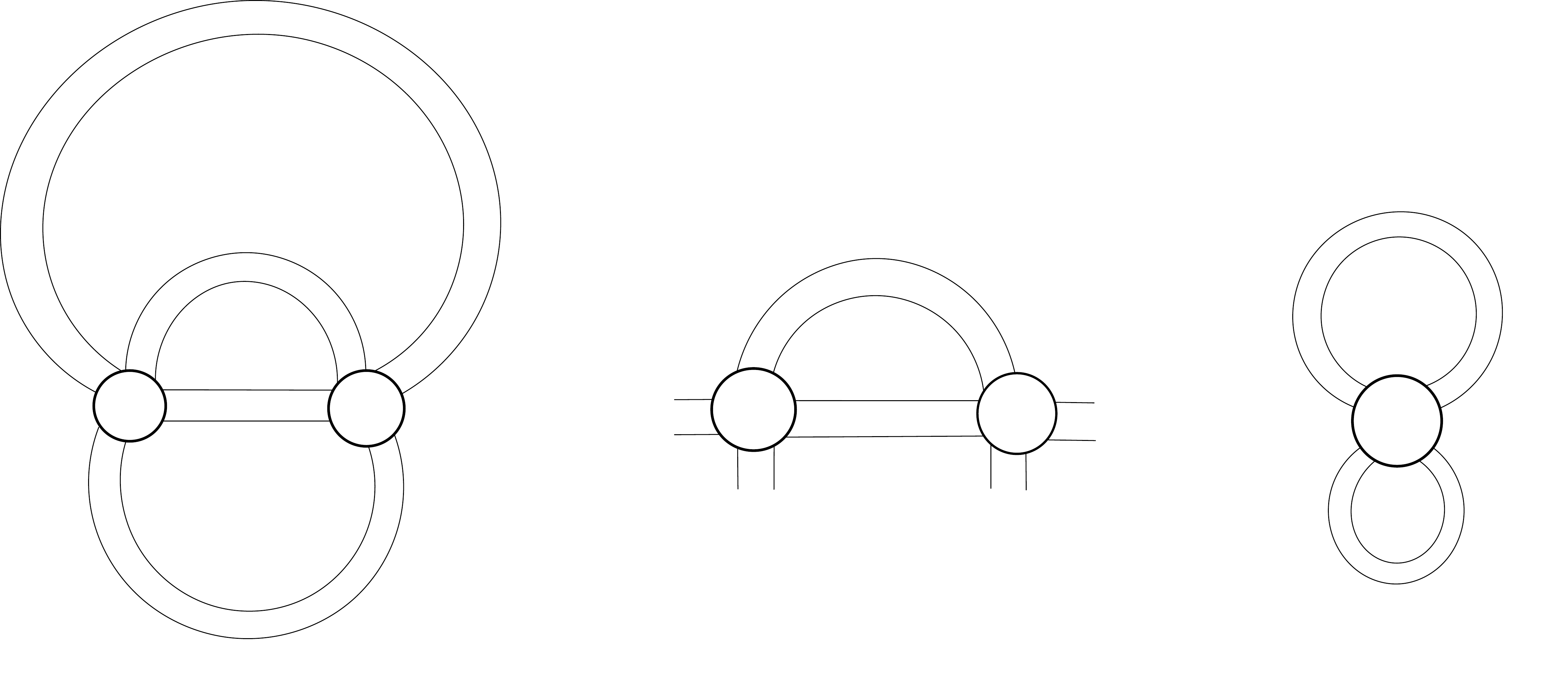
}
\caption{\small  A ribbon graph $\cG$ and one of its subgraph $S$: $\cF_{\inter}(S)=\{f_1\}$, $\cF'_{\inter}(\cG, S)=\{f_2,f_3\}$,  and $\cF''_{\inter}(\cG, S)=\{f_4\}$;
 $f'_{2,3} \in \cF_{\ext}(S)$; $ \cF'_{\inter}(\cG,S)/S = \{f''_2,f_3''\}$
and  $\cF''_{\inter} (\cG, S)/S=\{f_4\}$. 
For $\cG/S$, $\cF_{\inter}(\cG/S)=\{f_2'',f_3'',f_4\}$.
}
\label{fig:contractSsimple}  
\end{minipage}
\end{figure}

$\bullet$ For connected $S$, the contracted graph $\cG/S$ 
is a graph obtained from the full contraction of the lines in $S$
 (see an illustration in Fig.\ref{fig:contractSsimple}).
 If $S$ is non connected, one must apply 
the same procedure to each connected component.

- Consider $S \subset \cG$, strictly speaking, $\cG/S$ is not 
a subgraph of $\cG$.  The only point
which prevents to regard $\cG/S$ as a subgraph of $\cG$
is the fact that it might contain one or several vertices which are not included
in $\cG$. These vertices come  from the contraction of 
$S$. 
One notices that, by definition, $\cL(\cG/S)=\cL(\cG)\setminus \cL(S)$.

\medskip 

Let us introduce notations for subsets of $\cF_{\bullet}(\cG)$,
$\bullet = \inter\!,\ext$. 

\begin{definition}[Sets of faces]
For all $S\subset \cG$,

- $\cF_{\bullet}^*(S)$ is the set of $\bullet$-faces in $S$ having all their lines  lying only in $S$, i.e.
$\forall f \in \cF_{\bullet}^*(S)$, $\forall l \in f$,
$l \in \cL(S)$. 

-$\cF_{\bullet}'(\cG,S) $ is the subset of $\bullet$-faces  of $\cG$
passing through at least one line of $S$ and also through 
at least one line or vertex in $\cG/S$.  
 We have for this category of faces, $\forall f\in \cF_{\bullet}'(\cG,S)$,
$\exists (l,l') \in f\times f$ such that  $l \in \cL(\cG/S)$
and $l' \in \cL(S)$;

- $\cF''_{\bullet}(\cG,S)  = \cF_{\bullet}(\cG) \setminus (\cF_{\bullet}^*(S)
\cup \cF_{\bullet}'(\cG,S))$.

- $\cF_{\bullet}'(\cG,S) /S$ denotes the set of
$\bullet$-faces in $\cG$,  also in $\cG/S$,
coming from $\cF_{\bullet}'(\cG,S) $ 
and which are shortened after the contraction of $S$. 

- $ \cF^*_{\ext}(S)/S$ is the set of external faces in $\cG$,
 also in $\cG/S$, 
resulting from $\cF^*_{\ext}(S)$ after the contraction 
of $S$. 

- Given $e \in f$, we denote $f/e$ (resp. $f-e$) the face resulting from $f$ after the contraction (resp. the deletion) of $e$ in $\cG$ yielding $\cG/e$ (resp. $\cG-e$). 
Given a subgraph $S\subset \cG$, we denote $f/S$ the face resulting from $f$ in $\cG$ after successive contractions 
of all edges of $S$. 

\end{definition}

Some sets of faces as defined above for a ribbon graph $\cG$ and 
one of its subgraph $S$  have been illustrated in Fig.\ref{fig:contractSsimple}.

Few remarks can be spelled out:

- It is true that $\cF_{\inter}^*(S)= \cF_{\inter}(S)$,
however, $\cF_{\ext}^*(S) \neq \cF_{\ext}(S)$ as
a general external face in $S$ might have other lines in
the larger graph $\cG$ or might even close in $\cG$. 
Moreover, there are
external faces which do not contain any lines. 
These are generated by strands in vertices which 
are not connected to any lines.  
For this type of faces, we impose $f\in \cF^*_{\ext}(S)$
if the vertex attached to $f$ is in $\cV(S)$.

- If the external face $f$ does not pass through
any lines, we say that $f \in \cF_{\ext}''(\cG,S)$ if
the vertex touching $f$ belongs to $\cV(\cG)\setminus \cV(S)$. 

-  We define $\cF_{\bullet}(\cG)/S=\cF_{\bullet}(\cG/S)$.

The following statement will be useful (the symbol $\equiv$ below
means ``one-to-one''). 

\begin{lemma}[Sets of faces decomposition]\label{lem:facdecomp}
Consider a subgraph  $S$ of a graph $\cG$.
We have
\beq\label{fintdec}
\cF_{\bullet}(\cG) = \cF_{\bullet}^*(S) 
\cup \cF_{\bullet}' (\cG,S)\cup \cF_{\bullet}''(\cG,S) \,, \qquad
\bullet = \inter\!, \ext\,.
\eeq
The subsets $\cF_{\bullet}^*(S)$, $\cF_{\bullet}' (\cG,S)$ and $\cF_{\bullet}''(\cG,S)$ are pairwise disjoint. 
Furthermore, 
\bea
&&
 \cF'_{\bullet}(\cG,S)  \equiv \cF'_{\bullet}(\cG, S)/S \,, 
\label{fpfprim}\\
&&
 \cF_{\inter}(\cG/S) = (\cF'_{\inter}(\cG,S) /S) \cup \cF_{\inter}''(\cG,S) \,, 
\label{fgso}\\
&&
\cF_{\ext}(\cG/S) = (\cF^*_{\ext}(S)/S) \cup (\cF'_{\ext}(\cG,S) /S) \cup \cF_{\ext}''(\cG,S) \,.
\label{fintgs}
\eea

\end{lemma}
 \proof  The soft contraction of a line in $S$ only shortens
faces. No faces can be created or destroyed by such a move.
The number of faces must be conserved at the end of
the soft contraction of all lines in $S$. 
Moreover, the  ``internal'' or ``external'' nature of faces
is preserved during the procedure. The result of
a hard contraction can be inferred from this point. 

We will focus on \eqref{fpfprim} and on \eqref{fgso},
since the rest of the relations falls quite from the definitions. 

- To prove \eqref{fpfprim}, one must notice that we 
can associate with each element $f\in \cF_{\bullet}'(\cG,S)$ 
a line $l_f$ in $\cG/S$ which is not touched by
the (hard) contraction of $S$.  This line ensures the one-to-one
correspondence between an element in $\cF_{\bullet}'(\cG,S)$ 
and an element in $\cF_{\bullet}'(\cG,S)/S$ after (hard) contraction. 
Indeed, take $f\in \cF_{\bullet}'(\cG,S)$, and $\exists l_f \in \cL(\cG/S)$
such that $l_f\in f$. Then (hard) contract $S$, then $l_f \in f/S$ and
$f/S\in \cF_{\bullet}'(\cG,S)/S$. Reciprocally, take $f\in \cF_{\bullet}'(\cG,S)/S$,
then, by definition $\exists f_0\in \cF_{\bullet}'(\cG,S)$ such that $f_0/S = f$
and $f_0$ is not empty, since by definition there must exist $l \in \cL(S)$
and $l\in f_0$. Note also that \eqref{fpfprim}
does not depend on the type of contraction.

- To achieve \eqref{fgso}, one notes that, after the complete hard contraction of
all lines in $S$,  $\cF_{\inter}(S)$ is mapped to the empty set. Indeed, a closed face
$f$ in $S$ either becomes shorter and shorter after (hard or soft) 
contraction whenever there still exists a line $l \in f$. At some point, $f$ reaches a stage where it must generate a disc after soft contraction of its last line. Using hard contraction, this disc does not occur. 

\qed

We focus now on the scaling properties of the polynomials $U^{\od/\ev}$
and $W$.  Consider $S$ a  subgraph of $\cG$. 
 Rescaling by $\rho$ all variables $t_l$ such that $l \in \cL(S)$, 
one gets from $U^{\od/\ev}_{\cG}$ a new polynomial in $\rho$.
 We call 
$U^{\od/\ev;\, \ell}_{\cG}$ the sum of terms with minimal 
degree in the expansion of $U^{\od/\ev}_{\cG}$, and
$U^{\od/\ev;\, \ell(\rho)}_{\cG}$ the analogue sum for
the minimal degree in $\rho$ in the 
rescaled polynomial.  
Note that it is immediate to realize that 
\beq\label{min1}
U^{\od;\, \ell(\rho)}_{S} = \rho^{F_{\inter}(S)}\;U^{\od;\,\ell}_{S}\,,
\qquad
U^{\ev;\, \ell(\rho)}_{S} = 1 =U^{\ev;\,\ell}_{S}\,.
\eeq

The following statement holds. 
\begin{lemma}[Factorization of leading polynomials]
\label{lem:polyfactor}
Consider a graph $\cG$ and a subgraph $S$ of $\cG$. 
Under rescaling $t_l \to \rho t_l$, $\forall l \in \cL(S)$, we have    
\bea
&&
U^{\od;\, \ell(\rho)}_{\cG} = U^{\od;\, \ell(\rho)}_{S} \,U^{\od}_{\cG/S}\,,
\label{uodmin}\\
&&
U^{\ev;\, \ell(\rho)}_{\cG} =  U^{\ev}_{\cG/S} 
 \,. \label{uevmin}
\eea
Performing a Taylor expansion in $\rho$ around 0 
of $W_{\cG}(\{m_f\};\{\rho t_l\}_{l \in \cL(S)};
\{t_l\}_{l \in \cL(\cG)\setminus\cL(S)} )$ and taking 
$W^{\ell(\rho)}_{\cG}$  as the lowest order in $\rho$, we have 
\bea
W^{\ell(\rho)}_{\cG}(\{m_f\};\{t_l\}) =W_{\cG/S}(\{m_f\};\{t_l\}) \,. 
\label{vmin}
\eea
\end{lemma}
\proof  Computing 
the amplitude of $\cG/S$, one must simply put to 0 some
of the variables $\alpha_l$ $l \in \cL(S)$  in \eqref{amfa}  and do not integrate
over them. This expansion involves 
$U^{\od}_{\cG/S}$ defined with $\cF_{\inter}(\cG /S) $
as given by \eqref{fintgs} in Lemma \ref{lem:facdecomp}.

On the other hand, using \eqref{fintdec} in Lemma \ref{lem:facdecomp}, 
we can  write the following expression for a partially rescaled polynomial 
$U^{\od}_{\cG}$, 
\bea
&&
U^{\od}_{\cG}(\{\rho t_l\}_{l \in \cL(S)};
\{t_l\}_{l \in \cL(\cG)\setminus\cL(S)} ) 
 =  \Big[\prod_{f \in \cF_{\inter}(S)} \dots \Big] 
\Big[\prod_{f \in \cF_{\inter}' } \dots \Big] 
\Big[\prod_{f \in \cF_{\inter}'' } \dots \Big] 
\cr\cr
&&
=   \Big[\prod_{f \in \cF_{\inter}(S)} (\rho \sum_{l \in f}t_l + \dots )\Big] 
  \Big[\prod_{f \in \cF_{\inter}'} \Big\{
(\sum_{l \in f \cap \cL(\cG / S)  } t_l
 + 
\rho \sum_{l \in f \cap \cL(S)  }  t_l) + \dots \Big\}\Big]
\crcr
&& \qquad \qquad \qquad \times\Big[\prod_{f \in \cF_{\inter}'' } \dots \Big]  \,. 
\eea
At the smallest order in $\rho$, we collect from the first bracket $U^{\od;\, \ell(\rho)}_S$ 
and from the two remaining brackets, after putting $\rho =0$ (this
is similar to put $\alpha_l =0$, for $l \in \cL(S)$, in \eqref{amfa})  
the polynomial $U^{\od}_{\cG /S}$. Thus \eqref{uodmin} holds. 

In order to find the second equality for $U^{\ev;\ell(\rho)}_{\cG}$ \eqref{uevmin}, we use the same decomposition \eqref{fintdec} of Lemma \ref{lem:facdecomp} and \eqref{min1}. 

We now perform a Taylor expansion around $\rho=0$ of
the following expression (in suggestive though loose notations):
\bea&&
W_{\cG}(\{m_f\};\{\rho t_l\}_{l \in \cL(S)};
\{t_l\}_{l \in \cL(\cG)\setminus\cL(S)} )
 =  \Big[\prod_{f \in \cF^*_{\ext}(S)} \dots \Big] 
\Big[\prod_{f \in \cF_{\ext}' } \dots \Big] 
\Big[\prod_{f \in \cF_{\ext}'' } \dots \Big] 
(U^{\ev}_{\cG})^D \crcr
&&
 =\Big[\prod_{f \in \cF^*_{\ext}(S)} (1+\rho \dots )\Big] 
  \Big[\prod_{f \in \cF_{\ext}'} \{(1+ \rho \dots) \prod_{l \in f /\,
l \in \cL(\cG/S)} (\frac{1-t_l}{1+t_l})^{|m_f|}\}\Big]  
\Big[\prod_{f \in \cF_{\ext}'' } \dots \Big] 
(U^{\ev}_{\cG})^D\,,\crcr
&&
\eea
where we used  \eqref{fintdec} in Lemma \ref{lem:facdecomp}. 
Now at minimal degree in $\rho$, we infer 
\bea
W^{\ell(\rho)}_{\cG}(\{m_f\};\{t_l\}) 
= \Big[\prod_{f \in \cF_{\ext}'}\prod_{l \in f /\;
l \in    \cL(\cG/S)  } (\frac{1-t_l}{1+t_l})^{|m_f|}\Big]  
\Big[\prod_{f \in \cF_{\ext}'' } \dots \Big] 
(U^{\ev;\, \ell(\rho)}_{\cG})^D 
\eea
and one concludes using:  
(a) \eqref{uevmin} to 
map $U^{\ev;\, \ell(\rho)}_{\cG}$ onto $U^{\ev}_{\cG/S}$, 
(b) $\cF_{\ext}(\cG/S)$ from \eqref{fintgs}
in Lemma \ref{lem:facdecomp}, and finally (c) 
observe that $\cF^*_{\ext}(S)/S\subset \cF_{\ext}(\cG/S)$
are external faces in
the contracted graph $\cG/S$ which do not pass through
any lines and by convention $A^{\ext}_f =1$ \eqref{convfaext}.

 \qed

The preliminary factorization properties addressed in Lemma
\ref{lem:polyfactor} will allow us to understand the most diverging
part of the amplitude. However, in some cases, there exist 
subleading divergences which need to be renormalized as well. 
 In particular, these kinds of divergences occur in the two-point
function and the factorization must be extended up to higher
orders in the scale parameter $\rho$. This is our next goal. 

 Consider a diverging subgraph $S$ with internal lines with parameters $t_l \in \cL(S)$ which should be such that $t_l \ll t_{l}'$ for any $t_l' \in \cL(\cG/S)$. 
This condition simply suggests that any internal propagator 
 line is of higher energy than any external lines, as required in 
ordinary renormalization procedure. In this way, from the point of view of external legs the internal subgraph appears local. 
We will then perform a Taylor expansion on 
the variables $t_l \in \cL(S)$ but only on the 
faces which are in $\cF_{\bullet}^{'}(\cG,S)$
(that we simply denote henceforth $\cF_{\bullet}'$)
which links the subgraph $S$ to the rest of
the graph and we will prove that, at each order, 
the result factorizes for small $t_l$.

Consider the notations:
$U''^{\od/\ev}_{\cG; S}= \prod_{f\in \cF''_{\inter} } A^{\od/\ev}_f$
and $U'^{\od/\ev}_{\cG;S} = \prod_{f\in \cF_{\inter}'  } A^{\od/\ev}_f $.

\begin{lemma}[Factorization of a $N$-point subgraph]
\label{lem:ampfactor}
Consider a graph $\cG$ of a rank $d$ model 
and a subgraph $S$ of $\cG$
with external legs. 
For small $t_l$, $\forall l \in \cL(S)$, we have    
\bea
&& 
A_{\cG}(\{m_f\};D) 
= 
c\lambda_{\cG} \int\; 
\Big[\prod_{  l' \in \cL(\cG/S)} dt_l'
{\left( 1- t_{l'} \right)^{\mu -1} \over \left( 1 + t_{l'}\right)^{\mu +1} }\Big] \tilde{W}_{\cG/S}(\{t_{l'}\})
\left[\frac{U^{\ev}_{\cG/S}}{ U^{\od}_{\cG/S} }  ( \{t_{l'}\} ) \right]^D
\crcr
&&
\int 
\Big[\prod_{ l \in \cL(S)} d t_l  {\left( 1- t_l \right)^{\mu -1} \over \left( 1 + t_l\right)^{\mu +1} } \Big] \Big[\prod_{f\in \cF^*_{\ext}(S)}
 \prod_{l \in f } \left({ 1 - t_l \over 1 + t_l }\right)^{|m_{f}|} \Big]
\left[\frac{ U_{S}^{\ev} }{ U_{S}^{\od} } ( \{t_l\} ) \right]^D\crcr
&& 
\Big[1 - 2 \sum_{f\in \cF'_{\ext}}|m_{f}| R_{S,f}(\{t_l\})
+ D\textstyle
\sum_{f\in \cF'_{\inter} }
M_f(\{t_{l'}\})  \,R_{S,f}(\{t_l\}) + O(t^2_l) \Big]\,, \crcr
&&
R_{S,f}(\{t_l\}) = \sum_{l\in f \cap \cL(S)} t_l\,, 
\qquad \quad
M_f(\{t_l\}_{l \in \cL(\cG/S)}) = \frac{ (A_{f/S}^{\od})^2 -  (A^{\ev}_{f/S})^2}{ A^{\ev}_{f/S} A^{\od}_{f/S} }(\{t_l\})\,,  
\label{amplfactor}
\eea
where $O(t_l^2)$ is a big-O function of all possible products $t_lt_{l''}$, 
 for $l,l''\in \cL(S)$.

\end{lemma}

\proof 
Consider a graph  $\cG$ and fix one of its $N$-point subgraphs 
$S$. We write:
\bea
&&
A_{\cG}=c\lambda_{\cG} \int\; 
\Big[\prod_{ l'\in \cL(\cG/S)} dt_{l'}
{\left( 1- t_{l'} \right)^{\mu -1} \over \left( 1 + t_{l'}\right)^{\mu +1} }\Big]\Big[\prod_{f \in \cF''_{\ext}} 
\prod_{l' \in f } \Big(\frac{1- t_{l'} }{ 1+t_{l'} }\Big)^{|m_{f}|} \Big]
\left(\frac{ U''^{\ev}_{\cG;S} }{ U''^{\od}_{\cG; S} } ( \{t_{l'}\} ) \right)^D 
\crcr
&&
\int 
\Big[\prod_{ l \in \cL(S)} d t_l  {\left( 1- t_l \right)^{\mu -1} \over \left( 1 + t_l\right)^{\mu +1} } \Big] 
\Big[\prod_{f\in \cF^*_{\ext}(S)}
 \prod_{l \in f } \left({ 1 - t_l \over 1 + t_l }\right)^{|m_{f}|} \Big]
\Big[\prod_{f\in \cF'_{\ext}}
 \prod_{\ell \in f } \left({ 1 - t_\ell  \over 1 + t_\ell  }\right)^{|m_{f}|} \Big] \crcr
&& 
\left(\frac{ U'^{\ev}_{\cG;S} }{ U'^{\od}_{\cG;S} } ( \{t_\ell\} ) \right)^D
\left(\frac{ U_{S}^{\ev} }{ U_{S}^{\od} } ( \{t_l\} ) \right)^D.
\label{eq:defaex0}
\eea 
We used $\ell$ for a generic line label in $\cL(\cG)$. 
Now, we perform a Taylor expansion on part of the 
factor $\prod_{f\in \cF'_{\ext}}
 \prod_{l \in f }(\dots)$ for small $t_l$ only if $l \in \cL(S)$.
We obtain the contribution:
\bea\label{aeins}
 \prod_{f\in \cF'_{\ext}}
 \prod_{l \in f \cap \cL(S)} \left({ 1 - t_l \over 1 + t_l }\right)^{|m_{f}|}
 &=& 
 1 - 2 \sum_{f\in \cF'_{\ext}}|m_{f}| R_{S,f}(t_l) + O(t^2_l)\,,
\crcr
R_{S,f}(t_l) &=&  \sum_{l \in f \cap  \cL(S)}t_l \,,
\eea
where $l$ in $O(t_l^2)$ only refers to the lines in $\cL(S)$. 
Note that the remaining factors compile to  
\beq\label{boutxt}
\prod_{f\in \cF'_{\ext}}
 \prod_{l \in f \cap \cL(\cG/S)} \left({ 1 - t_l \over 1 + t_l }\right)^{|m_{f}|} 
 = \prod_{f\in \cF'_{\ext}/S}
 \prod_{l \in f } \left({ 1 - t_l \over 1 + t_l }\right)^{|m_{f}|}.
\eeq
Focusing on the factor $ U'^{\ev}/U'^{\od}$, we have 
\bea
&&
U'^{\ev/\od}_{\cG;S}
 = 
  \prod_{f\in \cF'_{\inter}} ( A^{\ev/\od}_{f/S}+ \sum_{l\in f \cap \cL(S)} t_l A_{f/S}^{\od/\ev} + O(t_l^2))\crcr
&&
=   \prod_{f\in\cF'_{\inter}} A^{\ev/\od}_{f/S} 
+ \sum_{f\in \cF'_{\inter}} [\prod_{f'\in\cF'_{\inter}/f'\neq f} A^{\ev/\od}_{f'/S}]A_{f/S}^{\od/\ev} R_{S,f}(t_l)+ O(t_l^2)\,.
\eea
Thus, the ratio behaves like
\bea
&&
\frac{U'^{\ev}_{\cG;S}}{U'^{\od}_{\cG;S}} 
=\textstyle
\frac{1}{\prod_{f\in\cF'_{\inter}} A^{\od}_{f/S} }
\Big( \prod_{f\in\cF'_{\inter}} A^{\ev}_{f/S}  - \frac{\prod_{f\in\cF'_{\inter}} A^{\ev}_{f/S} }{\prod_{f\in\cF'_{\inter}} A^{\od}_{f/S} }
 \sum_{f\in \cF'_{\inter}} [\prod_{f'\in\cF'_{\inter}/f'\neq f} A^{\od}_{f'/S}]A_{f/S}^{\ev} R_{S,f}(t_l) \crcr
&& \textstyle
\qquad \qquad + 
\sum_{f\in \cF'_{\inter}} [\prod_{f'\in\cF'_{\inter}/f'\neq f} A^{\ev}_{f'/S}]A_{f/S}^{\od} R_{S,f}(t_l)
+ O(t_l^2)\Big)  \cr\cr
&&=
\frac{U'^{\ev}_{\cG;\cG/S }}{ U'^{\od}_{\cG;\cG/S} } \textstyle
\Big( 1  + 
\sum_{f\in \cF'_{\inter} }
\big[\frac{ (A_{f/S}^{\od})^2 -  (A^{\ev}_{f/S})^2}{ A^{\ev}_{f/S} A^{\od}_{f/S} } \big]R_{S,f}(t_l)  + O(t_l^2)\Big)\,,
\label{upp}
\eea
where we define $U'^{\od/\ev}_{\cG;\cG/S }:= \prod_{f\in \cF'_{\inter}} A^{\od/\ev}_{f/S}$.
 One must use the bijection relation  \eqref{fpfprim} in 
Lemma \ref{lem:facdecomp} to map $ \cF'_{\inter}(\cG;S)$ to 
 $ \cF'_{\inter}(\cG;S)/S$ and we can write $U'^{\od/\ev}_{\cG;\cG/S }:= \prod_{f\in \cF'_{\inter}/S} A^{\od/\ev}_{f}$.
We insert \eqref{aeins}, \eqref{boutxt} and \eqref{upp} in \eqref{eq:defaex0}, 
and get the expansion
\bea
&&
A_{\cG} =  \crcr
&&
c\lambda_{\cG} \int\;\Big[\prod_{ l'\in \cL(\cG/S)} dt_{l'} 
{\left( 1- t_{l'}\right)^{\mu -1} \over \left( 1 + t_{l'}\right)^{\mu +1} }\Big]\Big[\prod_{f \in \cF''_{\ext} \cup \cF'_{\ext}/S} 
\prod_{l' \in f } \Big(\frac{1- t_{l'} }{ 1+t_{l'} }\Big)^{|m_{f}|} \Big]
\left[\frac{ U'^{\ev}_{\cG;\cG/ S}}{U'^{\od}_{\cG;\cG/S}} \frac{ U''^{\ev}_{\cG ; S} }{ U''^{\od}_{\cG ; S} } ( \{t_{l'}\} ) \right]^D
\crcr
&&
\int 
\Big[\prod_{ l \in \cL(S)} d t_l  {\left( 1- t_l \right)^{\mu -1} \over \left( 1 + t_l\right)^{\mu +1} } \Big] \Big[\prod_{f\in \cF^*_{\ext}(S)}
 \prod_{l \in f } \left({ 1 - t_l \over 1 + t_l }\right)^{|m_{f}|} \Big]
\left(\frac{ U_{S}^{\ev} }{ U_{S}^{\od} } ( \{t_l\} ) \right)^D\crcr
&& 
\Big[1 - 2 \sum_{f\in \cF'_{\ext}}|m_{f}| R_{S,f}(t_l) 
+ D\textstyle
\sum_{f\in \cF'_{\inter} }
\big[\frac{ (A_{f/S}^{\od})^2 -  (A^{\ev}_{f/S})^2}{ A^{\ev}_{f/S} A^{\od}_{f/S} } \big]R_{S,f}(t_l) + O(t^2_l) \Big] .
\label{eq:defaex1}
\eea 
Now using \eqref{fintgs} in Lemma \ref{lem:facdecomp}, 
we see that the complementary of $\cF''_{\ext} \cup \cF'_{\ext}/S$ 
in $\cF_{\ext}(\cG/S)$ is $\cF^*_{\ext}/S$.  But for any $f\in \cF^*_{\ext}/S$, $A^{\ext}_f =1$,
 so  \eqref{amplfactor} becomes immediate.

\qed

We can now interpret Lemma \ref{lem:ampfactor}:

- At the leading 0th-order the amplitude $A_{\cG}(\{m_f\})$
factorizes as (in loose notations) 
\beq\label{0th}
A^{0}_{\cG}(\{m_f\}) =\left( \int [dt_l]_{\cL(S)}\, \tilde{A}_{S}(\{t_l\}, \{m_f\})
\right)   A_{\cG/S}(\{m_f\})
\eeq
where $\tilde{A}_{S}(\{t_l\}, \{m_f\})$ is not exactly the integrand of the amplitude of $S$, namely $A_{S}(\{m_{f}\})$. Their set of external faces and set of external momenta  do not always match. However, the diverging or converging behavior of $\int \tilde{A}_{S}$ and $A_{S}$ are identical.  Thus  \eqref{0th} means that, at this order of perturbation, the amplitude
$A_{\cG} \simeq A^0_{\cG}$ can be computed by evaluating the
amplitude of $\cG/S$, where we insert a counter-term (generalized) vertex  obtained after the contraction of $S$, times the amplitude $\int \tilde{A}_{S}$ of
the divergent subgraph $S$.

- Up to 
the first order of perturbation, focusing on the internal variables
associated with $t_l$, $l\in \cL(S)$, we re-express the 
types of contributions appearing in \eqref{amplfactor} as:
\bea
&&
A^{1}_{\cG}(\{m_f\}) =
\Big[\int [dt_l]_{\cL(S)}\, \tilde{A}_S(\{t_l\};\{m_f\})\Big( 1+  B_{S}(\{t_l\})\cO_{S}(\{m_f\}) \Big)\Big]
A_{\cG/S}(\{m_f\}) \\
&&\qquad\qquad\quad +  
\sum_{f\in \cF_{\inter}'/S} 
\Big( \int [dt_l]_{\cL(S)}\,  \tilde{A}_S(\{t_l\};\{m_f\})  C_{S,f}(\{t_l\}) \Big)
\crcr
&&\qquad\qquad\quad \times  \int [dt_{l'}]_{\cL(\cG/S)} M_{f}(\{t_{l'}\}) {\tilde{ A}}_{\cG/S}(\{m_f\}, \{t_{l'}\})\,, 
\nonumber
\eea 
where ${\tilde{ A}}_{\cG/S}$ is the integrand of $A_{\cG/S}$.

$\sim$ For a 2-point subgraph $S$, the term $\int \tilde{A}_{S} B_S\cO_{S}$
of the form $\int \tilde{A}_{S} (\sum_{f} |m_f| \cdot R_{S,f})$ contributes to a wave function renormalization
counter-term. In general, it is well-known
that this counter-term $|m_f|\int \tilde{A}_{S}  R_{S,f}$ might have a different value for different external faces $f$. Therefore, it is not always
true that $\int \tilde{A}_{S} (\sum_{f} |m_f| \cdot R_{S,f})
= (\sum_{f} |m_f|)(\int \tilde{A}_{S} \cdot R_{S,f})$ where
the last factor should be independent of $f$. In order to achieve a final wave function renormalization, we carefully sum symmetric
graph contributions.

$\sim$ The remaining term  
$\sum_f \int \tilde{ A}_{S} C_{S,f} \times \int M_f \tilde{A}_{\cG/S}$ 
 will be subleading compared to $\int \tilde{A}_S \times A_{\cG/S}$. 
Still in the case of $N=2$ and for the just-renormalizable models
 \eqref{renmod}, this term may lead to a mass sub-leading divergence. 

At this order of perturbation  $A_{\cG} \simeq  A^{1}_{\cG}$
and its expansion organizes as follows. For a $N$-point divergent subgraph $S$, one inserts in $\cG/S$, two types of counter-terms or operators: 
a generalized ``wave-function'' vertex with amplitude 
$|m_f|\int \tilde{A}_S R_{S,f}$ associated with each external face $f$
of this vertex, and the vertex with amplitude 
 $\int \tilde{A}_S (1+ \sum_f  C_{S,f})\sim \int \tilde{A}_S$.

- At higher order terms, the contributions will be sub-leading in the
same way as explained above. For the
remaining analysis, the study of higher order terms will only be sketched.

\subsubsection{Meromorphic structure of the regularized amplitudes}
\label{subsect:div}

In this section, we consider a fixed graph $\cG$ and some
of its subgraphs. We simplify
notations and omit the dependency in the largest graph $\cG$ 
in integrands and several expressions when no confusion might occur, such that  $L = L(\cG)$, $F_{\inter} = F_{\inter}(\cG)$ and so forth.

Take a Hepp sector $\sigma$ such that 
\beq
0 \leq t_1 \leq t_2 \leq \dots \leq  t_{L}\,,
\eeq
and perform the following change  of variables
\beq\label{changof}
\forall l =1,\dots, L\,, \qquad t_l = \prod_{k=l}^{L} x_k\,.
\eeq
Consider the subgraph $G_i$ of $\cG$ defined by 
the lines associated with the variables $t_j$, $j=1,\dots, i$. 
We denote $L(G_i)=i$, $F_{\inter}(G_i)$ the number
of lines and internal faces, respectively, of $G_i$. 
The amplitude \eqref{eq:defa} of Proposition \ref{prop:propamp}
in the sector $\sigma$ in terms of the variables $x_l$ finds a new form;
\bea
A^{\sigma}_\cG (\{m_f\};  D)&=&
 \lambda_{c,\cG} \int_{[0,1]^{L}} \; \Big[\prod_{l =1}^L d x_l  \frac{\left( 1- \prod_{k=l}^{L} x_k \right)^{\mu -1} }
{\left( 1 + \prod_{k=l}^{L} x_k\right)^{\mu +1} } \Big]
 \prod_{f \in \cF_{\ext} } \prod_{l \in f} \left( 
\frac{1 - \prod_{k=l}^{L} x_k }{ 1 + \prod_{k=l}^{L} x_k} \right)^{|m_f|}\crcr
&& \times 
\Big[\prod_{i=1}^{L} x_i ^{L(G_i)-1}\Big]
\left[  \frac{\tilde{U}^{\ev}_\cG ( \{ x_l \} ) }{\tilde{U}^{\od}_\cG ( \{ x_l\} ) } \right]^{D}, 
\label{eq:ampli2}
\eea
where $\tilde{U}^{\od/\ev}$ are new polynomials obtained
from $U^{\od/\ev}$ after the substitution of Eq.~\eqref{changof}. For the moment, 
$D$ is real positive. In order to recover the full amplitude 
$A_{\cG}$ \eqref{eq:defa}, one sums over all possible 
Hepp assignments: $A_{\cG} = \sum_{\sigma}A_{\cG}^{\sigma}$. 
Hereunder, we will focus on $A_{\cG}^{\sigma}$
and the last sum over $\sigma$ will be
addressed later.

Focusing on the denominator
of the last line of \eqref{eq:ampli2},
we want to extract the term of minimal degree in $x_l$ in the polynomial. 
The term of minimal degree in $t_l$ any face amplitude $A^{\od}_f$ is nothing but $\sum_{l \in f} t_l$. However, this term is not yet the term with minimal 
degree in $x_k$'s. To obtain the monomial of minimal 
degree in $x_k$, one picks $t^0_f = t_{l^0_f}$ with 
$l^0_f  = \max_{l \in f} l$. We have 
\beq
\tilde{U}^{\od}_\cG ( \{ x_l\} ) = \prod_{f\in \cF_{\inter}} 
A^{\od}_f(\{x_l\}) 
 = \prod_{f\in \cF_{\inter}} \Big[\prod_{\alpha= l^0_f}^L 
x_\alpha \Big]\Big( 1+  A^{\rm r}_f(\{x_k\}) \Big)\,, 
\label{urpim}
\eeq
where $A^{\rm r}_f$ is the rest of the face amplitude
after the factorization. Focusing on the first factor, it 
recasts as
\bea
\prod_{f\in \cF_{\inter}} \Big[\prod_{\alpha= l^0_f}^L 
x_\alpha \Big] = 
\prod_{\alpha= 1}^L \prod_{f\in \cF_{\inter}/ \alpha \geq l^0_f} 
x_\alpha  
=\prod_{\alpha= 1}^L x^{|\{ f\in \cF_{\inter}\,/\, \alpha \geq l^0_f\}|}_\alpha  \,. 
\eea
An internal face $f$ in $\cF_{\inter}(G_i)$ is an internal face of $\cG$ such that its most higher index $l_f^0$ among $l \in f$ must be lower than $i$. 
We can conclude that $|\{ f\in \cF_{\inter}\,/\, i\geq l^0_f\}|= F_{\inter}(G_i)$ and it is direct to get: 
\bea
A^{\sigma}_\cG (\{m_f\},D)&=&
 \lambda_{c,\cG} \int_{[0,1]^{L}} \; \Big[\prod_{l =1}^L d x_l  \frac{\left( 1- \prod_{k=l}^{L} x_k \right)^{\mu -1} }
{\left( 1 + \prod_{k=l}^{L} x_k\right)^{\mu +1} } \Big]
 \prod_{f \in \cF_{\ext} } \prod_{l \in f} \left( 
\frac{1 - \prod_{k=l}^{L} x_k }{ 1 + \prod_{k=l}^{L} x_k} \right)^{|m_f|}\crcr
&& \times 
\Big[\prod_{i=1}^{L} x_i ^{L(G_i)-1 - D F_{\inter}(G_i)}\Big]
\left[  \frac{\tilde{U}^{\ev}_\cG ( \{ x_l \} ) }{(1+ U'_{\cG}(\{ x_l\} )) } \right]^{D} ,
\label{eq:ampli3}
\eea
with $U'_{\cG}$ readily obtained from \eqref{urpim}. 
The quantity 
\beq\label{convdeg}
\omega_{\rm d}(\cG) = (L - D F_{\inter})(\cG)
\eeq
is called the convergence degree of the graph amplitude. 

Before considering complex valued variables involved in this 
object, we will discuss better its constituents. 
In particular, the number of internal faces $F_{\inter}(\cG)$ must be
dissected. 
This number $F_{\inter}(\cG)$ of a connected
graph $\cG$ has been worked out in \cite{Samary:2012bw}. 
We have for a connected graph $\cG$ and in our case:

\noindent$\bullet$ In rank $d\geq 3$, introducing $d^-=d-1$,
 \beq
\label{faceinter3d}
F_{\inter}(\cG) = -\frac{2}{(d^-)!}(\omega(\cexG) - \omega(\bG)) - (C_\bG -1)
- \frac{d^-}{2} N_{\ext} + d^- - \frac{d^-}{4}(4-2n)\cdot V\,,
\eeq
where  $\cexG$ the so-called colored extension of $\cG$
in the sense of Subsection \ref{subsect:graphs},
$\bG$ its boundary with number $C_\bG$  of connected components, $V_k$ its number of vertices of coordination $k$, $V= \sum_{k} V_k$ its total number of vertices,  $n \cdot V = \sum_{k} k V_k$ its number of half-lines exiting from vertices, $N_{\ext}$ its number of external legs. 
We call $\omega(\cexG)=\sum_{J}g_{\tJ}$ the degree of $\cexG$, 
$\tJ$ is the pinched jacket associated with $J$ a jacket of $\cexG$, 
$\omega(\bG)=\sum_{\bJ}g_{\bJ}$ is the degree of $\bG$. 
Specifically, in rank $d=3$, $\omega(\bG)=g_{\bG}$, since
the boundary graph is a ribbon graph. 

\medskip 

\noindent$\bullet$ In rank $d=2$, using the Euler characteristics, the
following holds
\bea\label{faceinter2d}
F_{\inter}(\cG)
= - 2g_{\tilde \cG}  - (C_{\bG}-1) - \frac12 (N_{\ext}-2)
 - \frac12 (2 -n )\cdot V \,,
\eea
where $\tilde\cG$ is the closed (pinched) graph associated with $\cG$
and we used the relation $2L=  n \cdot V - N_{\ext}$. 

Thus, one can write both \eqref{faceinter3d} and \eqref{faceinter2d}
under the form 
\beq\label{faceinternes}
F_{\inter}(\cG) = -\frac{2}{(d^-)!}{\mathbf{\Omega}}(\cG) - (C_\bG -1)
+ d^- \tilde{F}_{\inter}(\cG) \,,  
\qquad \tilde{F}_{\inter}(\cG)= \frac{1}{2}\Big(2- N_{\ext}  +(n-2)\cdot V \Big),
\eeq
where ${\mathbf{\Omega}}(\cG) = \omega(\cexG) - \omega(\bG)$ if $d=3$, and ${\mathbf{\Omega}}(\cG) = g_{\tilde{\cG}}$ if $d=2$.

Note that the number of internal faces does not depend
on $D$ but only on the combinatorics of the graph itself. 
From \eqref{faceinternes}, 
a formula for $\omega_{\rm d}(\cG) $ can be easily 
obtained after substituting this expression in \eqref{convdeg}. However, in the following we are interested only in bounds involving directly the degree of convergence. 
It is a non-trivial fact that, for any graph in this category of models
one has (see \cite{BenGeloun:2011rc} and its addendum \cite{Geloun:2012fq}),  
\beq
{\text{either}} \;\; 
\Om(\cG) = 0 \,,\;\;\;\;
{\text{or}}\;\;
 \frac{2}{(d^-)!} \Om(\cG)  \geq d-2\geq 0\,.
\label{looseb}
\eeq
In a renormalization program, we are mainly interested in graphs with external legs. These are graphs with boundary, in other words
graphs satisfying $C_\bG \geq 1$. Therefore, for any connected graph, the following is true  for $d^- \geq 1$,
\beq
\label{boundface}
F_{\inter}(\cG) \leq
d^- \tilde{F}_{\inter}(\cG)\,.
\eeq
It is also a known fact that in any rank $d\geq 3$, the so-called
melonic graphs defined such that $\omega(\cexG)=0$ with a melonic boundary, i.e. $\omega(\bG)=0$, and with a unique connected component on the boundary saturate this bound. Therefore, the melonic graphs have a dominant power counting and this shows that \eqref{boundface} is an optimal bound.  Matrix models are
similar. The dominant amplitudes in power counting are those
with a maximal number of internal faces. These are
planar graphs with 
$g_{\tilde \cG}=0$ and $C_{\bG}=1$. Hence \eqref{boundface} 
 is again saturated. 

We now discuss possible interesting complexifications of
the amplitude $A_{\cG}(\{m_f\})$. 
So far, we have two  parameters which 
are the dimension $D$ of the group and the theory
rank $d$. A priori, from \eqref{eq:ampli3}, we can 
define a complex integral 
$
A_{\cG}(\{m_f\}, D,d)$, for $D,d \in \C$. 
However, the non trivial dependency 
in $d$ in the amplitude makes the study of this 
function drastically complicated.
We will only achieve a complexification in
the standard way, i.e. by considering a complex 
dimension  $A_{\cG}(\{m_f\}, D)$ for $D\in \C$,
and will undertake the dimensional regularization
of an arbitrary amplitude in this variable.

\medskip 

\noindent{\bf Domain of  analyticity.}
The analysis of $A_{\cG}(\{m_f\}, D)$ can be undertaken as follows. 
From the fact that $\tilde{U}^{\ev}_\cG = 1+ U''_{\cG}$,
the last factor $\tilde{U}^{\ev}_\cG/(1+ U'_{\cG}(\{ x_l\} ))$ in \eqref{eq:ampli3}
can be bounded by a constant $k_{\cG}=\tilde{U}^{\ev}_\cG(\{x_k=1\})$
depending on the graph. It is immediate to infer from \eqref{eq:ampli3} that, 
in the UV regime $x_i \to 0$,  
\bea
&&
\text{if}\;\; \forall i=1,\dots, L, \;\;\Re( \omega_{\rm d}(G_i)) >0 \,,\qquad \text{then the amplitude converges;} \crcr
&&
\text{if}\;\; \exists i=1,\dots, L,\;\; \Re( \omega_{\rm d}(G_i)) \leq 0 \,,\qquad \text{then the amplitude diverges.}
\label{condi}
\eea
Consider the subgraphs $G_i$ associated with Hepp sectors, with positive numbers of vertices $V(G_i) \geq 1$ and lines $L(G_i) \geq 1$. The
convergence of the amplitude is then guaranteed
(sufficient condition) if we have
\beq
\Re(D) < D_0^{\sigma}  = \inf_{i}   \frac{L(G_i) }{F_{\inter}(G_i)} \,.
\eeq 
Note that if $L=0$, the graph is actually 
either empty or formed by disconnected vertices and so, 
$F_{\inter}=0$ and there are no divergences.
On the other hand, setting $F_{\inter}=0$ means
already that we have no divergences. 
We are led to the following bound, $\forall i$,
\beq
\Re(D) <  \frac{1}{d^-}  \leq\frac{2L}{d^-( 2L +  2(1-V))}(G_i) \leq 
\frac{L(G_i) }{d^-\tilde{F}_{\inter}(G_i)} 
\leq  \frac{L(G_i) }{F_{\inter}(G_i)} 
 \eeq
where uses have been made of \eqref{boundface}, and
the fact that either $V(G_i)=1$ or $V(G_i)>1$
and so $\frac{2L}{d^-( 2L +  2(1-V))}(G_i)  > \frac{2L}{d^-( 2L)}(G_i).$

We infer that, the amplitude $A_{\cG}(\{m_f\}, D)$ is convergent 
and analytic in $D$ in the strip
\beq\label{streep}
\mathcal{D}^{\sigma} = \left\{ D \in \C \;|\;   0< \Re(D) < \frac{1}{d^-}  \right\}.
\eeq
At the first sight, increasing the rank of the theory induces a reduction
of the  analyticity strip of the amplitude. Also, as a recurrent feature,
this  analyticity domain is again restricted to a strip the real part of which is bounded by half of the dimension of the group manifold. 

Now, we proceed further and extend $A_{\cG}(\{m_f\}, D)$ to a complex function of $D$ in the strip  
$  1/d^-  \leq \Re(D) \leq \bdel$,
where $\bdel$ plays the role of the initial dimension of the group 
that is either $\bdel =1$ for the models $_1\Phi^{4}_3$
and $_1\Phi^{2k}_2$,  or $\bdel=2$ for $_2\Phi^{4}_2$. 
\begin{theorem}[Extended domain of  analyticity]\label{theo:extan}
Consider a tensor model $_{\bdel}\Phi_{d}^{k_{\max}}$,
$(\bdel,d,k_{\max})$  $\in \{(1,3,4),$ $(2,2,4), (1,2,2n)\}$,
for $n \geq 2$. 
Let $\cG$ be one of its graphs and define $\sigma$ 
an associated Hepp sector of $\cG$.
For $_{\bdel}\Phi_{d}^{k_{\max}}$,
if one of the following conditions is fulfilled,  
\bea
(a) && \forall i \,,\;\; N_{\ext}(G_i) > k_{\max}\,,\crcr
(b1)&& \text{for}\;\, d=2, \;\;\forall i \,,\;\; 
\;\;\Om (G_i) > 0 \,,\crcr
(b2)&& \text{for}\;\, d=3,\;\; \forall i \,,\;\; 
\{N_{\ext}(G_i) >2 \,,\;\; \Om (G_i) > 0\} \;\; or\;\;
\{N_{\ext}(G_i) >0 \,,\;\; \Om (G_i) > 1\}  \,,\crcr
(c)&&\forall i \,,\;\;  C_{\partial G_i} > 1  \,,
\eea
and, specifically for $_{1}\Phi_{2}^{2n}$, $n \geq 2$, 
if 
\bea
(d) 
&& \forall i \,, \;\; V(G_i) >1\,, \crcr
(e) &&\forall i \,, \;\;
N_{\ext} (G_i)= k_{\max}\,,
\eea
then $A_{\cG}(\{m_f\}, D)$ converges and is analytic in the strip 
\beq\label{convstrip}
\tilde{\mathcal{D}}^{\sigma} = \left\{ D \in \C \;|\;   0< \Re(D) < 
\bdel+\varepsilon_{\cG}  \right\} ,
\eeq
for $\varepsilon_{\cG} $  a small positive constant depending on the graph. 
\end{theorem}

 Let us comment that although in the following proof of 
Theorem \ref{theo:extan}, the main
variables $k_{\max}$, $d^-$ are always fixed and are
expressed simply, we first perform general calculations
and then sometimes use  the specific values of $k_{\max}$
and $d^-$. Foreseeing the generic dimensional 
regularization of the other tensor models in higher ranks, the method
used and the several relations generated will remain valid. As such,
these are worth to be listed. 

\proof[Proof of Theorem \ref{theo:extan}]
$\bullet$ We shall start by the models 
$_1\Phi^{4}_3$ and $_2\Phi^{4}_2$. 
First, one notices that the following relations hold:
\beq\label{base}
\bdel d^- - 1 >0 \,,
\qquad 
 (\bdel d^- - 1)k_{\max} - 2  \bdel d^- 
 = 0 \,, 
\eeq
where $k_{\max}=4$ stands for the maximal 
valence allowed for vertices in these models. 

Consider the subgraphs $G_i$ associated with Hepp sectors, with positive numbers of vertices $V(G_i) \geq 1$
and lines $L(G_i) \geq 1$ such that (a) holds, i.e., for all $i$, $N_{\ext}(G_i) > k_{\max}$
holds. We define $q(G_i) = N_{\ext}(G_i) - k_{\max}>0$, 
$V_{<} = \sum_{k=2}^{k_{\max}-2} V_k= V_2$ and $n\cdot V_{<}
= \sum_{k=2}^{k_{\max}-2} k V_k=2V_2$  and  write
\bea
&&\textstyle
\frac{L(G_i)}{F_{\inter}(G_i)}
 \geq \frac{L(G_i)}{d^- \tilde{F}_{\inter}(G_i)}
 \geq \frac{ (n \cdot V - N_{\ext})(G_i)}{d^- (2- N_{\ext} + (n-2)V)(G_i)} 
\cr\cr
&&\textstyle 
\geq \frac{ (k_{\max} (V_{k_{\max}} -1) +  n \cdot V_{<} - q)(G_i)}{d^- (2- k_{\max} -q + (k_{\max}-2)V_{k_{\max}} )(G_i)} 
\geq  
\frac{ (k_{\max} (V_{k_{\max}} -1)  - q)(G_i)}{d^- ( (k_{\max}-2)(V_{k_{\max}}-1)  -q  )(G_i)} \,,
\eea
where we used  $(n-2)V_{<}=0$ in an intermediate step. 
Two cases may occur:
(A) if $V_{k_{\max}} -1 =0$, this means that 
the graph is formed with a unique vertex (forgetting mass vertices). This is a tadpole
graph and certainly, $N_{\ext} \leq k_{\max}$,
which is inconsistent with our  initial assumption. 
(B) We consider then $V_{k_{\max}} -1 >0$
and obtain
\begin{align}
\textstyle{\frac{L(G_i)}{F_{\inter}(G_i)}}
> \frac{1}{d^-} ( 1+ \frac{2}{k_{\max}-2}
+ \tilde\varepsilon_{\cG} )
\geq \frac{1}{d^-} (  \frac{k_{\max}}{k_{\max}-2}
+ \tilde\varepsilon_{\cG} )  
> \bdel \geq \Re(D) \,,&\crcr
0 <\tilde{\varepsilon}_{\cG} < 
 \textstyle{ \inf_{i}\frac{2q(G_i)}{[ (k_{\max}-2)(V_{k_{\max}}-1)  -q  ]
(k_{\max}-2) (G_i) }} , &
\end{align}
where \eqref{base} has been used to get ${\boldsymbol{\delta}} $. 
We  can then introduce $\varepsilon_{\cG} = \tilde{\varepsilon}_{\cG}
/d^-$ so that \eqref{convstrip} holds under the condition (a). 

Now, we focus on the conditions (b1)-(b2). Assume $\mathbf{\Omega}(G_i)>0$
for all $i$. It immediately implies that $2\mathbf{\Omega}(G_i)/(d^-)!
\geq d^- - 1$ \eqref{looseb}. However, this bound
is quite loose. For $d=2$, clearly 
$2\mathbf{\Omega}(G_i)=2g_{\tilde\cG}>1$ while $d^- -1=0$.
Meanwhile, for $d=3$, it is still a good bound as
$2\mathbf{\Omega}(G_i)/(d^-)! = \mathbf{\Omega}(G_i) 
\geq d^- - 1=1$. Hence for generic $d=2,3$, assuming $\mathbf{\Omega}(G_i)>0$, we shall use
$2\mathbf{\Omega}(G_i)/(d^-)!\geq 1$. One finds a bound on the
number of internal faces as
\beq\label{finfin}
F_{\inter}(G_i) \leq  -1 + d^- \tilde{F}_{\inter}(\cG)\,.
\eeq
We then obtain new bounds on the ratio  
\beq
\frac{L(G_i)}{F_{\inter}(G_i)} \geq 
\frac{ \frac12(n\cdot V - N_{\ext})(G_i)}{-1 + d^- \tilde{F}_{\inter}(G_i)} 
\geq
\frac{ \frac12(n\cdot V - N_{\ext})(G_i)}{-1 +\frac{d^-}{2}(2-N_{\ext} + (n-2)\cdot V)(G_i)}  \,.
\eeq
The last expression exhibits a different behavior for
$d=2$ and $d=3$. We discuss them separately. 

- If $d=2$,  for all $N_{\ext}(G_i)>0$ and $\Om(G_i) >0$ , we get the bound
\bea\label{ran2}
&&
\textstyle
\frac{L(G_i)}{F_{\inter}(G_i)}
\geq
\frac{ (n\cdot V - N_{\ext})(G_i)}{((n-2)\cdot V-N_{\ext}  )(G_i)} 
 \geq  1+ \frac{2V}{((n-2)\cdot V-N_{\ext}  )(G_i)} \cr\cr
&& \textstyle
\geq  1+ \frac{2V_{k_{\max}}}{((k_{\max}-2)V_{k_{\max}}-N_{\ext}  )(G_i)} 
 > 1 + \frac{2 }{k_{\max}-2} + \tilde{\varepsilon}_{\cG} \geq
1 + 1 + \tilde{\varepsilon}_{\cG} >  {\boldsymbol{\delta}} \geq 
\Re(D)\,,\cr\cr
&& \textstyle
0 < \tilde{\varepsilon}_{\cG} < \inf_i \frac{2 N_{\ext} (G_i)}{[(k_{\max}-2)V_{k_{\max}}-N_{\ext}  ](k_{\max}-2)(G_i)} \,.
\eea

- If $d=3$, assume that $N_{\ext}(G_i) - 2 >0$ and 
$\Om(G_i) >0$, and the following bound is instead valid
\bea
&&
\textstyle
\frac{L(G_i)}{F_{\inter}(G_i)}
\geq
\frac{ (n\cdot V - N_{\ext})(G_i)}{d^-[((n-2)\cdot V-N_{\ext} )(G_i)+1]} 
 \geq \frac{1}{d^-}( 1+ \frac{2V(G_i)-1}{((n-2)\cdot V-N_{\ext}  )(G_i)+1}) \cr\cr
&& \textstyle
\geq \frac{1}{d^-}(  1+ \frac{2V_{k_{\max}}(G_i)-1}{((k_{\max}-2)V_{k_{\max}}-(N_{\ext}-2)  )(G_i)-1} )
>  \frac{1}{d^-}(  1+ \frac{2V_{k_{\max}}(G_i)-1}{((k_{\max}-2)V_{k_{\max}} )(G_i)-1} +  \tilde{\varepsilon}_{\cG}  )\crcr
&& 
>
\frac{1}{d^-}( 1 + 1 + \tilde{\varepsilon}_{\cG} )>  {\boldsymbol{\delta}} \geq 
\Re(D) \,,\cr\cr
&& \textstyle
0< \tilde{\varepsilon}_{\cG}  < \inf_i \frac{[ (2V_{k_{\max}} -1)(N_{\ext}-2)] (G_i)}{[ (k_{\max}-2)V_{k_{\max}} -1][((k_{\max}-2)V_{k_{\max}}-(N_{\ext}-2)  )-1](G_i)}\,.
\eea

- For the last case of the rank $d=3$ imposing that $N_{\ext}(G_i)>0$ and $\Om(G_i) >1$, one has a better bound $F_{\inter}(G_i) \leq  -2 + d^- \tilde{F}_{\inter}(\cG)$, and the rest of the proof is similar to
\eqref{ran2}. 

Hence, setting $\varepsilon_{\cG}= \tilde{\varepsilon}_{\cG}/d^-$,
one recovers \eqref{convstrip} under the statements (b1)-(b2).

Now, we assume that (c) holds i.e. $C_{\partial G_i} >1$ for all $i$. 
 The number of internal faces is again bounded
in the same way as \eqref{finfin}. 
For the rank $d=2$, the analysis can
be redone in the same way as (b1) and allows us to conclude. 
For the rank $d=3$, we have another
piece of information on the number of external 
legs: $N_{\ext}(G_i)\geq 2 C_{\partial G_i} > 2$
(an important fact to notice is that each component of the boundary passes through necessarily at least two external legs in a complex model). 
Thus the analysis of the above case (b2) applies once
again and leads to the same conclusion. 
This achieves the proof of the analyticity domain
of amplitudes in the models $_{\bdel=2}\Phi^{4}_{2}$ and
 $_{\bdel=1}\Phi^{4}_{3}$. 

$\bullet$ Let us discuss the tower of matrix
models $_1\Phi^{2n}_2$. 
We shall first prove that the  analyticity domain 
extends when (d) $V(G_i)>1$ holds. 
It is direct to achieve this by noting
\begin{align}
\textstyle 
\frac{L(G_i)}{F_{\inter}(G_i)}
\geq  \frac{ (n \cdot V -  N_{\ext})(G_i) }{ (2-N_{\ext} + (n-2)\cdot V)(G_i)  }
\geq   1+ \frac{2(V-1)(G_i)}{ ( 2-N_{\ext} + (n-2)\cdot V)(G_i)   }  >  1+ 
\varepsilon_{\cG} > \bdel \geq  \Re(D)\,, & \crcr
\textstyle 
0 <  \varepsilon_{\cG}  < \inf_i  \frac{2(V-1)(G_i)}{ ( 2-N_{\ext} + (n-2)\cdot V)(G_i)   } \,. &
\end{align}
Then the analyticity domain of $A_{\cG}$ extends to \eqref{convstrip}.

The fact that the domain extends under the assumption (d)
has a consequence for the same study now under the 
 conditions (a) and (e). Meanwhile, the reason that under (b1) and (c)
the domain extends as well is derived simply from 
the similar situation of the model $_2\Phi^4_2$. 

- Consider all graphs $G_i$ such that
 (a) or (e) holds then $N_{\ext}(G_i) \geq k_{\max} $.
It  implies that either we are using 
a number of vertices larger than 1 or
a unique vertex with the maximal valency
$k_{\max}= N_{\ext}(G_i)$ and
no lines are present in the graph. Clearly, both situations 
lead to a convergent amplitude. 

- Consider now (b1). For all $i$, $\Om(G_i)>0$ such that
 \eqref{finfin} holds now. 
The calculations are similar to the previous case (b1). We have, for all $N_{\ext}(G_i) >0$,  
\begin{align}\label{2v}
&&
\textstyle
\frac{L(G_i)}{F_{\inter}(G_i)}
\geq 
\frac{ (n\cdot V - N_{\ext})(G_i)}{((n-2)\cdot V-N_{\ext}  )(G_i)} 
\geq 1+ \frac{2V}{((n-2)\cdot V-N_{\ext}  )(G_i)} 
>  1+\varepsilon_{\cG}  > \bdel \geq \Re(D)\,, \crcr
&& \textstyle
0 < \varepsilon_{\cG} < \inf_i \frac{2 V (G_i)}{((n-2)\cdot V-N_{\ext})(G_i)}
\,.
\end{align}

- Finally, assuming (c) so that $C_{\partial G_i}>1$, 
we use again \eqref{2v} to complete the proof. 

\qed

\noindent{\bf Meromorphic structure.} The next task is to prove
the meromorphic structure of the amplitudes $A_{\cG}(\{m_f\}, D)$ on
the  domain $\tilde{\mathcal D}^{\sigma}$ when all conditions
listed in Theorem \ref{theo:extan} are dropped. 

From Theorem \ref{theo:extan}, the only 
cases which lead to divergent amplitudes
can be listed as follows: 

\begin{enumerate}
\item[(di)]  In the  $_1\Phi^4_3$ model,  graphs 
with  $N_{\ext}(\cG)\leq 4=k_{\max}$ with $C_{\bG}=1$ and

a-  $\Om(\cG) =0$ (melonic with melonic boundary) and $N_{\ext}(\cG)=4$;

b- $\Om(\cG) =0$ (melonic with melonic boundary) and $N_{\ext}(\cG)=2$;

c- $\Om(\cG) =1$ (non melonic with melonic boundary) and $N_{\ext}(\cG)=2$.

\item[(dii)] In the  $_2\Phi^4_2$ model,  graphs with 
$N_{\ext}(\cG)\leq 4=k_{\max}$ with $C_{\bG}=1$ and

a- $\Om(\cG) =0$ (planar) and $N_{\ext}(\cG)=4$;

b- $\Om(\cG) =0$ (planar) and $N_{\ext}(\cG)=2$.

\item[(diii)] In the  $_1\Phi^{2n}_2$ model,  graphs with 
$V=V_{2k}=1$, $k\leq n=k_{\max}/2$, $N_{\ext}(\cG) < 2k$ with $C_{\bG}=1$ and $\Om(\cG) =0$ (planar).

\end{enumerate}

The above list of primitively divergent graphs matches with the one issued in \cite{Geloun:2013saa}. 

We come back to the integrand of the  amplitude \eqref{eq:ampli3}
and focus on the following function:
\bea\label{II}
&&
I^{\sigma}_{\cG} (\{x_l\}, \{m_f\}, D) =\crcr
&& \Big[\prod_{l =1}^L   \frac{\left( 1- \prod_{k=l}^{L} x_k \right)^{\mu -1} }
{\left( 1 + \prod_{k=l}^{L} x_k\right)^{\mu +1} } \Big]
 \prod_{f \in \cF_{\ext} } \prod_{l \in f} \left( 
\frac{1 - \prod_{k=l}^{L} x_k }{ 1 + \prod_{k=l}^{L} x_k} \right)^{|m_f|}
\left[  \frac{\tilde{U}^{\ev}_\cG ( \{ x_l \} ) }{(1+ U_{\cG}'(\{ x_l\} )) } \right]^{D}.
\eea
Since all $x_l$ are positive, $I^{\sigma}_{\cG}$ is a continuous function in the $x_l$ variables and admits a simultaneous Taylor expansion 
in the $x_l$'s around $x_l= 0$. 

At this point, we use a different strategy from the one 
introduced in \cite{Bergere:1977ft}. We do not perform
the generic Taylor expansion in all the $x_l$'s before
integrating all $x_l$'s and getting the poles and 
meromorphicity conditions on the amplitude using diverging 
subgraphs. 
The reason motivating our present study is that, 
given a divergent primitively graph $S$ (with characteristics listed above)
of a graph $\cG$, 
we  know exactly on which variables we must perform
the Taylor expansion for extracting the divergence. 
Precisely, by Lemma \ref{lem:ampfactor}, we know
that the variables in which one must perform a Taylor series
are the ones which ``tie'' $S$ and $\cG/S$. 
One must then prove that, 
the first order of the Taylor expansion is large enough
to ensure  analyticity at a higher order.

Consider then a primitively divergent subgraph $S \subset \cG$,
 and a decomposition in  Hepp sectors $\sigma'$,
$t_1\leq t_2 \leq \dots \leq t_{L(S)}$ of the lines of $S$
and introduce the usual change 
of variables in $x_k$ as in \eqref{changof}. 
Because $S$ is primitively divergent, for all subgraphs $G_i$
except $S= G_{L(S)}$, we have $\Re( \omega_{\rm d}(G_i)) >0$
and  $\Re( \omega_{\rm d}(G_{L(S)})) \leq 0$.
Lemma \ref{lem:ampfactor} ensures, using \eqref{amplfactor}
and writing $ \cF_{\bullet}'(\cG,S)= \cF_{\bullet}'$,
 that
\bea
A_{\cG} & =& 
\int [dt_l]_{\cL(S)}  \; \tilde{A}_{S}\, \Big[ 1+ \sum_{f\in \cF_{\ext}' }R_f(t_l) \cO_S\Big]
A_{\cG/S} \crcr
&& 
+\int [dt_l]_{\cL(S)}  \; \tilde{A}_{S}\, \Big[\int [dt_{l'}]_{\cL(\cG/S)}\sum_{f\in \cF_{\inter}'}
R_f(t_l) \cO'_{S}(t_{l'})  + O(t_l^2)\Big]   \tilde{A}_{\cG/S}   
\cr\cr
 &=&   \sum_{\sigma'}
 A_{\cG}^{\sigma'} \,, 
\eea 
with some appropriate operator insertions $\cO_{S}$ and $\cO'_{S}$.
Now, we perform the change of variables $t_l \to x_k$, 
and, after denoting $x_{L(S)}=x$ and singling out the integration in
$x$, one gets
\bea\label{chic}
A_{\cG}^{\sigma'}& =& \int dx\; 
x^{\omega_{\rm d}(G_{L(S)})-1}
\Big( A_{\cG/S}\,   {\mathcal{J}}^{\sigma'}_S+ x\, \Big\{\cR_S(x,\{m_f\}) A_{\cG/S} \crcr
&& 
 + \int [dt_l']_{l\in \cL(\cG/S)}\cR'_S (x,\{m_f\},\{t'_l\}) \tilde{A}_{\cG/S}
\Big\}
+ O(x^2)\Big)\,, 
\\
{\mathcal{J}}^{\sigma'}_S &=& \int [ \prod_{k\neq L(S)} dx_k]\,
\tilde{I}^{\sigma'}_{S} (\{x_l\}, \{m_f\})\,, \crcr
x\cR_S(x,\{m_f\}) 
& =& \int [ \prod_{k\neq L(S)} dx_k]\,
\tilde{I}^{\sigma'}_{S}(\{x_l\}, \{m_f\}) \sum_{f\in \cF_{\ext}'} \tilde{R}_{f}(x_l) \cO_S \,,
 \crcr
&&
\crcr
x\cR'_S(x,\{m_f\},\{t'_l\}) 
& = &\int [ \prod_{k\neq L(S)} dx_k]\,
\tilde{I}^{\sigma'}_{S}(\{x_l\}, \{m_f\})
\sum_{f\in \cF_{\inter}'} \tilde{R}_{f}(x_l) 
\cO'_S(t'_l) \,,
\nonumber
\eea
 where $\tilde{R}_{f}(x_l)$ are  obtained from  $R_f$
after the change of variables and 
where $\tilde{I}^{\sigma'}_{S}$ is of form 
 $(\prod_{k\neq L(S)} x_k^{\omega_{\rm d}(G_k)-1})I^{\sigma'}_{S}$
and $I^{\sigma'}_{S}$ is given by \eqref{II} except that the product
over external faces must be restricted to $\cF^*_{\ext}$.
Interestingly, one notes that we have traded a big-O function in
$t_l^2$ for the same function in $x^2$ without having
lost any terms.   

The Taylor expansions of 
$\cR_S(x,\{m_f\}) $ and of $\cR'_S(x,\{m_f\},\{t'_l\}) $
around $x=0$ involve the expansions of 
$\tilde{I}^{\sigma'}_{S}(\{x_l\}, \{m_f\})$ and of $R_f(x_l)$ 
(as a simple polynomial). This yields extra $x^{p}$ factors, $p\geq 0$. 
Thus, the convergence of the integral of each term in the expansion
is expected to improve for the simple reason that $\Re(\omega_{\rm d}(G_k))+p\geq \Re(\omega_{\rm d}(G_k))$, for 
$p \geq 0$. 

Assume that the subgraph $S$ of $\cG$ obeys one of the conditions (di)--(diii). 
The following cases might occur:  

- For a 4-point subgraph under conditions (dia) or (diia),
or a
$2k'$-point subgraph obeying (diii), the integration over $x$
yields the divergent part of $A^{\sigma'}_{\cG}$ like
the first term of \eqref{chic}: 
\bea
{\rm A}_{\rm 4pt/2k'}(D) = \frac{c_{\lambda}}{\omega_{\rm d}(G_{L(S)})}
 = \frac{c_{\lambda}}{ L - D F_{\inter}(S)}\,,
\eea
for some constant $c_{\lambda}$ which incorporates the
couplings and other constant factors. We get a pole
at
\beq
D_0 =   \bdel\,,
\eeq
with the further condition that, in the $_{\bdel =1}\Phi^4_3$ and
$_{\bdel =2}\Phi^4_2$ models, 
$S$ does not contain mass vertices $V_2(S)=0$. If $V_2(S)\ne 0$,
in these models, we discover poles at the rational values
\beq\label{d1pol}
D_1 = \frac{2}{d^-}[1 + \frac{V_2}{2(V_4-1)}] =  \bdel + \frac{V_2}{d^-(V_4-1) }\,, \qquad 
V_4 -1 >0\,.
\eeq
The last condition on $V_4$ is imposed since
we want $N_{\ext}=4=k_{\max}$ (as previously 
discussed, this is only possible if there is a number
of 4-valent vertices strictly greater than 1). 
Note also that \eqref{d1pol} means that for a fixed graph $S$, $V_2$ 
and $V_4$ are certainly fixed, thus $V_2/[d^-(V_4-1)]$ cannot be very large
otherwise $D_1$ would lie outside the strip $0<\Re(D)<\bdel + \varepsilon_S$. Therefore, this gives us an estimate
like $0<\varepsilon_S< \frac{V_2}{d^-(V_4-1) }$ to get
a convergent integral whenever $V_2\neq 0$.

Now we inspect a generic term in the Taylor expansion. 
This situation is similar to the case when $V_2\neq 0$. 
 Under the current conditions, let
us assume that we integrate another term
$\int x^{\omega_{\rm d}(G_{L(S)}) + p-1}$, $p >0 $,
in the Taylor series. 
We get a pole at $D'_1(p) = \bdel + (V_2+ p)/[d^{-}(V_4-1)]$,
for the models $_\bdel\Phi^4_3$ and $_\bdel\Phi^4_2$.
For $_\bdel\Phi^{2n}_2$, one obtains a pole 
at $D''_1(p)= \bdel + p/[d^-(k-k')]$, 
where $2k$ is the vertex valence 
and $N_{\ext}=2k'<2k$. 
Using the same previous argument,  
this also means that $p+V_2$ (resp. $p$) cannot be very large
otherwise $D'_1(p)$ (resp. $D''_1(p)$) would lie outside the strip $0<\Re(D)<\bdel + \varepsilon_S$. 
 It is immediate that this term certainly converges for a given 
estimate of $\varepsilon_S$ that one can choose as
$0<\varepsilon_S<\frac{(V_2+1)}{d^-(V_4-1) }$
(resp. $0<\varepsilon_S<\frac{1}{d^-(k-k') }$). 
If we allow $D$ to explore in the whole complex plane
(and wander outside of the strip $0<\Re(D)<\bdel + \varepsilon_S$), 
then we discover that the function is meromorphic
with poles at $D_0$, $D_1$, $D_1'(p)$ and $D''_1(p)$ 
which are rationals. 

- For a 2-point subgraph of $_{\bdel =1}\Phi^4_3$  under assumption (dic), such that $V_2 =0$, one obtains again a pole at $D=\bdel$. 
However, when $V_2 \neq 0$, the poles shift to
the rational values
\beq
D_1''' = \bdel + \frac{V_2}{2V_4-1}\,.
\eeq
The fact that the remaining Taylor terms converge
in a precise strip  follows from the same arguments
invoked above.

- For a 2-point function respecting conditions 
(dib) or (diib), we get the following contributions
to the amplitude of $A^{\sigma'}_{\cG}$
\bea
{\rm A}_{\rm 2pt}(D) = \frac{c_{1;\lambda}}{\omega_{\rm d}(G_{L(S)})}
 + \frac{c_{2;\lambda}}{\omega_{\rm d}(G_{L(S)}) + 1}\,, 
\eea
for some constants $c_{1;\lambda}$ and $c_{2;\lambda}$. 
For $V_2=0$, we have two types of poles located at rational values
\beq\label{d0d2}
D_0 = \bdel \,,  \qquad 
D_2 = \bdel - \frac{1}{d^- V_4}\,,
\qquad 
V_4 >0\,,
\eeq
where the last condition $V_4 >0$ refers to the trivial fact
that a 2-pt subgraph  with $V_4=0$ cannot
diverge. If $V_2\neq 0$, the poles shift to 
\beq
D_0' = \bdel + \frac{V_2}{d^- V_4 }\,,
\qquad 
D_2' = \bdel + \frac{V_2-1}{d^- V_4}\,,
\qquad 
V_4 >0\,.
\eeq
 As performed previously, the case $V_2\neq 0$
and the integration of higher order Taylor terms at
 $p \geq 2$ with poles like 
$D_3(p) = \bdel + (V_2-1+p)/[d^- V_4]$, 
lead to a similar conclusion as above.

One remark must be made at this stage. In the above 
pole equations including the variable $V_2$, 
the value of $V_2$ cannot be large as compared
to $V_4$. Indeed, if $V_2$ grows faster 
than $V_4$, most of the above equations
yield $\Re(\omega_{\rm d}(S))>0$, immediately
leading to convergence. One must also remember
that the analyticity domain is only extended up to
a small enough parameter $\varepsilon_{S}$
such that $0<\Re(D)< \bdel + \varepsilon_{S}$.
The appearance of $V_2$ if not trivial
must be simply considered as a mild modification
of the case when $V_2 =0$. The later case
simplifies the formalism and one finds only  two 
types of poles $D_0 = \bdel$ and $D_2 = \bdel - 1/d^-V_4$
\eqref{d0d2}. 
They certainly lie in the strip $0<\Re(D)< \bdel + \varepsilon_{S}$ 
and are quite reminiscent of the poles found
 in \cite{gurau4}. 

We have listed all poles of the amplitude $A^{\sigma'}_{\cG}$
related to  a primitively divergent subgraph $S$ of $\cG$. 
These poles are always located at rational values.

We are in position to achieve the main statement on
the meromorphic structure of $A_{\cG}$. 
Let us now come back to the beginning and consider $\sigma$
a Hepp sector of $\cG$ and its associated amplitude
expansion $A_{\cG} = \sum_{\sigma}A^{\sigma}_{\cG}$.
Consider all primitively divergent subgraphs $S$ of $\cG$. 
Each subgraph $S$ must appear at least in one
sector $\sigma$, and therefore $S$ must be included in some $G_{i}$
with $i \geq L(S)$. We can conclude that the amplitude $A_{\cG}$
is convergent, defines an analytic
function in $D$ in the strip ${\mathcal{D}}$ \eqref{streep} 
and admits an analytic continuation
as a meromorphic function in the strip $\tilde{\mathcal{D}}^{\sigma} = \left\{ D \in \C \;|\;   0< \Re(D) < 
\bdel+\varepsilon_{\cG}  \right\}$, with  
\beq
0 < \varepsilon_{\cG} < \inf_{S \subset \cG; \; S \text{ primitively divergent}} \varepsilon_{S}\,. 
\eeq

In fact, the above bound on $\varepsilon_{\cG} $ can 
be improved in a more useful way using Hepp sectors.  
A crucial observation is that the set of the $G_i$'s is totally
ordered under inclusion, $G_i \subset G_{i+1}$. 
Precisely, the set of all of their connected components $G^{(k)}_i$
of all $G_i$'s is partially ordered and forms an abstract tree with nodes the $G^{(k)}_i$'s. 
This is the Gallavotti-Nicol\`o tree \cite{galla}. The set
$\{G^{(k)}_i\}_{k;i}$ also defines the set of quasi local (or  high) subgraphs in the
formulation of \cite{Rivasseau:1991ub}. 
The introduction
of such tree becomes extremely useful  for the
treatment of the so-called overlapping divergences appearing
in ordinary renormalized expansion in the coupling constants. 
The point is that divergences in some sector $A^{\sigma}_{\cG}$
are now indexed by disconnected subgraphs 
organized into a forest. 

Thus, given two primitively divergent subgraphs $S$ and $S'$
of $\cG$, the only relevant case is when $S\cap S' = \emptyset$.
They form connected and disjoint subgraphs in 
the same Hepp sector $\sigma$, 
$t_{1}\leq  \dots \leq t_{L(\cG)}$.  
Let us assume, without loss of generality, that 
the lines of $S$ are indexed from $1$ to $L(S)$
and those of $S'$ indexed from $p+1$ to $p+L(S')$, with $p > L(S)$.
In other words, consider the ordering of lines of $\cG$ like 
\beq
t_{1}\leq  \dots \leq t_{L(S)} \leq \dots \leq 
t_{p} \leq t_{p+1} \leq \dots \leq t_{p+ L(S')} \leq \dots \leq t_{L(\cG)}\,.
\eeq
Then there exist two independent variables $x_{S}=x_{L(S)}$ and $x_{S'}
= x_{p+L(S')}$ that we can use to perform distinct
integrals of the same form as \eqref{chic} with this time
the final factor as $\tilde{A}_{(\cG/S)/S'}$. 
The above reasoning applies (note that
the order of the integrations does not matter either
if we solve the most nested divergence which corresponds 
to $S$ and then the second associated with $S'$ or the other way around). We proceed by induction on the rest of
primitively divergent graphs in this sector. 
In the end, the amplitude $A^{\sigma}_{\cG}$ 
is meromorphic in the strip $0<\Re (D) < \bdel +\varepsilon^{\sigma}_{\cG}$
with poles at rationals, where 
\beq
0<\varepsilon^{\sigma}_{\cG}<\inf_{S \subset \mathfrak{F}^{\sigma}}\varepsilon_{S}
\eeq
where $\mathfrak{F}^{\sigma}$ is the forest of connected primitively divergent subgraphs of $\cG$ related to the Hepp sector $\sigma$. Summing over all Hepp sectors
$\sigma$, we simply have to infer that $A_{\cG}$ defines
a convergent integral and a meromorphic function 
in the strip $0<\Re (D) < \bdel +\varepsilon_{\cG}$ with 
\beq
0 <\varepsilon_{\cG} < \inf_{\sigma}\varepsilon^{\sigma}_{\cG}\,.
\eeq
We have finally achieved the following statement:
\begin{theorem}[Meromorphic structure of the amplitudes]
\label{theo:merom}
The amplitude $A_{\cG}(\{m_f\}, D)$ of the model $_\bdel\Phi^{k_{\max}}_{d}$ (listed above) is a meromorphic function
in $D$ on the strip 
\beq
\tilde{\mathcal{D}}^{\sigma} = \left\{ D \in \C \;|\;   0< \Re(D) <  \bdel+\varepsilon_{\cG}  \right\},
\eeq 
for $\varepsilon_{\cG}$ a small positive quantity depending
on the graph $\cG$.  
\end{theorem}

\subsection{Renormalization}
\label{subsect:renorma}

From this point, the standard definition
of the subtraction operator \cite{Bergere:1977ft}
can be applied. The discrepancy between the present
study and the formalism therein is that we are considering a radically different set of primitively divergent subgraphs. We will only sketch
the definition of the subtraction operator (details
can be found in the above reference). 

We introduce the operator $\tau$ as the generalized Taylor
operator defined as follows: let $f(x)$ be a function
such that $x^{-\nu} f(x)$ is infinitely differentiable
at $x=0$, $\nu$ might be complex. One defines
\beq
\tau^n_{x} f(x) = x^{-\xi -\eps } T_x^{n + \xi} (x^{\xi + \eps} f(x))\,,
\quad 
T_x^{m \geq 0} (f) = \sum_{k=0}^{m} \frac{x^k}{k!} f^{(k)}(0)\;,
 \quad T_x^{m <0}(f) =0\,, 
\eeq
where $\xi$ is an integer obeying $\xi \geq - E(\nu)$,
 $E(\nu)$ is the smallest integer $\geq \Re(\nu)$,
and $\eps = E(\nu) - \nu $.  

Consider a subgraph $S\subset \cG$ and a function $f(\{t_l\})$ on 
the graph $\cG$, $l \in \cL(\cG)$. We associate $S$ with the
following operator:
\beq
\tau_{S}^{n}\, f(\{t_l\}) = 
\Big[\tau^n_{\rho} f(\{\rho t_l\}_{l \in \cL(S)}; 
\{t_l\}_{l \in \cL(\cG/S)})\Big]\Big|_{\rho =1}\,.
\eeq
Finally, one defines the subtraction operator
acting on amplitudes as
\beq\label{rop}
R = 1 + \sum_{\mathfrak{F}} \prod_{S \in \mathfrak{F}}
\big(-\tau_{S}^{-L(S)}\big)\triangleright\,,
\eeq
where the sum is performed over the set $\mathfrak{F}$
of all forests of primitively divergent subgraphs,
and the symbol $\triangleright$ refers to the fact
that the operator $R$ must act on the integrand
of a given amplitude. 
Another important remark is that any
subtraction operator is defined by two actions:
(1) the action
of a Taylor operator on the amplitude integrand targeting
specifically the variables which are associated with a primitively divergent subgraph $S$ (using in particular scaling properties of Lemma \ref{lem:polyfactor}) and (2) the subtraction of the pole induced
by the diverging part for each term in the Taylor expansion.
The procedure is completely standard at this stage for our
models \eqref{renmod}.
Thus, 
\bea
R A_{\cG} = A^{\rm ren}_{\cG}
\eea
is a finite integral and an analytic function in $D$ on the strip $\tilde{\mathcal{D}}^{\sigma}$.

Finally, let us mention that written in the way \eqref{rop}, the operator $R$ might not seem to be related to a Hepp decomposition.
It is simply a subtraction operator acting on the
amplitude which will prove to lead to a convergent
integral. There is of course a way to make this operator
compatible with Hepp sectors and 
has been defined in the context of TGFT models \cite{Carrozza:2013mna}.

\section{Polynomial invariants}
\label{sect:poly}

We study now in  details the polynomials obtained in the
parametric amplitudes \eqref{eq:defa}. $U^{\od/\ev}_{\cG}$ 
 will be referred to the first Symanzik polynomial associated
with the model amplitude. Since $W_{\cG}$ is not a polynomial, it cannot be directly called the second Symanzik polynomial. Nevertheless, we can study its properties as well. 
It is worth emphasizing that the following analysis does not
specially focus on the models listed in \eqref{renmod}. 
The amplitude \eqref{eq:defa} is completely 
general for a generic rank $d\geq 2$ Abelian model with particular
linear kinetic term. Therefore, the following
study is valid for any model of this kind using
tensor invariant vertices (in the sense of definition \eqref{intergen}) 
and rank $d$ stranded lines.  Furthermore, as one can realize
in a straightforward manner, the definitions of the polynomial $U^{\od/\ev}_{\cG}$ and function $W_{\cG}$ can be extended
to the larger class of rank $d$ colored tensor graphs as defined in Subsection \ref{subsect:graphs}. 
This  means that these generalized 
polynomials should appear in the parametric amplitudes
of dynamical Abelian colored tensor models \cite{Geloun:2013zka}. 
One must simply observe that, in the definition of
the polynomials 
$U^{\od/\ev}_{\cG}$ and $W_{\cG}$, 
 the factorization in faces and the 
bi-coloring of strands play the main roles.
The following analysis only 
relies on these ingredients which are in both
models (the unitary tensor invariant and rank $d$ colored
models).
In the following, we will not distinguish the study
between these frameworks. Any graphs which 
might come from these models are simply referred to 
rank $d$ color tensor graphs.  
Finally, the particular case of $d=2$
might generate some peculiarities that we will often address
in a separate discussion. For higher rank $d>2$ illustrations,
we will restrict ourselves to $d=3$ which is 
already not trivial. The higher rank case can be deduced
from the $d=3$ case.

The usual Symanzik polynomials 
must satisfy some invariance properties under specific operations
on their graphs. In scalar quantum field theory, 
it is well-known  that such polynomials
satisfy a contraction/deletion rule, hence, by a famous
universality theorem, define Tutte polynomials \cite{riv}. 
For the GW model in 4D, the polynomials on 
ribbon graphs discovered in  the parametric representation of this model \cite{gurau3} were deformed versions 
of the Bollob\`as-Riordan polynomial \cite{bollo3}\cite{bollo2}. 
The recurrence relation obeyed by these invariants is however much more involved \cite{Krajewski:2010pt} (a four-term recurrence  using Chmutov partial duality \cite{schmu}). Our remaining task is to investigate 
the types of relations which are satisfied by the identified
functions $U^{\od/\ev}_{\cG}$ and $\tilde{W}_{\cG}$
($W_{\cG}$ will satisfy relations which can be inferred
from these points). 

The rest of the section is divided into  three parts. 
The first part focuses on the study of $U^{\od/\ev}_{\cG}$ and $\tilde{W}_{\cG}$ and the type of modified relation that they satisfy. 
In rank $2$, a connection with the work by Krajewski et al. \cite{Krajewski:2010pt} is rigorously established in the second point. 
Motivated by the two initial discussions, in the third part of this section, 
we  identify a polynomial that we call of the second kind, $\cU_{
\cG}$, which is stable under a contraction reduction. To the best of our knowledge, it is for the first time that
such a rule without referring to the deletion operation can be defined on a graph polynomial invariant. As an intriguing object worth to be exemplified, we list its properties and include several illustrations. The definition of the new polynomial is however totally  abstract and, of course, it remains an open question if there exists a quantum field theory having such a polynomial appearing in its parametric amplitudes.  

Few remarks must be made at this stage. 
The cut of an edge in a tensor invariant theory
is performed in the same way as is done in the 
colored case as discussed
in Subsection \ref{subsect:graphs} (see Fig.\ref{fig:cutstr}). 
However, the contraction of an edge
 in a graph in an invariant tensor model
must be understood as
the contraction of a stranded line of color 0
with the same rule explained in Subsection \ref{subsect:graphs}
(see Fig.\ref{fig:strcont}). It turns out that
our final statements are always independent on the type
of models either tensor invariant or colored. 
Finally, in the following a rank $d$ graph can either be 
a ribbon graph with half-ribbons or a rank $d>2$
colored tensor graph 
(either in the sense of Section \ref{sect:sgraphs}
or coming from the gluing of rank $d$ 
unitary tensor invariants). 

\subsection{Polynomials of the first kind}
\label{subsect:poly1}

The objects of interest are the polynomials
$U^{\od/\ev}_{\cG}$ and $\tilde{W}_{\cG}$. These
polynomials are called of the first kind. 

Ordinary operations of contraction and deletion 
of edges of a graph $\cG$ have been defined
in Section \ref{sect:sgraphs}.  
We recall some terminology and give precisions:

-  Given an edge $e$ and a face $f$, we write $e \in f$ when the face $f$ passes through $e$ (we also say that ``$e$ belongs to $f$'').  If $f$ passes through $e$ exactly $\alpha$ times,
we denote as $e^{\alpha}\in f$. Note that $ 0\leq \alpha \leq 2$. From now, $e^{1} \in f \equiv e \in f$. 

- In the rank $d>2$, the theory is colored and
always $e^{\alpha}\in f$, $\alpha$ is necessarily 1. 

 - In this section, ``contraction'' always refers to soft contraction.

- Given $e^{\alpha} \in f$, we denote $f/e$ (resp. $f-e$, $f\vee e$) the face resulting from $f$ after the contraction (resp. the deletion, the cut) of $e$ in $\cG$ yielding $\cG/e$ (resp. $\cG-e$, $\cG\vee e$).

The following statement holds.

\begin{lemma}[Face contraction]\label{lem:facontr}
Let $e$ be an edge of $\cG$ a rank $d\geq 2$ graph, which is not a loop and consider $e^{\alpha}\in f$, $f \in \cF_{\inter}$. 
We have
\begin{enumerate}
\item[(i)] If $\alpha=1$, then 
\bea
A^{\od}_f = t_e \, A^{\ev}_{f/e} + A^{\od}_{f/e} \,,
\qquad
A^{\ev}_f  = t_e \, A^{\od}_{f/e} + A^{\ev}_{f/e} \,,
\eea
\item[(ii)] If $\alpha=2$, then 
\bea
A^{\od}_f  = 2t_e \, A^{\ev}_{f/e} + (t_e^2+1)A^{\od}_{f/e} \,,
\qquad
A^{\ev}_f  = 2t_e \, A^{\od}_{f/e} + (t_e^2+1)A^{\ev}_{f/e} \,.
\eea
\end{enumerate}
When $e$ is a trivial untwisted (resp. twisted) loop,
then one can only have  $e\in f$ (resp. $e^2 \in f$) and
(i) (resp. (ii)) holds. 
\end{lemma}
\proof Clearly, the even face polynomial and odd face polynomial 
play a symmetric role. We shall prove the claims for the odd
case, from this, the even case can simply be inferred.  

Let us assume that $e\in f$. This means that the factor $t_e$
appears just once in $A^{\od}_f$ so that 
\beq
A^{\od}_f(\{t_l\}) =
\Big[\sum_{
\shortstack{ $_{A \subset f}$ \\ $_{|A| \; {\rm odd} ; \; e \in A}$}}
+\sum_{
\shortstack{ $_{A \subset f}$ \\ $_{|A| \; {\rm odd} ; \; e \notin A}$}
}\Big]  \;\prod_{l \in A} t_l\,.
\eeq
 The subsets $A \sset f$ such that $e\notin A$ correspond
exactly to subsets $A' \sset f/e$. This shows that 
$\sum_{
\shortstack{ $_{A \subset f}$ \\ $_{|A| \; {\rm odd} ; \; e \notin A}$}
}\prod_{l \in A} t_l
= A^{\od}_{f/e}$. Meanwhile, the subsets $A \sset f$ such that $e \in A$
have a common factor $t_e$. This simply factorizes and yields
the even monomials generated by $A \sset f/e$. 

Assume now that $e^2 \in f$. The terms $(1+t^2_e)$ and $2t_e$ must
appear in $A^{\od}_f$. We have 
\beq
A^{\od}_f(\{t_l\}) =
\Big[\sum_{
\shortstack{ $_{A \subset f}$ \\ $_{|A| \; {\rm odd} ; \; e \in A}$}}
+\sum_{
\shortstack{ $_{A \subset f}$ \\ $_{|A| \; {\rm odd} ; \; e^2 \in A}$}
} + 
\sum_{
\shortstack{ $_{A \subset f}$ \\ $_{|A| \; {\rm odd} ; \; e \notin A}$}
}\Big] 
\;\prod_{l \in A} t_l\,.
\label{face2}
\eeq
One notices that factoring out $t_e^2$ common in all 
monomials in the middle sum, the odd monomials generated by 
$A \sset f$, such that $e^2 \in A$ and $l \in A, l \neq e,$
are precisely those generated by $A \sset f$ such that $e \notin A$.  
Moreover, the last sum coincides with $A^{\od}_{f/e}$ for the same
reason invoked above (in the case $e^1 \in f$). Then the two last sums are nothing
but $(t_e^2+1) A^{\od}_{f/e}$. In the first sum in \eqref{face2}, 
after factoring out $2t_e$, for the same reason as previously stated, 
we obtain exactly the even monomials  generated by the contracted
face $f/e$.
 
The last point on trivial loops can be inferred in the similar
way.

\qed

We are in position to investigate the recurrence rules obeyed
by $U^{\od/\ev}_\cG$ in rank 2.

\begin{proposition}[Broken recurrence rules for $U^{\od/\ev}$ in rank 2]
\label{theo:cdel}
Let $\cG$ be a ribbon graph with half-ribbons,  $\cF_{\inter;\cG}$
and $\cF_{\ext;\cG}$ be its sets of internal and external faces,
respectively, and $e$ be a regular edge of $\cG$. Then: 

\begin{enumerate}
\item[(i)] 
If $e$ belongs only to external faces then  
\beq
\label{ext2}
U^{\od/\ev}_{\cG}  = U^{\od/\ev}_{{\cG}/e} \,.
\eeq
Furthermore, if the deletion of the edge $e$ does not generate a new internal face 
$U^{\od/\ev}_{\cG}  = U^{\od/\ev}_{{\cG}/e} = U^{\od/\ev}_{{\cG} - e} $.
If it generates a new internal face $f$, then
$A^{\od/\ev}_{f} U^{\od/\ev}_{\cG} = U^{\od/\ev}_{\cG-e}$.
\item[(ii)] If $e \in f$  and $e \in f'$, $f \in \cF_{\inter;\cG}$ and 
$f' \in \cF_{\ext;\cG}$, we have $U^{\od/\ev}_{\cG\vee e} = U^{\od/\ev}_{\cG -e}$ 
and 
\beq
\label{extint}
U^{\od/\ev}_{\cG} 
= t_e \; A^{\ev/\od}_{f/e} \; U^{\od/\ev}_{\cG  - e} + U^{\od/\ev}_{\cG/e} \,.
\eeq
\item[(iii)] If $e \in f$  and $e \in f'$, $f \ne f'$, 
and $f,f'  \in \cF_{\inter;\cG}$, we get 
\beq\label{fnf}
U^{\od/\ev}_{\cG} 
= t_e \; U^{\od}_{\cG - e} + U^{\od/\ev}_{\cG/e} + t_e^2 \; A^{\ev/\od}_{f/e} \; A^{\ev/\od}_{f^\prime/e} \; U^{\od/\ev}_{\cG \vee e}\;.
\eeq
\item[(iv)] If  $e^2 \in f$, $ f \in \cF_{\inter;\cG}$, then 
two cases occur:
\subitem(a) the deletion of $e$ yields two internal faces $f_1$ and $f_2$,
then 
\beq\label{e2ff}
 \left\{ \begin{array}{c} 
U_\cG^{\od}\\
U_{\cG}^{\ev}\end{array}
\right. =
 \left\{ \begin{array}{c}
\left(1 + t_e^2\right)  \; U^{\od}_{\cG/e}
+
2  t_e \, U^{\od}_{\cG-e} 
+
2 t_e \, A^{\ev}_{f_1} A^{\ev}_{f_2} \, U^{\od}_{\cG  \vee e}
 \\
\left(1 + t_e^2\right)  \; U^{\ev}_{\cG/e}
+
2  t_e( A^{\ev}_{f_1} \; A^{\od}_{f_2}
+
 A^{\od}_{f_1} \; A^{\ev}_{f_2}) \, U^{\ev}_{\cG  \vee e} 
\end{array} \right.   \, .
\eeq
\subitem(b)
the deletion of $e$ yields one internal face $f_{12}$, 
\beq
\label{e2f}
U^{\od /\ev}_{\cG} 
=( 1 + t_e^2 ) \, U^{\od /\ev}_{\cG/e} + 2 t_e \, A^{\ev / \od}_{f_{12}} \, U^{\od /\ev}_{\cG \vee e} \,.
\eeq

\end{enumerate}
\end{proposition}

\proof  See Appendix \ref{app:theocdel}. 

\qed 

For special edges, the above proposition is still 
valid but simplifies drastically: 

- if $e$ is a bridge, then
under condition (i), \eqref{ext2} and the following relations 
are all valid and, under condition (iv.a), \eqref{e2ff} holds.
These are the only possibilities for a bridge. 

- if $e$ is a  trivial untwisted loop, under condition (i), 
\eqref{ext2} is valid and since the contraction of a such a loop
cannot create a new internal face, we always have
$U^{\od/\ev}_{\cG}  = U^{\od/\ev}_{{\cG}/e} = U^{\od/\ev}_{{\cG} - e}.$
Assuming (ii), \eqref{extint} holds as well. 
Now, under (iii), one proves that \eqref{fnf} is true. 
We do not have any further choices.  

- if $e$ is a trivial twisted  loop, assuming (i) holds,
\eqref{ext2} is valid and $U^{\od/\ev}_{{\cG}/e} = U^{\od/\ev}_{{\cG} - e}$ by definition. Under (iv.b), \eqref{e2f} holds.
These are the only cases valid for a trivial twisted  loop.

\begin{proposition}[Broken recurrence rules for $U^{\od/\ev}$ in rank $d>2$]
\label{theo:cdeltensor}
Let $\cG$ be a rank $d>2$ colored tensor model graph. Let $e$ be
an edge of $\cG$ and $N \in [0,d]$ be the number of internal faces that pass through the edge $e$. Then
\beq \label{utensor}
U^{\od/\ev}_{\cG} 
=\left\{ \begin{array}{c}
U^{\od/\ev}_{\cG/e}  
+
U^{\od/\ev}_{\cG \vee e} 
\sum_{K \in [1, N]/K\ne \emptyset} 
\Big [
t_e^K 
\big( \prod_{i \in K} A^{\ev/\od}_{f_i/e} \big)
\big(\prod_{i \in K^c} A^{\od/\ev}_{f_i/e} \big)
\Big ]\, \quad {\rm for} \; N \ge 1\,,\crcr
U^{\od/\ev}_{\cG/e} 
\hspace{10.8cm} {\rm for} \; N = 0\,.
\end{array} \right.
\eeq
\end{proposition}
\proof
Let us assume that $N= 0$, no internal faces pass through $e$.
The result $U^{\od/\ev}_{\cG} = U^{\od/\ev}_{\cG/e}$ is direct. 
Let us call $f_i$, $i=1,\dots,N$ the internal face passing through $e$. 
We start by writing, using Lemma \ref{lem:facontr},
\bea
U^{\od /\ev}_{\cG}
&=&
\prod_{ f_i \in \cF_{\inter}} A^{\od /\ev}_{f_i} 
\prod_{\shortstack{ $_{{\rm f} \in \cF_{\inter}}$ \\$_{{\rm f} \ne f_i}$ }} A^{\od /\ev}_{\rm f} =
\prod^N_{i = 1} (t_e \, A_{f_i/e}^{\ev / \od} + A_{f_i/e}^{\od/\ev}) 
\prod_{\shortstack{ $_{{\rm f} \in \cF_{\inter}}$ \\$_{{\rm f} \ne f_i}$ }} A^{\od/\ev}_{\rm f}
\crcr
&=&\Big[
\sum_{K \subset [1, N]} \prod_{i \in K} ( t_e \, A_{f_i/e}^{\ev/\od}) \prod_{i \in K^c} A_{f_i/e}^{\od/\ev}
\Big]
\prod_{\shortstack{ $_{{\rm f} \in \cF_{\inter}}$ \\$_{{\rm f} \ne f_i}$ }} A^{\od/\ev}_{\rm f} \crcr
&=&
U^{\od/\ev}_{\cG \vee e} 
\sum_{K \subset [1, N]/K\ne \emptyset} 
\Big [
t_e^K 
\big( \prod_{i \in K} A^{\ev/\od}_{f_i/e} \big)
\big(\prod_{i \in K^c} A^{\od/\ev}_{f_i/e} \big)
\Big ]
+ U^{\od/\ev}_{\cG/e} \,.
\eea

\qed

One notices that Proposition \ref{theo:cdeltensor} generalizes Proposition \ref{theo:cdel} in rank $d=2$ if the internal faces do not
pass more than once through $e$. In particular, \eqref{ext2},
\eqref{extint} and \eqref{fnf} can be recovered from \eqref{utensor}. 
Meanwhile,  for trivial tensor loop edges,  the result again holds. 
For a bridge, in a colored or invariant tensor model, all faces are necessarily external
\cite{avohou}, and we have $U_{\cG}^{\od/\ev} =  U^{\od/\ev}_{\cG/e} $.

\begin{remark}
We notice that the polynomials $U^{\od/\ev}_\cG$ do not obey
stable contraction/deletion/cut rules on ribbon graphs with flags like the Tutte and Bollob\`as-Riordan polynomials. The insisting appearance of the face polynomials $ A^{\ev/\od}_f$, in the above broken recurrence relations, suggests the existence of a more general polynomial. We will introduce an extended version of $U^{\od/\ev}$ in Subsection \ref{subsect:poly2}.
\end{remark}

\begin{proposition}[Modified recurrence rules for $\tilde W$ in rank 2]
\label{theo:W} Let $\cG$ be a ribbon graph with half-ribbons, $\cF_{\inter; {\cG}}$
and $\cF_{\ext; \cG}$ be its sets of internal and external faces,
respectively, and $e$ be a regular edge of $\cG$. We write $T_{l,f} = \left({1 - t_l \over  1 + t_l}\right)^{|m_f|}$. 
\begin{enumerate}
\item[(i)] 
Consider $e$ belongs only to external faces, $f$ and $f'$. 
Then 
\beq\label{www}
{\tilde W}_{\cG} = T_{e, f} T_{e, f'} {\tilde W}_{\cG/e} \,.
\eeq
Furthermore,
\subitem(a) if either $\{f\neq f'\}$ or $\{ f=f'$ $(e^2 \in f)$ and  the deletion of $e$ does not generate any new internal faces$\}$, then
\beq
{\tilde W}_{\cG/e} =  {\tilde W}_{\cG \vee e}= {\tilde W}_{\cG-e}\,,
\label{eq:Wi}
\eeq
\subitem(b) $f=f'$ $(e^2 \in f)$ and the deletion of $e$ generates a new internal face $f''$,
\beq 
{\tilde W}_{\cG/e} =  {\tilde W}_{\cG \vee e} = \Big(\prod_{l \in f''} T_{l,f''}\Big)  {\tilde W}_{\cG-e}\,.
\label{eq:Wii}
\eeq
\item[(iii)]
If $e \in f$  and $e \in f'$, $f \in \cF_{\inter;\cG}$ and 
$f' \in \cF_{\ext;\cG}$, 
\beq
{\tilde W}_{\cG} = T_{e,f} {\tilde W}_{\cG/e}\,,\qquad
 T_{e,f'} {\tilde W}_{\cG \vee e} = T_{e,f'}  {\tilde W}_{\cG - e} = \Big ( \prod_{\shortstack{ $_{l \in f}$ \\ $_{l \ne e }$} } T_{l,f} \Big) {\tilde W}_{\cG}\,.
\label{eq:Wiiicut}
\eeq
\item[(iv)]
If $e \in f$  and $e \in f'$, and $f,f'  \in \cF_{\inter;\cG}$
\beq
{\tilde W}_{\cG} = {\tilde W}_{\cG/e} = {\tilde W}_{\cG - e}\,.
\label{eq:Wiv}
\eeq
\subitem(a)
Furthermore, if $f \ne f'$, 
\beq
{\tilde W}_{\cG \vee e} 
= 
\Big ( \prod_{\shortstack{ $_{l \in f}$ \\ $_{l \ne e }$} } T_{l, f} \Big)
\Big ( \prod_{\shortstack{ $_{l \in f'}$ \\ $_{l \ne e }$} } T_{l,f'}\Big )
{\tilde W}_{\cG} \,.
\label{eq:Wiva}
\eeq
\subitem(b)
If  $f = f'$, $(e^2 \in f)$
\beq
{\tilde W}_{\cG \vee e} 
= 
\Big ( \prod_{\shortstack{ $_{l \in f}$ \\ $_{l \ne e }$} } T_{l,f}\Big )
{\tilde W}_{\cG} \,.
\label{eq:Wivb}
\eeq
\end{enumerate} 
\end{proposition}
\proof
We will concentrate on the cases which can only occur
in rank $d=2$. These cases include $e^2 \in f$, for some face
$f$, or when the deletion $\cG-e$ can be performed. 
All the remaining relations will be
a corollary of the next Proposition \ref{theo:Wtensor}. 

By cutting an external 
face ($f\vee e$), or by contracting it ($f/e$), then
$\prod_{l\in f\vee e} T_{l,f\vee e} = \prod_{l\in f/ e} T_{l,f/e} $.
Using this, one proves that in \eqref{eq:Wi} and \eqref{eq:Wii},
${\tilde W}_{\cG/e} =  {\tilde W}_{\cG \vee e}$. 

Proving ${\tilde W}_{\cG \vee e}= {\tilde W}_{\cG-e}$ \eqref{eq:Wi}, one must  observe that $\prod_{l\in f\vee e} T_{l,f\vee e} = \prod_{l\in f- e} T_{l,f-e} $, where $f-e$ is the external face resulting from $f$ in $\cG-e$. 

Focusing on \eqref{eq:Wii}, the cut graph $\cG \vee e$ contains an additional external face compared to $\cG -e$ (in fact, this additional
external face becomes a closed face in $\cG-e$). The same external 
face of $\cG\vee e$ generates the additional factor.

Now \eqref{eq:Wiiicut} holds for almost the same reasons
 mentioned above: cutting $e$ or removing it, from
the graph $\cG$ cannot be distinguished by $\tilde{W}$.
The set of lines in $f-e$
union the set of lines in $f'-e$ is one-to-one with 
the set of lines in $f\vee e$ union the set of lines $f'\vee e$. 

Concerning \eqref{eq:Wiv}, one must
pay attention that, either in $\cG-e$ or in $\cG/e$,
the faces passing through $e$ are internal after the operation. 

We focus on \eqref{eq:Wivb} and note that the set
of lines in $f$ subtracted by $e$ coincides with 
the set of lines of $f \vee e$ in $\cG \vee e$. 
Thus, after cutting $e$ in $\cG$, $\tilde{W}_{\cG \vee e}$ possesses 
an extra factor coming from the set of lines resulting from
the cut of $f$.  

\qed

\begin{proposition}[Modified recurrence rules for $\tilde W$ for rank $d>2$]
\label{theo:Wtensor} 
Let $\cG$ be a colored tensor graph of rank $d>2$, $\cF_{\inter;\cG}$
and $\cF_{\ext; \cG}$ be its sets of internal and external faces,
respectively, and $e$ be a regular edge of $\cG$. Let $\cF_{\ext;e}$
(resp. $\cF_{\inter;e}$)
be the set of external (resp. internal) faces going through $e$. 
Then, 
\beq
{\tilde W}_{\cG} = \left \{
\begin{array}{cc}
{\tilde W}_{\cG/e} & \,{\rm for} \; \cF_{\ext;e} = \emptyset \,,\\
\big(\prod_{f \in \cF_{\ext;e}} T_{e, f} \big) {\tilde W}_{\cG/e}&  \; {\rm for} \; \cF_{\ext;e} \ne \emptyset  \,.
\end{array}
\right.
\label{eq:Wtensorcont}
\eeq
and
\beq
\big(\prod_{f\in \cF_{\ext;e} }T_{e, f}\big) {\tilde W}_{\cG \vee e} 
=\Big( 
\prod_{f\in \cF_{\inter;e} }  \prod_{l \in f/l\ne e} T_{l,f} 
\Big){\tilde W}_{\cG} \,.
\label{eq:Wtensordelete} 
\eeq
\end{proposition}
\proof
Noting that the operation of contraction preserves the number 
of external (resp. internal) faces in $\cG$, then in $\cG/e$, we only lose the variables associated with $e$. Then, \eqref{eq:Wtensorcont} follows.

For the cut operation, one must pay attention to the fact that the internal faces in $\cG$ become external faces in $\cG \vee e$, whereas external faces in $\cG$ generate only more external faces in $\cG \vee e$.
Then, \eqref{eq:Wtensordelete} follows.

\qed

Some comments are in order: 

- One can check now that all statements except those
involving $e^2 \in f$ or $\cG -e$ in Proposition \ref{theo:W}
can be recovered from Proposition \ref{theo:Wtensor}. 

- Notice that $\tilde{W}_{\cG}$ is a polynomial in $\{T_{l,f}\}$
which always satisfies a well defined recurrence relation
under contraction operation. To be clearer, $\tilde{W}_{\cG}$ is
stable under contraction or cut rule.

- Discussing special edges (bridges, trivial loops), one can check that 
the above propositions specialize but are still valid.

\subsection{Relations to other polynomials}
\label{subsect:relaQ}

The type of graphs we are treating here
have been discussed in several works. However, the only polynomial
that we find related to $U^{\od}_\cG$
is provided by Krajewski et al. \cite{Krajewski:2010pt}
in the context of ribbon graphs with flags. 
We do not see, at this stage, any relationships between the 
polynomial on rank 3 colored graphs as worked out by Avohou et al. \cite{avohou} and the polynomials of the present
work. 

In this section, we will concentrate on the relationship between the polynomial $U^{\od}_\cG$ and polynomials discussed in \cite{Krajewski:2010pt}.
As an outcome of this discussion, we will motivate the introduction 
of a new invariant $\cU_{\cG}$ in the next section. 
We mention that this section is devoted exclusively to matrix model case or ribbon graphs. 
Henceforth, when there is no possible confusion, 
we simply refer ribbon graphs (possibly with flags) to graphs.

First, one must clarify the setting in which two (Hyperbolic) polynomials
 $\HU_{\cG}(t)$ and $\widetilde{\HU}_{\cG}(t)$ 
by Krajewski et al. are found. 
The model considered is the GW model in $D$ dimensions.
The corresponding parametric amplitudes  have been computed
and give, as expected,  generalized Symanzik polynomials. The first
Symanzik polynomial is $\HU_{\cG}(\Omega,t)$. Such an object
has two kinds of variables $t=\{t_l\}$ and $\Omega=\{\Omega_l\}$ which are line or edge
variables ($\Omega_l$ is a new parameter 
important for ensuring renormalizability through
the cure of the so-called UV/IR mixing).  

The key  relation that $\HU_{\cG}$\footnote{The expression of $\HU_{\cG}$ can be found in \cite{Krajewski:2010pt}. For the rest of the discussion, we only need the
recurrence relation that this polynomial satisfies. }
satisfies is a four-term recurrence relation of the
form (omitting boundary conditions i.e. vertices
with only flags and terminal forms), for a regular edge $e$, 
\beq
\HU_{\cG} 
= t_e \HU_{\cG-e} + t_e \Omega^2 {\rm HU}_{\cG \vee e} + \Omega_e {\rm HU}_{\cG^e - e} + \Omega_e t_e^2 {\rm HU}_{\cG^e \vee e},
\label{eq:HUrecurrence}
\eeq
where $\cG^e$ stands for the so-called Chmutov partial dual \cite{schmu}
of $\cG$ with respect to the edge $e$. This operation can simply be explained as follows: one must cut all lines in $\cG$ except
$e$, then perform a dual on the pinched graph $\tilde\cG$,
and glue black all edges previously cut. The interest of introducing such partial dual reflects on the contraction operation:
 $\cG/e = \cG^e - e$.

It turns out that the GW model can be expressed as well as
a matrix model \cite{Grosse:2004yu}. Moreover, at the limit when $\Omega\to 1$,
the amplitudes of this model are of the form \eqref{eq:defa}.
To be precise, the rank $d$ must be fixed to $2$, and since
the summation over the matrix indices in the GW model 
are performed over $\N^2$, one obtains a modified definition
of $W_\cG = \tilde{W}_\cG$. Finally, after this re-adjustment, 
we have the same amplitude up to a constant (a power of 2)
depending on the graph. This constant has been incorporated
in the definition of the polynomial but for the ensuing discussion
this factor is totally inessential. 

The problem as raised by authors, to the best of our
understanding, is how to relate
the new first Symanzik polynomial $\widetilde{\HU}_{\cG}(t)$ 
\footnote{ Note that in \cite{Krajewski:2010pt}, $\widetilde{\HU}$ is denoted $\HU$ again, see Eq.(6.8)
therein. For avoiding confusion, 
we use a different notation here.}
obtained in this matrix base and the limit 
$\HU_{\cG} (1,t)$.   
We emphasize a series of subtleties in the comparison 
procedure which will make clear our next point:

- First, the polynomial $\widetilde{\HU}_{\cG}(t)$
was computed in an amplitude involving a closed graph 
i.e. a ribbon graph without flags. In fact, it directly extends
to the case of a ribbon graph with flags provided one still
performs a product over closed faces. Hence, 
$\widetilde{\HU}_{\cG} = U^{\od}_{\cG}$, up
to a constant, on ribbon graphs with flags. 

- Second, in order to relate $\widetilde{\HU}_{\cG}(t)$
and $\HU_{\cG}$ the authors introduce another polynomial 
 called $U_{\cG}$ (Eq.(6.12), p. 532). 
This polynomial is defined as
\beq
U_{\cG}(t) := \sum_{g\, \in\, \underline{\check{\rm Odd}}(\cG^*)}
\prod_{l \in \cHE(g)} t_l \,,
\label{eq:Udefine}
\eeq
where $\cG^*$ is the dual of $\cG$, 
$\underline{\check{\rm Odd}}(\cG)$ is the set 
odd (colored) cutting spanning subgraphs of a graph $\cG$.
In the previous sentence, we put in parenthesis colored
because precisely, the coloring refers to the bi-coloring 
of vertices. It has the effect of introducing a prefactor
$2^{v(\cG)}$ which is inessential in our discussion. An odd graph is a graph 
with all degrees of its vertices of odd parity. 
 An odd cutting spanning subgraph $g\in \check{\rm Odd}(\cG^*)$ is a spanning 
subgraph of a graph $\cG$ (having all its vertices), 
obtained by choosing $\cH(g)\subset \cH(\cG)$ and $\cFL(\cG)\subset \cH(g)$ and such that $g$ is odd.

The issue is that $U_{\cG}(t)$  is defined on open and
closed graphs. And, as proved in the above reference, this quantity
always coincides with $\HU_{\cG}(1,t)$ and so satisfies
the same four-term recurrence rule \eqref{eq:HUrecurrence}. 
On closed graphs, $U_{\cG}(t)=\widetilde{\HU}_{\cG}(t)$
and so matches with $U^{\od}_{\cG}$. However, on open graphs
it is not true that $U_{\cG}(t)$ is equal 
to $\widetilde{\HU}_{\cG}(t)=U^{\od}_{\cG}$.
The reason why there certainly is a discrepancy 
is because $U^{\od}_{\cG}$ meets another formula.
Indeed, since a closed face in $\cG$ corresponds
to a vertex in $\cG^*$ which does not have any flags, 
we partition the vertices of $\cG^*$ in two  distinct subsets: 
$\cV(\cG^*) = \cV'(\cG^*) \cup  \cV''(\cG^*)$, where
$v\in \cV'(\cG^*)$ is without flags. Now considering 
the cutting subgraph $S(\cG^*)$ of $\cG^*$ having a
set of vertices $\cV ( S(\cG^*) ) = \cV'(\cG^*)$ and a set of
edges $\cE( S(\cG^*) )$ containing all edges
from $ \cV'(\cG^*)$  to $ \cV'(\cG^*)$ and cutting
all edges from $ \cV'(\cG^*)$  to $ \cV''(\cG^*)$.
We write
\beq
U^{\od}_{\cG}
= \prod_{v^* \in V'(\cG^*) }
\Big \{\sum_{\shortstack{
$_{A \sset \cHE(v^*)}$ \\
$_{|A| \; {\rm odd}}$}
}
\prod_{l \in A} t_l
\Big \}
=
\sum_{g \in {\rm Odd}^{\natural}  (S(\cG^*)) }
\prod_{l \in \cH(g)} t_l= U^{\natural}_{S(\cG^*)}\,,
\label{eq:Uoddgp}
\eeq
where $\cHE(v^*)$ stands for the set of half-edges incident
to $v^*$, and in $\cH(g)$, flags are labeled with the same label
of the edges where they come from.
$ {\rm Odd}^{\natural}(\cG)$ is set of odd cutting spanning subgraphs
of a second kind: $g \in  {\rm Odd}^{\natural}(\cG)$ is
defined such that $\cH(g) \subset \cH(\cG)$. 
Hence, $U^{\od}_{\cG} \neq U_{\cG}$ and
$U^{\natural}_{S(\cG^*)}$ is the closest expression that
we have found related to $U_{\cG}$. 

\medskip 
We fully illustrate now the above discussion by examples.

\medskip 

\noindent{\bf Example 1: Triangle with flags.} 
\begin{figure}[h]
 \centering
     \begin{minipage}[t]{.8\textwidth}
      \centering
\def\svgwidth{0.8\columnwidth}
\tiny{
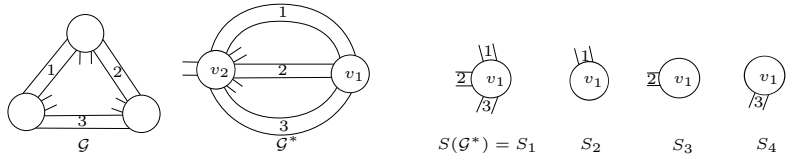
}
\caption{\small Triangle graph with flags $\cG$, $\cG^*$,  $S_1$, $S_2$, $S_3$, and $S_4$ which define ${\rm Odd}^{\natural}(S(\cG^*))$. }
\label{fig:triangle}
\end{minipage}
\end{figure}
Consider the graph $\cG$ as a triangle with one flag on each vertex, 
all in the same face (see Fig.\ref{fig:triangle}). In \cite{Krajewski:2010pt}, $\HU_{\cG}(1,t)= U_{\cG}(t)$ was already computed and it gives
\beq
\HU_{ \cG}(1,t) = 4 (t_1 + t_2 + t_3 + t_1 t_2 t_3) (1 + t_1 t_2 + t_1 t_3 + t_2 t_3).
\eeq
Computing $U^{\od}_{\cG}$ directly from the face amplitude
formula, one has:
\beq
U^{\od}_{ \cG}(t) = t_1 + t_2 + t_3 + t_1 t_2 t_3.
\eeq
Clearly, for open graphs the polynomials do not agree. 

Let us now explain the expansion \eqref{eq:Uoddgp}. 
Consider the dual $\cG^*$ of $\cG$ in Fig.\ref{fig:triangle}. 
First $\cV'(\cG^*)=\{v_1\}$ (the vertex without flags), $S(\cG^*)$ is the graph 
made with $v_1$ with three flags labeled by $1,2,3$
in the same way of the lines $l_1$, $l_2$ and $l_3$
and are associated with variables $t_1,t_2$ and $t_3$. 
We obtain four cutting spanning subgraphs in 
\beq
{\rm Odd}^{\natural}(S(\cG^*)) = \{ \{l_1\}, \{l_2\},\{l_3\},\{l_1,l_2,l_3\} \}
\eeq
 as in  Fig.\ref{fig:triangle}
with contributions $t_1$,  $t_2$, $t_3$ and $t_1 t_2 t_3$, respectively. 
On the other hand, 
\bea
&&
\check{{\rm Odd}}(\cG^*) = \{ \{l_1; l_1,l_2\}, \{l_1; l_1, l_3\},
\{l_1; l_2,l_3\},
\{l_2;  l_2, l_1\} ,\{l_2; l_2, l_3\}, \{l_2; l_1, l_3\}, \{l_3;l_3, l_1\}, \crcr
&&\{l_3;l_3, l_2\},\{l_3;l_1, l_2\},\{l_1,l_2,l_3;l_1,l_2\}, \{l_1,l_2,l_3;l_1,l_3\}, \{l_1,l_2,l_3;l_2,l_3\},\crcr
&&
\{l_1; \emptyset \}, \{l_2; \emptyset \}, \{l_3; \emptyset \}, 
\{l_1,l_2,l_3; \emptyset\}\}\,,
\eea
where on each side the semi-colon in the brackets, we collect half-edges on each vertex $v_1$ and $v_2$.

\medskip

\noindent{\bf Example 2: Pretzel without flags.}
\begin{figure}[h]
 \centering
     \begin{minipage}[t]{.8\textwidth}
      \centering
\def\svgwidth{0.8\columnwidth}
\tiny{
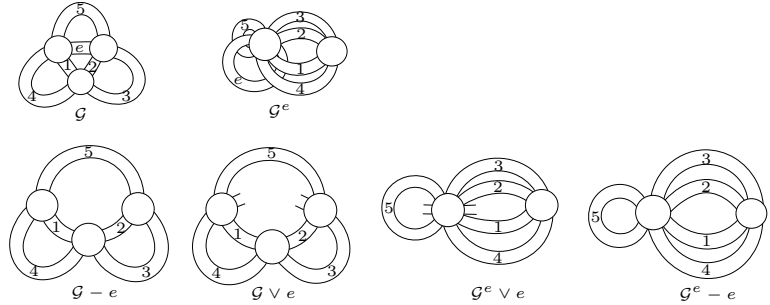
}
\caption{\small The pretzel graph $\cG$, its partial dual with respect to $e$, $\cG^e$, and its operated graphs.}
\label{fig:pretzelwithoutflags}
\end{minipage}
\end{figure}
Consider the graph $\cG$ drawn in Fig.\ref{fig:pretzelwithoutflags}. 
We also illustrate $\cG^{e}$, $\cG -e$, $\cG \vee e$, $\cG^{e} \vee e$, and $\cG^{e} - e$ in that picture.
We call $C(t)= (t_2 + t_3) (t_1 + t_4)(t_2 + t_3)(t_3 + t_4 + t_5 + t_3 t_4 t_5)$, and we evaluate
\bea
U^{\od}_{\cG} &=& (t_e + t_1 + t_2 + t_e t_1 t_2) (t_e + t_5) C(t)
\crcr
&=& [t_e^2 + t_e (t_1 + t_2 + t_5 + t_1 t_2 t_5) + t_e t_1 t_2 + t_1 t_5 + t_2 t_5]C(t)\,, \crcr
U^{\od}_{\cG - e} &=& [ t_1 + t_2 + t_5 + t_1 t_2 t_5]C(t)\,,\crcr
U^{\od}_{\cG \vee e} &=& C(t)= U^{\od}_{\cG^{e} \; \vee e} \,,\crcr
U^{\od}_{\cG^{e} \; - e} &=& t_5 (t_1 + t_2) C(t) = (t_1 t_5 + t_2 t_5)C(t)\,.
\eea
From this point, by observing the term 
$t_et_1 t_2$ in $U^{\od}_{\cG}$, one can readily check that there exist
no polynomials function $p_i(t_e)$ in $t_e$ variable such that 
a relation of the type
\beq
p_1(t_e) U^{\od}_{\cG}
 =p_2(t_e) U^{\od}_{\cG - e}  + p_3(t_e)U^{\od}_{\cG \vee e}
+ p_4(t_e)U^{\od}_{\cG^e - e}  + p_5(t_e)U^{\od}_{\cG^e \vee e} 
\eeq
is satisfied. Thus $U^{\od}_{\cG}$ does not obey 
the same relation as $\HU_{\cG}$. Therfore, it is not
a ${\mathcal Q}_{\cG}$ polynomial in the general 
sense of Krajewski et al.

We understand now that $U^{\od}_{\cG}$ on the class of
open graphs does not obey any known recurrence relations. 
In the tensor situation, things become worse:
we do not have any clear notion of duality at the level 
of graphs and the notion of Chmutov dual is still
not defined. This urges us to find another
path to understand this object $U^{\od/\ev}_{\cG}$. A natural route that
we will investigate is the understanding
of the notion of face amplitude that
we observe to be at the heart of this theory.
We will introduce an extended  
framework, where a generalized version of $U^{\od/\ev}_{\cG}$
makes sense and turns out to satisfy a proper
invariance rule. This is the purpose of the
next section.

\subsection{Polynomial of the second kind}
\label{subsect:poly2}

First recognizing that the polynomials are sensitive to the properties of faces, we will exploit this face-structure by defining
a new polynomial $\cU_{\cG}$. 
This object generalizes $U^{\od/\ev}_{\cG}$ and obeys a novel recurrence relations based only on contraction operation. 
We call it of the second kind. An extension of $\tilde{W}_{\cG}$
will not be discussed for two main reasons: first, $\tilde{W}_{\cG}$ is already stable under contraction and, second, the notion of parity  in $U^{\od/\ev}_{\cG}$ which is at the core of the next developments does not appear at all in $\tilde{W}_{\cG}$. Finally, 
most of the ingredients used in the following have been 
 introduced in Subsection \ref{subsect:poly1}. 

Let $\cG^\star$ be the set of isomorphism classes of rank $d$ tensor
graphs (including ribbon graphs with half-ribbons) $\{\od, \ev\}$ be the set of parities (in obvious notations). Let $\cG \in \cG^\star$  with a set of internal faces $\cF_{\inter;\cG}$ and $\cP_{\inter;\cG}$ be the power set of $\cF_{\inter;\cG}$.

\begin{definition}[Generalized polynomial]
\label{def:genpoly}
Consider an element 
$(\cG, \cF, \bcF, \eps, \eps')\in  \cG^\star \times (\cP_{\inter;\cG} )^{\times 2}$ $ \times \{\od, \ev\}^{\times 2}$ 
such that $\cF \cup \bcF =\cF_{\inter;\cG}$ and $\cF \cap \bcF = \emptyset$.
We define a generalized polynomial associated with 
$(\cG, \cF, \bcF, \eps, \eps')$ as
\beq
\cU^{\,\eps,\,\eps'}_{\cG;\, (\cF, \, \bcF)} (\{t_l\})= \Big[\prod_{{\rm f} \in \cF} A^{\eps}_{\rm f}(\{t_l\}_{l \in {\rm f}}) \Big] \Big[\prod_{{\rm f} \in \bcF}
A^{\eps'}_{\rm f}(\{t_l\}_{l \in {\rm f}}) \Big] .
\label{eq:defcu}
\eeq 
\end{definition}
Note that from the definition of $U_\cG$ \eqref{eq:defu}, it is immediate to have (when using subscripts, we write $Q_{\cF_{\inter;\cG}}=Q_{\cF_{\inter}}$, for any quantity $Q$)
\bea
&& 
\cU^{\,\eps,\,\epsilon'}_{\cG;\, (\cF_{\inter},\,\emptyset)} (\{t_l\}) =
\cU^{\,\eps',\,\eps}_{\cG;\, (\emptyset,\cF_{\inter})} (\{t_l\}) =
U^{\eps}_{\cG}(\{t_l\}) \,, \qquad \quad \eps = \od,\ev\,;
\crcr
&&
\forall \cF \in \cP_{\inter;\cG}\,, \;\;\;
\cU^{\,\eps,\,\eps}_{\cG;\,  (\cF, \, \bcF)} (\{t_l\}) 
= U^{\eps}_{\cG}(\{t_l\}) \,, \qquad \quad \eps = \od,\ev\,,
\eea
for any value of $\eps'$. 
Furthermore $\cU_{\cG}$ is symmetric under the flips: 
\beq
\cU^{\,\epsilon,\,\epsilon'}_{\cG;\, (\cF,\bcF)} (\{t_l\}) = \cU^{\,\epsilon',\,\epsilon}_{\cG;\,(\bcF,\cF)} (\{t_l\})\,.
\eeq
From these properties, the only case of interest is of $\cU^{\od,\ev}_{(\cdot)}$.

 As a  convention, for the empty graph $\cG=\emptyset,$
\beq
\label{eq:convention1}
\cU^{\od,\ev}_{\emptyset; \, (\emptyset, \, \emptyset)} = 1 
\eeq  
and, 
on the bare vertex graph $\cG=o$ with a unique closed face $f$, according to \eqref{convface}, we have
\beq
\label{eq:convention2}
\cU^{\od,\ev}_{o; \, (\{f\}, \, \emptyset)} = 0\,, 
\qquad
\cU^{\od,\ev}_{o; \, (\emptyset, \, \{f\})} = 1 \,. 
\eeq 
Now, if it occurs that $\cG \neq \emptyset$ and $\cF_{\inter;\cG}=\emptyset$, then $\cF= \bcF=\emptyset$, so that
\beq
\cU^{\od,\ev}_{\cG; \, (\emptyset, \, \emptyset) } = 1\,.
\label{eq:conventionext}
\eeq

The following proposition follows from definitions. 

\begin{proposition}[Disjoint union operations]
\label{coro:disjointunion}
Let $\cG_1$ and $\cG_2$ be two rank $d$ graphs,
let $\cG_1 \sqcup \cG_2$ be their disjoint union. 
Then, 
\beq
\cU^{\,\eps,\,\eps'}_{\cG_ 1  \sqcup \cG_2 ;\, (\cF, \, \bcF)} =\cU^{\,\eps,\,\eps'}_{\cG_ 1 ;\, (\cF_1, \, \bcF_1)} \, \cU^{\,\eps,\,\eps'}_{\cG_ 2 ;\, (\cF_2, \, \bcF_2)} 
\label{eq:sqcup}
\eeq 
where $\cF_1 \cup \cF_2 = \cF$ and $\bcF_1 \cup \bcF_2 = \bcF$.  
\end{proposition}

\subsubsection{A new recurrence rule: Regular edges}
\label{sec:regularedges} 

We shall drop the subscript $\cG$ in the subsequent notations
for sets. For instance, $\cL$ and $\cF_{\inter/\ext}$ 
will denote the set of lines and set of faces of $\cG$.

 - We introduce the following definition: Given a 
subset $\cF $ of internal faces, we define $\cF/e$ to be
the subset of faces corresponding to $\cF$
after the contraction of $e$ in $\cG$. 

The following statement holds.

\begin{theorem}[Generalized contraction rule for $\cU^{\epsilon, \bar \epsilon}$ in rank $d=2$]
\label{theo:gencdel}
Let $\cG$ be a ribbon graph with half-ribbons with $\cL$ set of lines,
$\cF_{\inter}$ set of internal faces, $(\cF,\bcF)$ a 
pair of disjoint subsets of 
$\cF_{\inter}$ 
with 
$\cF \cup\bcF=\cF_{\inter}$.

Let $e$ be a regular edge of $\cG$, we have four disjoint 
cases:
\begin{enumerate}
\item[(0)] If $e$ passes through only external faces, then
\beq\label{fexfex} 
\cU^{\epsilon,\bar\epsilon}_{\cG;\, (\cF,\,\bcF)} 
 = 
\cU^{\epsilon,\bar\epsilon}_{\cG/e;\, (\cF,\,\bcF)} \,,
\eeq
where $(\cF,\,\bcF)= (\cF/e,\,\bcF/e)$. 

\item[(i)] If $e\in f$, for a unique internal face $f \in \cF$
(the other strand of $e$ is external), then  
\beq\label{fexfin} 
\cU^{\epsilon,\bar\epsilon}_{\cG;\, (\cF,\,\bcF)} 
 = 
\cU^{\epsilon,\bar\epsilon}_{\cG/e;\, (\cF/e,\,\bcF)} 
+ t_e\,
\cU^{\epsilon,\bar\epsilon}_{\cG/e;\, ((\cF/e) \setminus\{f/e\},\,\bcF\cup \{f/e\})}\,,
\eeq
where $\bcF=\bcF/e$.

\item[(ii)] If $e^2\in f$ with $f \in \cF$, then 
\beq\label{f2fin} 
\cU^{\epsilon,\bar\epsilon}_{\cG;\, (\cF,\,\bcF)} 
 = (1+t_e^2)\, 
\cU^{\epsilon,\bar\epsilon}_{\cG/e;\, (\cF/e,\,\bcF)} 
+ 2 \,t_e \,
\cU^{\epsilon,\bar\epsilon}_{\cG/e;\, ((\cF/e)\setminus\{f/e\},\,\bcF\cup \{f/e\})} \, ,
\eeq
where $\bcF=\bcF/e$.

\item[(iii)] If $e\in f_1$ and $e\in f_2$, $f_1 \neq f_2$, 
then 
\begin{enumerate}
\item[(a)] if $f_i \in \cF$, then 
\bea \label{f1f2in} 
\cU^{\epsilon,\bar\epsilon}_{\cG;\, (\cF,\,\bcF)} 
&=&  \cU^{\epsilon,\bar\epsilon}_{\cG/e;\, (\cF/e,\,\bcF)} \crcr
& +& t_e \,\Big(\,\cU^{\epsilon,\bar\epsilon}_{\cG/e;\, ((\cF/e)\setminus\{f_1/e\},\,\bcF\cup \{f_1/e\})}  + (1\leftrightarrow 2) \, \Big) 
\crcr
&+&  t_e^2 \,
\cU^{\epsilon,\bar\epsilon}_{\cG/e;\, ((\cF/e)\setminus\{f_1/e,f_2/e\},\,\bcF\cup \{f_1/e,f_2/e\})}\,,
\eea
where $\bcF=\bcF/e$;

\item[(b)] if $f_1 \in \cF$ and $f_2 \in \bcF$, then 
\bea \label{f1nf2in} 
\cU^{\epsilon,\bar\epsilon}_{\cG;\, (\cF,\,\bcF)} 
&=& \cU^{\epsilon,\bar\epsilon}_{\cG/e;\, (\cF/e,\,\bcF/e)} \crcr
&+& \, t_e \,\Big(\cU^{\epsilon,\bar\epsilon}_{\cG/e;\, ((\cF/e)\setminus \{f_1/e\},\,\bcF/e \cup \{f_1/e\} )} 
+
\cU^{\epsilon,\bar\epsilon}_{\cG/e;\, ((\cF/e)\cup \{f_2/e\},\,(\bcF/e) \setminus \{f_2/e\} )}    
\Big) 
\crcr
&+& \,  t_e^2 \,
 \cU^{\epsilon,\bar\epsilon}_{\cG/e;\, ([(\cF/e)\setminus\{f_1/e\}]\cup \{f_2/e\} ,\,[(\bcF/e)\setminus\{f_2/e\}]\cup \{f_1/e\})} \,.
\quad 
\crcr
&&
\eea
\end{enumerate}
\end{enumerate}

\end{theorem}

\proof  See Appendix \ref{app:theogencdel}. 

\qed

Theorem \ref{theo:gencdel} expresses the reduction of 
the polynomial $\cU_{\cG}$ only in terms of edge contractions. It
is a new feature of an polynomial invariant on a graph. 
As a function depending on a partition of the 
set of internal faces, one must pay attention that in 
each expression involving
$\cU^{(\cdot)}_{\cG/e;\, (\cF,\,\bcF )}$ in the r.h.s of 
the equations \eqref{fexfin}-\eqref{f1nf2in}, 
$\cF$ and $\bcF$  
always define a partition of the set $\cF_{\inter}$ of internal faces of  $\cG/e$. 
The invariant $\cU^{\eps,\beps}_{\cG;\,(\cF,\, \bcF)}$ is a multivariate
polynomial distinct from the Bollob\`as-Riordan polynomial
\cite{bollo}.

In rank $d=2$, seeking a state sum formula for $\cU^{\od,\ev}_{\cG;\,(\cF,\, \bcF)}$, we have using \eqref{eq:Uoddgp}, 
\beq
\cU^{\od,\ev}_{\cG;\,(\cF,\, \bcF)}= 
\sum_{(g_1,g_2) \in {\rm Odd}^{\natural}  (S_1(\cG^*))
\times {\rm Even}^{\natural}  (S_2(\cG^*))  }
\Big[\prod_{l \in \cH(g_1)} t_l\Big]
\Big[\prod_{l \in \cH(g_2)} t_l\Big]
\label{summa0}
\eeq
where the definition of $ {\rm Even}^{\natural}(\cdot) $ 
can be deduced from ${\rm Odd}^{\natural} (\cdot)$ by 
replacing ``odd'' by ``even'', $S_1(\cG^*) \cup S_2(\cG^*)$ are
defined through a partition of the vertices of the subgraph $S(\cG^*)$. 
 
Let us comment now special edges. Considering
first the bridge case, relations (0) and (ii) in the above
theorem are valid. For the trivial untwisted loop, (0), (i)
and (iii) are true. Finally, for the trivial twisted loop (0)
and  (ii) hold. Thus, once again special edges are evaluated
from the same theorem. This brings the following important
question: ``Can we find a closed formula for any polynomial $\cU^{\eps,\beps}_{\cG}$ 
on any graph $\cG$ using only the recurrence relation and a finite list of boundary conditions?''. In other words, given a graph, its number of
internal and external lines, its number of bridges, loops, etc., 
is there a unique polynomial solution of the above recurrence relations 
expressed simply as a function of these numbers?
If the answer to this question is yes, 
then the above polynomial will prove to be a very neat
and computable invariant simpler than the Bollob\`as-Riordan polynomial. 
However, a notion captured by the Bollob\`as-Riordan
polynomial is the genus of the ribbon graph and of its spanning subgraphs. The polynomial $\cU^{\eps,\beps}_{\cG}$,  in its present form, does not explicitly exhibit this genus. 
It would be interesting to investigate if $\cU^{\eps,\beps}_{\cG;\,(\cF,\, \bcF)}$ could be provided 
with another variable associated with 
the genus of the ribbon graph. For the moment, as a naive
example,  if we consider 
a closed graph $\cG$, and add a new set of variables $\{x_{\rm f}\}$ associated with the faces, we can define
\beq
\widetilde{\cU}^{\,\epsilon,\,\epsilon'}_{\cG;\, (\cF, \, \bcF)} (\{  t_l\}; \{ x_{\rm f}\})=  \Big[\prod_{{\rm f} \in \cF} x_{\rm f} A^{\epsilon}_{\rm f}(\{t_l\}_{l \in {\rm f}}) \Big] \Big[\prod_{{\rm f} \in \bcF} x_{\rm f}
A^{\epsilon'}_{\rm f}(\{t_l\}_{l \in {\rm f}}) \Big] . 
\label{eq:genus}
\eeq 
Thus this polynomial computes to $\widetilde{\cU}^{\,\epsilon,\,\epsilon'}_{\cG;\, (\cF, \, \bcF)}= (\prod_{{\rm f} \in \cF_{\inter}}
x_{\rm f} ) \cdot \cU^{\,\epsilon,\,\epsilon'}_{\cG;\, (\cF, \, \bcF)}$ 
and should obey modified contraction rules from \eqref{fexfex}-\eqref{f1nf2in}. 
Under the rescaling $x_{\rm f} \to \rho x_{\rm f}  $,
we have 
\beq
\widetilde{\cU}^{\,\epsilon,\,\epsilon'}_{\cG;\, (\cF, \, \bcF)} (\{  t_l\}; \{\rho x_{\rm f}\})= \rho^{F^{\inter}}\widetilde{\cU}^{\,\epsilon,\,\epsilon'}_{\cG;\, (\cF, \, \bcF)} (\{  t_l\}; \{ x_{\rm f}\})\,.
\eeq
Then certainly, $\widetilde{\cU}_{\cG}$ knows about the 
(generalized) genus $\kappa$ of the closed ribbon graph since $F_{\inter}=2-\kappa - (V-E)$. Maybe to have a better picture and 
a  good  starting point for extracting information about the genus of the subgraphs, 
one can consider the expression \eqref{summa0}. 
This problem is left to a subsequent work.

\medskip 

Before addressing the tensor case, let us recall the definition of
a trivial loop in rank $d$. These have been called in \cite{avohou}
$p$-inner self-loops, $p=1,2,3,$ in the context $d=3$; this definition
extends in any $d$. A trivial loop is an edge in a rank $d$ colored
tensor graph such that after its contraction the number of connected
components is always $d$.

\begin{theorem}[Recurrence relation for $\cU^{\epsilon, {\bar \epsilon}}$ for rank $d>2$]
\label{theo:tensorrecurrence}
Let $\cG$ be a rank $d$ colored tensor graph and
$e$ one of its regular edges or trivial loops. 
Let $\cF_e$ be the set of internal faces passing through $e$
and denote $\cF^{\eps}_e = \cF_e \cap \cF $ and
$\cF^{\beps}_e = \cF_e \cap \bcF$.  
We have
\beq
\cU^{\epsilon, {\bar \epsilon}}_{\cG; \, (\cF, \, \bcF)}
=
\sum_{K\times L  \subset (\cF^{\eps}_e/e) \times  (\cF^{\beps}_e/e)}
t_e^{|K| +|L|} 
\;
\cU^{\epsilon, {\bar \epsilon}}_{\cG/e; \, (
[(\cF/e)\setminus (\cF^{\eps}_e/e)] \cup K^c \cup L, \,
[(\bcF/e) \setminus (\cF^{\beps}_e/e)] \cup K \cup L^c
)} \,,
\eeq
in particular, for $ \cF_e=\emptyset$
\beq
\cU^{\epsilon, {\bar \epsilon}}_{\cG;\, (\cF, \,\bcF)}
=
\cU^{\epsilon, {\bar \epsilon}}_{\cG/e; \, (\cF/e, \,\bcF/e)}\,.
\eeq
\end{theorem}

\proof
Consider $\cG$, a rank $d\geq 3$ colored tensor graph and $\cF,\bcF\subset \cF_{\inter}$ which satisfy the ordinary conditions
for defining $\cU^{\eps,\beps}_{\cG;\,(\cF,\,\bcF)}$.

Let us assume that $\cF_e= \emptyset$, namely there are no internal faces pass through $e$. The result 
$\cU^{\epsilon, {\bar \epsilon}}_{\cG; \, (\cF, \,\bcF)}
=
\cU^{\epsilon, {\bar \epsilon}}_{\cG/e; \, (\cF/e, \, \bcF/e)}$ is obvious. 
Now, we assume that $\cF_e  \neq \emptyset$.
Using Lemma \ref{lem:facontr}, one writes
\bea
&&
\cU^{\epsilon, {\bar \epsilon}}_{\cG; \, (\cF, \,\bcF)}
=
\prod_{f\in \cF^{\eps}_e} A^\epsilon_{ f} \; \prod_{ f \in \cF^{\beps}_e} A^{\bar \epsilon}_{ f} 
\prod_{f \in \cF\setminus \cF^{\eps}_e} A^\epsilon_{f} \; \prod_{ f \in \bcF\setminus \cF^{\beps}_e} A^{\bar \epsilon}_{ f} 
\crcr
&& = 
\prod_{f \in \cF^{\eps}_e} (t_e \, A_{f/e}^{\bar \epsilon} + A_{f/e}^\epsilon) 
\prod_{f \in \cF^{\beps}_e} (t_e \, A_{f/e}^\epsilon + A_{f/e}^{\bar \epsilon}) 
\prod_{ f \in \cF\setminus \cF^{\eps}_e }  A^\epsilon_{ f} \; \prod_{ f \in \bcF\setminus \cF^{\beps}_e } A^{\bar \epsilon}_{ f} \crcr
&&=
\Bigg[ 
\sum_{K \subset \cF^{\eps}_e } \prod_{f \in K} ( t_e \, A_{f/e}^{\bar \epsilon}) \prod_{f \in K^c} A_{f/e}^{\epsilon}
\Bigg]
\Bigg[ 
\sum_{L \subset  \cF^{\beps}_e } \prod_{f \in L} ( t_e \, A_{f /e}^\epsilon) \prod_{f \in L^c} A_{f /e}^{\bar \epsilon}
\Bigg]
\prod_{ f \in \cF\setminus \cF^{\eps}_e }  A^\epsilon_{ f} \; \prod_{ f \in \bcF\setminus \cF^{\beps}_e } A^{\bar \epsilon}_{ f} \crcr
&&= 
\sum_{K\times L  \subset \cF^{\eps}_e \times  \cF^{\beps}_e} 
t_e^{|K| +|L|} 
\;
\Big[\prod_{ f \in \cF\setminus \cF^{\eps}_e }  A^\epsilon_{ f} 
\prod_{f\in K^c} A_{f/e}^{\epsilon}
\prod_{f \in L} A_{f/e}^\epsilon \big]
\Big[\prod_{f \in \bcF\setminus \cF^{\beps}_e } A^{\bar \epsilon}_{ f}
\prod_{f \in L^c} A_{f/e}^{\bar \epsilon}
\prod_{f \in K} A_{f/e}^{\bar \epsilon} \big]\crcr
& &=
\sum_{K\times L  \subset \cF^{\eps}_e \times  \cF^{\beps}_e} 
t_e^{|K| +|L|} 
\;
\cU^{\epsilon, {\bar \epsilon}}_{\cG/e; \, (
(\cF\setminus \cF^{\eps}_e) \cup [K^c \cup L]/e, \,
(\bcF \setminus \cF^{\beps}_e) \cup [L^c \cup K]/e
)}  \crcr
&& =
\sum_{K\times L  \subset \cF^{\eps}_e/e \times  \cF^{\beps}_e/e} 
t_e^{|K| +|L|} 
\;
\cU^{\epsilon, {\bar \epsilon}}_{\cG/e;\, (
[(\cF/e)\setminus (\cF^{\eps}_e/e)] \cup K^c \cup L, \,
[(\bcF/e) \setminus (\cF^{\beps}_e/e)] \cup L^c \cup K
)}  \,,
\eea
where we used $\cF\setminus \cF^{\eps}_e
 =(\cF/e)\setminus (\cF^{\eps}_e/e)$. 

Now, given $K\times L  \subset \cF^{\eps}_e/e \times  \cF^{\beps}_e/e$, one must prove that 
$\tilde\cF=(\cF\setminus \cF^{\eps}_e) \cup K^c \cup L$
and $\tilde \bcF = (\bcF \setminus \cF^{\beps}_e) \cup L^c \cup K$,
are such that (1) $\tilde \cF \cup \tilde \bcF = \cF_{\inter}(\cG/e)$
and (2) $\tilde \cF \cap \tilde \bcF = \emptyset$. 

With a moment of thoughts one sees that the  statement (2) is true. 
Now the former is proved. First, one recognizes that 
$ \cF_{\inter}(\cG/e) = \cF_{\inter}(\cG)/e  =\{ [(\cF\setminus \cF^{\eps}_e )\cup \cF^{\eps}_e ]/e\} \;\cup\; (\cF \leftrightarrow \bcF)$. 
Furthermore, $(\cF\setminus \cF^{\eps}_e )/e=\cF\setminus \cF^{\eps}_e $,
and $\cF^{\eps}_e/e = K\cup K^c$, $\forall K\subset \cF^{\eps}_e$, and the same is true for 
$\cF^{\beps}_e/e = L \cup L^c$, $\forall L\subset \cF^{\eps}_e$.
We can conclude to the equality at this point. 

\qed

Again, we note here that Theorem \ref{theo:tensorrecurrence} is consistent with Theorem \ref{theo:gencdel} for rank $d=2$ models if we exclude the cases where the same face goes through the same edge more than once.

\medskip
It appears possible to further precise some relations
 and to introduce rules involving the deletion in the case of ribbon graphs. This question will be addressed
now. In particular, the interesting cases correspond to (0), ($ii$) 
and ($iii.a$) of Theorem \ref{theo:gencdel}. 
 Note that, in the ribbon graph case and for
a given subset of internal faces $\cF$, the notation
$\cF-e$ might not always make sense.
We  define $\cF-e$ as a set  of internal 
faces in $\cG - e$ as follows:

a)  $\cF-e=\cF$ if the 
removal of $e$ do not affect the faces in $\cF$;

b) $\cF -e = \cF\setminus \{f\}$, 

\begin{enumerate}
\item[b1)] if $e \in f$, $f\in \cF$, and 
if $\cF$ loses the internal face $f$ passing through $e$ and $f$ merges with an external face;

\item[b2)]
if the face $f\in \cF$ is such that $e^{2}\in f$ 
and $f$ does not split into two internal faces after
the removal of $e$; then $f-e$ 
 makes sense as a unique internal face;

\item[b3)]
if the face $f\in \cF$ splits into two faces $f_1$ and $f_2$
both internal after the 
removal of $e$, and in this case $\{f-e\}=\{f_1,f_2\}$;

\end{enumerate}

c) if $e$ passes through two different internal lines
$f_1$ and $f_2$, $f_{1,2}$ are in $\cF_{1,2}$ and the removal of $e$ merges these two lines in one,
 then $\cF_i-e = \cF_i\setminus \{f_1,f_2\}$.

\medskip 

Cases a), b2), b3) and c) are the ones under which we can recast
some polynomials $\cU^{\eps,\beps}_{\cG/e}$ in terms 
of the deleted graph $\cG-e$. The following
statement holds.

\begin{proposition}[Deletion relations]
\label{coro:deletion}
Let $\cG$ be a ribbon graph with half-ribbons and
$e$ be one of its edges.  
\begin{enumerate}
\item[(0)]
If $e$ belongs only to a unique external face, 
and if it does not generate any new internal faces after the deletion of $e$ in $\cG$, then 
\beq
\cU^{\epsilon,\bar\epsilon}_{\cG;\, (\cF,\,\bcF)}  = 
\cU^{\epsilon,\bar\epsilon}_{\cG/e;\, (\cF/e,\,\bcF/e)} 
= 
\cU^{\epsilon,\bar\epsilon}_{\cG-e;\, (\cF-e,\,\bcF-e)} \,, 
\eeq
with $\cF-e = \cF$ and $\bcF-e=\bcF$. 

\item[(i)] If $e^2\in f$ with $f \in \cF$,
\subitem(a) and if the removal of $e$ will result in one unique internal face $f-e$ from $f$, then
\beq \label{eq:f2fin++} 
\cU^{\epsilon,\bar\epsilon}_{\cG;\, (\cF,\,\bcF)}=
(1+t^2)\, \cU^{\epsilon,\bar\epsilon}_{\cG-e;\, ((\cF-e) \cup \{f-e\},\,\bcF)} 
+ 2t_e \,\cU^{\epsilon,\bar\epsilon}_{\cG-e;\, (\cF-e,\,\bcF\cup \{f-e\})} \, ,
\eeq  
where $\cF-e=\cF\setminus \{f\}$ 
and $\bcF-e=\bcF$; 
\subitem(b)  if the removal of $e$   produces  two internal faces $f_1$ and $f_2$ from $f$, then
\bea
\label{eq:f2fin+++} 
\cU^{\eps,\beps}_{\cG;\, (\cF,\,\bcF)} 
&=& 
\rho_{\beps,\eps}
\Big(\cU^{\eps,\beps}_{\cG-e;\, ((\cF-e)\cup \{f-e\},\,\bcF)} 
+
\cU^{\eps,\beps}_{\cG-e;\, ( \cF-e,\,\bcF\cup \{f-e\})} \Big) \crcr
&+& 
 \rho_{\eps,\beps}\Big( \cU^{\eps,\beps}_{\cG-e;\, ((\cF-e)\cup \{f_1\},\,\bcF\cup \{f_2\})} 
+  (1\leftrightarrow 2)  \Big) \,, 
\cr\cr
&& 
\rho_{\od,\ev}(t_e) := 1+t_e^2 \,, \qquad
\rho_{\ev,\od}(t_e) :=2t_e \,,
\eea
where, we denote $\{f_1,f_2\}:=\{f-e\}$, $\cF-e = \cF\setminus\{f\}$ 
and $\bcF-e= \bcF$.
\item[(ii)] If $e\in f_1$ and $e\in f_2$, $f_1 \neq f_2$,
\subitem(a) and if $f_{1,2} \in \cF$,
\bea \label{eq:f1f2++}  
\cU^{\epsilon,\bar\epsilon}_{\cG;\, (\cF,\,\bcF)}
 &=& 
\cU^{\epsilon,\bar\epsilon}_{\cG/e;\, (\cF/e,\,\bcF/e)}
+  t_e^2 \,
\cU^{\epsilon,\bar\epsilon}_{\cG/e;\, (\cF/e\setminus\{f_1/e,f_2/e\},\,\bcF\cup \{f_1/e,f_2/e\})}
\cr\cr
&+ &  \, t_e 
\Big( \delta_{\eps,\od}\,  \cU^{\epsilon,\bar\epsilon}_{\cG-e;\,( (\cF-e)\cup\{f\},\,\bcF)}  
+\delta_{\beps,\od}\, \cU^{\epsilon,\bar\epsilon}_{\cG-e;\, (\cF-e,\,\bcF\cup\{f\} )}   \Big)\,,
\eea
where we denote $f$ the unique resulting internal face in 
$\cG-e$ coming from the faces $f_1$ and $f_2$, 
and 
where we note that $\cF-e =\cF\setminus\{f_1,f_2\}$,  
and $\bcF-e =\bcF=\bcF/e$; 

\subitem(b) if $f_{1} \in \cF$ and $f_{2}\in \bcF$, then 
\bea \label{cbnn}  
\cU^{\epsilon,\bar\epsilon}_{\cG;\, (\cF,\,\bcF)}
 &=& 
\cU^{\epsilon,\bar\epsilon}_{\cG/e;\, (\cF/e,\,\bcF/e)}
+ t_e^2 \,
\cU^{\epsilon,\bar\epsilon}_{\cG/e;\, (\cF/e\setminus\{f_1/e\}\cup \{f_2/e\} ,\,\bcF/e\setminus\{f_2/e\}\cup \{f_1/e\})} 
\cr\cr
&+ &  \, t_e 
\Big( \delta_{\eps,\od}\, \cU^{\epsilon,\bar\epsilon}_{\cG-e;\,( (\cF-e)\,,\,(\bcF-e) \cup\{f\})}  
+\delta_{\beps,\od}\, \cU^{\epsilon,\bar\epsilon}_{\cG-e;\, ((\cF-e)\cup \{f\},\,(\bcF-e) )}  \Big)\,,
\eea
where we denote $f$ the unique resulting internal face in 
$\cG-e$ coming from the faces $f_1$ and $f_2$, 
  with $\cF-e =\cF\setminus\{f_1\}$,  
and $\bcF-e =\bcF\setminus\{f_2\}$. 

\end{enumerate}
\end{proposition}

\proof  
The first relation does not cause any trouble. We focus on
$(ia)$ and expand the polynomial, for $e^2\in f$ with $f \in \cF$,
and get
\beq\label{calU}
\cU^{\epsilon, \bar{\epsilon}}_{\cG; \,(\cF, \,\bcF)} 
=
\left((t_e^2 + 1) A^{\epsilon}_{f/e} + 2 t_e A^{\bar{\epsilon}}_{f/e}  \right)
\Big (\prod_{\shortstack{ $_{\rm{f} \in \cF}$ \\ $_{{\rm f} \ne  f}$} } A^{\epsilon}_{\rm f} \Big ) 
\Big (\prod_{\rm f \in \bcF} A^{\bar{\epsilon}}_{\rm f} \Big) \,.
\eeq
Because all the edges contained in $f/e$ and $f-e$ are the same,
we  can write:
\beq
A^{\eps}_{f/e} 
=
A^{\eps}_{f-e}\,, \quad \forall \eps\,.
\label{aepsu}
\eeq
By definition $\cF-e=\cF\setminus\{f\}$, we can conclude $(ia)$. 

One proves $(ib)$ by first observing that
\beq
\begin{cases}
A^{\od}_{f/e} 
=
A^{\od}_{f_1} A^{\ev}_{f_2}+A^{\ev}_{f_1} A^{\od}_{f_2}\\
A^{\ev}_{f/e} 
=
A^{\od}_{f_1} A^{\od}_{f_2}+A^{\ev}_{f_1} A^{\ev}_{f_2}\,,
\end{cases}
\label{AepsU}
\eeq
where $f_1$ and $f_2$ are generated by the deletion of $e$. 
We insert \eqref{AepsU}  in \eqref{calU}, 
and using the definition $\cF - e =\cF \setminus \{f\}$
and $\{f-e\} = \{f_1,f_2\}$, we arrive at the desired relations.

Let us now prove $(ii)$. One starts from the expansion of 
$\cU^{\epsilon, {\bar{\epsilon}}}_{\cG;\, \left( \cF, \, \bcF \right)} $ 
focusing on the two amplitudes of faces $f_i$ sharing $e$.
From Theorem \ref{theo:gencdel}, in particular \eqref{f1f2in}
and \eqref{f1nf2in}, know that the two contraction terms present in \eqref{eq:f1f2++}  and \eqref{cbnn}, respectively, have been shown true. 
We focus on the additional terms in \eqref{f1f2in} and
\eqref{f1nf2in} and prove that they involve contraction terms. For $(iia)$, the key relation is 
\beq
A^{\beps}_{f_1/e} A^{\eps}_{f_2/e} + A^{\eps}_{f_1/e} A^{\beps}_{f_2/e} 
= 
A^{\od}_{f}\,,
\label{eq:Aodparity}
\eeq
where $f$ is the face formed from $f_{1,2}$ after
the deletion of $e$. 
This leads us to choose  the parities of each 
sector $\cF$ and $\bcF$. 
Using this and the definition $\cF \setminus \{f_1, f_2\} = \cF-e $,
we write
\beq\label{patti}
A^{\od}_{f}
\Big (\prod_{ {\rm{f}} \in \cF-e} A^{\epsilon}_{\rm f} \Big ) 
\Big (\prod_{ {\rm{f}} \in \bcF } A^{\bar{\epsilon}}_{\rm f} \Big)
=
\begin{cases}
\cU^{\eps=\od, \,\beps=\ev}_{\cG-e; \, ((\cF-e)\cup \{f\}, \, \bcF)} \\
\cU^{\eps=\ev,\, \beps= \od}_{\cG-e; \, (\cF -e, \, \bcF\cup \{f\})} \,
\end{cases}
\eeq
leading to \eqref{eq:f1f2++}.
 Now if $f_1 \in \cF$ and $f_2 \in \bcF$, a counterpart relation
of \eqref{eq:Aodparity} is 
\beq
A^{\ev}_{f_1/e} A^{\ev}_{f_2/e} + A^{\od}_{f_1/e} A^{\od}_{f_2/e} 
= 
A^{\ev}_{f}\,,
\eeq
and one concludes \eqref{cbnn} with the
definitions $\cF-e = \cF \setminus\{f_1\}$
and $\bcF-e = \bcF \setminus\{f_2\}$.

\qed

\begin{figure}[h]
 \centering
     \begin{minipage}[t]{.8\textwidth}
      \centering
\def\svgwidth{0.6\columnwidth}
\tiny{
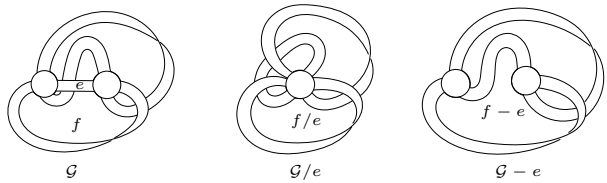
}
\caption{\small 
A graph $\cG$ obeying condition $(ia)$ of Proposition \ref{coro:deletion}.
}\label{fig:uniqueinternalface} 
\end{minipage}
\end{figure}

\begin{figure}[h]
 \centering
     \begin{minipage}[t]{.8\textwidth}
      \centering
\def\svgwidth{0.55\columnwidth}
\tiny{
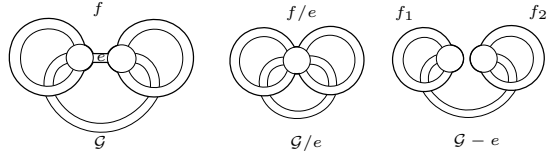
}
\caption{\small 
An example of graph $\cG$ satisfying condition $(ib)$ of Proposition~\ref{coro:deletion}. 
}\label{fig:pelvisgraph} 
\end{minipage}
\end{figure}

Let us comment that in the above statement, in the 
cases $(ia)$, $(ib)$ and $(iia)$, we 
assumed that the face $f$ or faces $f_i$ passing through $e$ 
are in $\cF$. It is simple to infer what happens if
they all belong to the other set $\bcF$. 
As an illustration of some of the configurations involved in Proposition \ref{coro:deletion}, 
we provide  Figures \ref{fig:uniqueinternalface} and \ref{fig:pelvisgraph}.


Proposition \ref{coro:deletion} establishes that 
some terms appearing in the recurrence relations of $\cU^{\eps, \beps}_{\cdot;\,(\cdot, \, \cdot)}$ as stated in Theorem \ref{theo:gencdel} may be re-expressed in terms of the polynomials involving a deletion of $e$. 
After such reductions in terms of contraction/deletion 
of an edge, the reader may wonder if 
the polynomial $\cU_{\cG}$ may be expressed in term
of the Tutte polynomial (the sole universal invariant satisfying
the contraction/deletion rule on a graph) or Bollob\`as-Riordan 
polynomial on ribbon graphs. The
answer to that question is definitely no because
there exist several cases for which the 
present rule fails to be a proper contraction/deletion relation
with exactly two terms: $\cU_{\cG}\neq \sigma_e\, \cU_{\cG-e}+\tau_e \,\cU_{\cG/e}$, for all $e$ regular, with $\sigma_e$ and $\tau_e$
functions of $t_e$.
 Thus $\cU_{\cG}$ is certainly not a Tutte polynomial 
and  therefore defines a new kind of invariant on its
enlarged space.

\subsubsection{Special edges}

We give a treatment of some of the special edges (or terminal forms)  when evaluating $\cU^{\epsilon, {\bar \epsilon}}_{\cG; \, (\cF, \, \bcF )}$. 
Terminal forms are crucial because they specify the boundary
conditions of the recurrence relations. 
Hence, the following study may help for the evaluation of 
the polynomial, when after a sequence
of reductions (contraction/deletions), 
the graph reaches some cases listed below. 

As commented after Theorems \ref{theo:gencdel} and
\ref{theo:tensorrecurrence}, terminal forms in any 
rank $d$ also satisfy special relations listed therein. Now the issue addressed  in the present section is
to show that, under particular circumstances,
these relations reduce and, sometimes, yield neat
factorizations. 

\medskip 
 \noindent{\bf Matrix case.}
We show that using
the disjoint union operation, some recurrence relations when applied to special edges lead to further simplification in terms
of subgraphs within the larger graph.

\begin{enumerate}

\item[(1)] 
\begin{figure}[h]
\centering
\begin{minipage}[t]{0.8\textwidth}
\centering
\def\svgwidth{0.8\columnwidth}
\tiny{
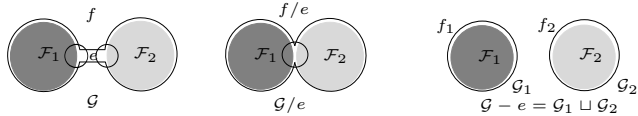
}
\caption{\small \label{fig:terminalform1} 
A bridge $e$: two blobs $\cF_1$ and $\cF_2$ which are 
collections of faces in subgraphs on each side of the bridge. Circles denote the
end-vertices of $e$. }
\end{minipage}
\end{figure}
We consider a graph $\cG$ with a bridge $e$ (Fig.\ref{fig:terminalform1}). We are
interested in the nontrivial configuration when $e$ belongs to a unique internal face $f \in \cF_{\inter}$ which corresponds to Theorem \ref{theo:gencdel} ($ii$). 
 Furthermore, we take $f \in \cF$.
Call $\cG_1$ and $\cG_2$ the two  disconnected subgraphs resulting from the deletion of  $e$, namely $\cG - e= \cG_1 \sqcup \cG_2$.
Call $\cF_i$ the set of internal faces in $\cG_i$.
Taking a partition $\cF\cup \bcF$ of the set of internal faces 
of $\cG$, four distinct cases can occur: (1) $\cF_1 \subset \cF$ and $\cF_2 \subset \bcF$, (2) $\cF_i \subset \cF$, (3) $\cF_i \subset \bcF$, and (4) $\cF_1 \subset \bcF$ and $\cF_2 \subset \cF$.
Only the case (1) will be discussed here, as the other ones can
be derived in a similar manner. We identify for the bridge graph,
\beq
\label{eq:1cdotsqcup} 
\qquad \cG-e = \cG_1 \sqcup \cG_2\,.
\eeq

The result of Proposition~\ref{coro:deletion} $(ib)$ still holds.
Then, assuming $\cF_1 \subset \cF=\cF_1 \cup \{f\}$ and 
$ \bcF=\cF_2$, and noting that $\cF-e = \cF_1 $ and $\bcF=\bcF - e= \cF_2$, we start from \eqref{eq:f2fin+++}. Apply repeatedly Proposition \ref{coro:disjointunion} and get:
\bea
\cU^{\eps,\beps}_{\cG;\, (\cF,\,\bcF)} 
&=& 
\rho_{\beps,\eps}
\Big(\cU^{\eps,\beps}_{\cG_1 \sqcup \cG_2;\, (\cF_1\cup \{f_1,f_2\},\,\cF_2)} 
+
\cU^{\eps,\beps}_{\cG_1 \sqcup \cG_2;\, ( \cF_1,\,\cF_2\cup \{f_1,f_2\})} \Big) \crcr
&+& 
 \rho_{\eps,\beps}\Big( \cU^{\eps,\beps}_{\cG_1 \sqcup \cG_2;\, \cF_1\cup \{f_1\},\,\cF_2\cup \{f_2\})} 
+  \cU^{\eps,\beps}_{\cG_1 \sqcup \cG_2;\, \cF_1\cup \{f_2\},\,\cF_2\cup \{f_1\})}    \Big)  
\cr\cr
&=&
 \rho_{\beps,\eps}
\Big(
\cU^{\eps,\beps}_{\cG_1;\, (\cF_1\cup \{f_1\},\,\emptyset)} 
\cU^{\eps,\beps}_{\cG_2;\, (\{f_2\},\,\cF_2)} 
+
\cU^{\eps,\beps}_{\cG_1 ;\, ( \cF_1,\,\{f_1\})}
\cU^{\eps,\beps}_{ \cG_2;\, ( \emptyset,\,\cF_2\cup \{f_2\})}
 \Big) \crcr
&+& 
 \rho_{\eps,\beps}\Big( 
\cU^{\eps,\beps}_{\cG_1;\, (\cF_1\cup \{f_1\},\,\emptyset)} 
\cU^{\eps,\beps}_{\cG_2;\, (\emptyset,\,\cF_2\cup \{f_2\})} 
+
 \cU^{\eps,\beps}_{\cG_1 ;\, (\cF_1,\,\{f_1\})}
\cU^{\eps,\beps}_{ \cG_2;\, (\{f_2\},\,\cF_2)}  
  \Big)  
\eea
and it partially factorizes.

\item[(2)] 
We now consider a trivial untwisted loop. This configuration divides into 
nontrivial cases where $e$ is shared between two internal faces 
(see Fig.{\ref{fig:untwistedselfloop2}})
or between one internal and one external faces.
The first case subdivides into two subcases determined
by the fact that the faces passing through 
$e$ may or may not belong to the same parity 
when the polynomial will be evaluated. 
 We focus on the situation described
by the condition $(iia)$ of Proposition \ref{coro:deletion}
while the same technique can be applied for all the
remaining cases. A basic relation is
\beq
\label{eq:2cdotsqcup}
\cG/e = \cG_1 \sqcup \cG_2\,.
\eeq
\begin{figure}[h]
\centering
\begin{minipage}[t]{0.8\textwidth}
\centering
\def\svgwidth{0.5\columnwidth}
\tiny{
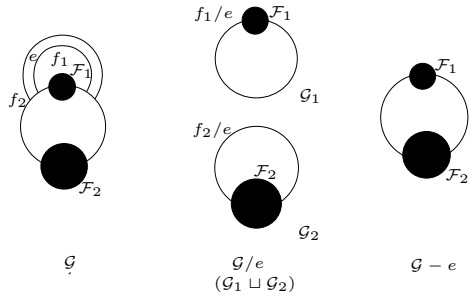
}
\caption{\small \label{fig:untwistedselfloop2}  A
trivial untwisted loop and the internal faces $f_i$ through $e$.
$\cF_i \cup \{f_i/e\}$ are the sets internal faces of the
subgraphs $\cG_i$ obtained by contraction of $e$.
}
\end{minipage}
\end{figure}

Consider a partition
of the set of internal faces of $\cG$ as $\cF\cup \bcF$.
In Fig.{\ref{fig:untwistedselfloop2}}, consider that 
the internal faces passing through $e$ are such that 
$f_1 \in \cF$ and $f_2  \in \cF$ (and $f_1 \ne f_2$). 
We contract $e$ in the original graph $\cG$  and call the resulting graphs as $\cG_1$ and $\cG_2$.
$\cG_1$ (resp. $\cG_2$) contains the set of internal faces $\mathcal{H}^*_1= \cF_1 \cup \{f_1/e\}$ (resp. $\mathcal{H}^*_2=\cF_2 \cup \{f_2/e\}$). Let us denote 
$\mathcal{H}_i  = \cF \cap \mathcal{H}^*_i$ and $\bar{\mathcal{H}}_i=\bcF \cap \mathcal{H}^*_i$.
We define $f$ the face obtained from $f_{1,2}$
after the deletion of $e$.

From \eqref{eq:f1f2++}, one writes, using the intermediate 
step \eqref{patti}, 
 \bea \label{bubu}  
\cU^{\epsilon,\bar\epsilon}_{\cG;\, (\cF,\,\bcF)}
 &=& 
\cU^{\epsilon,\bar\epsilon}_{\cG/e;\, (\cF/e,\,\bcF)}
+  t_e^2 \,
\cU^{\epsilon,\bar\epsilon}_{\cG/e;\, (\cF/e\setminus\{f_1/e,f_2/e\},\,\bcF\cup \{f_1/e,f_2/e\})}
\cr\cr
&+ &  \, t_e (A^{\eps}_{f_1/e}A^{\beps}_{f_2/e} + A^{\beps}_{f_1/e}A^{\eps}_{f_2/e})
\Big (\prod_{ {\rm{f}} \in \cF;  \; {\rm{f}}\ne f_i} A^{\epsilon}_{\rm f} \Big ) 
\Big (\prod_{ {\rm{f}} \in \bcF } A^{\bar{\epsilon}}_{\rm f} \Big)
\cr\cr
 &=& 
\cU^{\epsilon,\bar\epsilon}_{\cG_1 \sqcup \cG_2;\, (\mathcal{H}_1 \cup  \mathcal{H}_2,\,\bar{\mathcal{H}}_1 \cup  \bar{\mathcal{H}}_2 )}
+  t_e^2 \,
\cU^{\epsilon,\bar\epsilon}_{\cG_1 \sqcup \cG_2;\, (\mathcal{H}_1 \cup  \mathcal{H}_2\setminus\{f_1/e,f_2/e\},\,\bar{\mathcal{H}}_1 \cup  \bar{\mathcal{H}}_2\cup \{f_1/e,f_2/e\})}
\cr\cr
&+ &  \, t_e (A^{\eps}_{f_1/e}A^{\beps}_{f_2/e} + A^{\beps}_{f_1/e}A^{\eps}_{f_2/e})
\Big[\prod_{ {\rm{f}} \in \mathcal{H}_1 \cup  \mathcal{H}_2\setminus\{f_1/e,f_2/e\}} A^{\epsilon}_{\rm f} \Big]
\Big [\prod_{ {\rm{f}} \in  \bar{\mathcal{H}}_1 \cup  \bar{\mathcal{H}}_2  } A^{\bar{\epsilon}}_{\rm f} \Big],
\eea
where we recall that
\bea
&&
\cF/e = (\cF\setminus\{f_1,f_2\})\cup \{f_1/e,f_2/e\}
= \mathcal{H}_1 \cup  \mathcal{H}_2\,,
\qquad 
\cF-e =\cF\setminus\{f_1,f_2\},
\crcr
&&\bcF/e =  \bcF-e  = \bcF= \bar{\mathcal{H}}_1 \cup  \bar{\mathcal{H}}_2 \,. 
\eea 
We arrive at

\bea
\cU^{\epsilon,\bar\epsilon}_{\cG;\, (\cF,\,\bcF)}
&=&
\cU^{\epsilon,\bar\epsilon}_{\cG_1;\, (\mathcal{H}_1,\,\bar{\mathcal{H}}_1  )}
\cU^{\epsilon,\bar\epsilon}_{\cG_2;\, (  \mathcal{H}_2,\, \bar{\mathcal{H}}_2)}
\crcr
&+&  t_e^2 \,
\cU^{\epsilon,\bar\epsilon}_{\cG_1 ;\, (\mathcal{H}_1  \setminus\{f_1/e\},\,\bar{\mathcal{H}}_1 \cup  \{f_1/e\})}
\cU^{\epsilon,\bar\epsilon}_{ \cG_2;\, ( \mathcal{H}_2\setminus\{f_2/e\},\,\bar{\mathcal{H}}_2\cup \{f_2/e\})} 
 \crcr
&+& \, t_e \Big(
\cU^{\epsilon,\bar\epsilon}_{\cG_1;\,(\mathcal{H}_1,\,\bar{\mathcal{H}}_1  )}
\cU^{\epsilon,\bar\epsilon}_{\cG_2;\, (  \mathcal{H}_2\setminus \{f_2/e\},\, \bar{\mathcal{H}}_2\cup \{f_2/e\} )}\crcr
&+& 
\cU^{\epsilon,\bar\epsilon}_{\cG_1;\,(\mathcal{H}_1\setminus \{f_1/e\},\,\bar{\mathcal{H}}_1 \cup \{f_1/e\} )}
\cU^{\epsilon,\bar\epsilon}_{\cG_2;\, (  \mathcal{H}_2,\, \bar{\mathcal{H}}_2 )} \Big) \crcr
&=&
\Big(\cU^{\epsilon,\bar\epsilon}_{\cG_1;\, (\mathcal{H}_1,\,\bar{\mathcal{H}}_1  )} +t_e \cU^{\epsilon,\bar\epsilon}_{\cG_1;\,(\mathcal{H}_1\setminus \{f_1/e\},\,\bar{\mathcal{H}}_1 \cup \{f_1/e\} )}\Big)\crcr
&& \times \Big(\cU^{\epsilon,\bar\epsilon}_{\cG_2;\, (\mathcal{H}_2,\,\bar{\mathcal{H}}_2  )} +t_e\cU^{\epsilon,\bar\epsilon}_{\cG_2;\,(\mathcal{H}_2\setminus \{f_2/e\},\,\bar{\mathcal{H}}_2 \cup \{f_2/e\} )}\Big)
\eea
which is a factorized polynomial.

\item[(3)] 
We now consider a graph with a trivial twisted loop as in 
Fig.{\ref{fig:untwistedselfloop3}}.
This necessarily leads to unique face passing through the edge $e$.
This case has already been computed in  \eqref{f2fin} in
Theorem~\ref{theo:gencdel} and \eqref{eq:f2fin++} in Proposition~\ref{coro:deletion}. No factorization occurs and we have
the relations: 
 \bea
\cU^{\epsilon,\bar\epsilon}_{\cG;\, (\cF,\,\bcF)}&=&
2  t_e \, \cU^{\epsilon,{\bar \epsilon}}_{ \cG/e; \, (\cF/e \setminus \{f/e\}, \, \bcF \cup \{f/e\} )}  
+  \;
(t_e^2 + 1 )  \, \cU^{\epsilon,{\bar \epsilon}}_{ \cG/e; \, (\cF/e, \, \bcF )}
\crcr
&=& 2t_e \,\cU^{\epsilon,\bar\epsilon}_{\cG-e;\, (\cF-e,\,\bcF\cup \{f-e\})} + 
(1+t_e^2)\, \cU^{\epsilon,\bar\epsilon}_{\cG-e;\, ((\cF-e) \cup \{f-e\},\,\bcF)}  \, .
\eea
\begin{figure}[h]
\centering
\begin{minipage}[t]{0.8\textwidth}
\centering
\def\svgwidth{0.58\columnwidth}
\tiny{
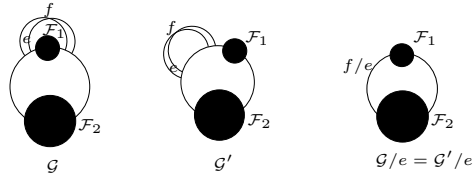
}
\caption{\small \label{fig:untwistedselfloop3}
A trivial twisted loop $e$ and $f$
the internal face passing through $e$. 
$\cG$ and $\cG'$  possess  the same polynomial. 
After contracting or deleting $e$, 
$\cG/e = \cG'/e = \cG - e = \cG'-e$. }
\end{minipage}
\end{figure}

\end{enumerate}

\medskip 

\begin{figure}[h]
\centering
\begin{minipage}[t]{0.8\textwidth}
\centering
\def\svgwidth{0.55\columnwidth}
\tiny{
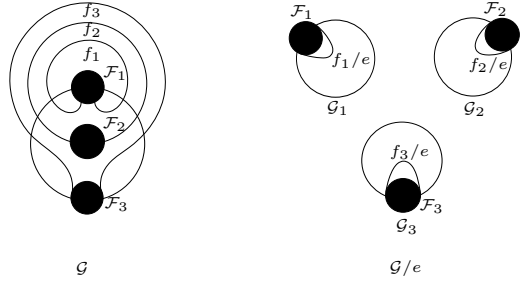
}
\caption{\small \label{fig:tensorterminalC} A terminal form for rank $3$ colored tensor model. $\cG$ is a trivial loop with the set $\cF_i$ of internal faces of the blob-subgraphs. Contracting $e$ gives three graphs $\cG_i$ with the set of faces $\cF_i \cup \{f_i/e\}$.
}
\end{minipage}
\end{figure}

\noindent{\bf Rank 3 colored tensor models.}
Theorem~\ref{theo:tensorrecurrence} is also valid in the case of the terminal forms. Particular classes of terminal forms have been 
discussed in \cite{avohou}. We will use one of these
terminal forms depicted in  Fig.\ref{fig:tensorterminalC}
(illustrated for rank $d=3$, but the idea generalizes easily). 
This graph is a higher rank generalization of a trivial
untwisted loop in the ribbon case.  
Each blob appearing in black is a subgraph 
of $\cG$ which is not connected (by any strand) to any other blobs. 
After contraction of the edge, the graph 
splits in $d$ disjoint subgraphs. 

Let us now restrict to $d=3$ and  
call the sets of internal faces contained in each blob
$\cF_i$. Assume the internal
faces $f_i$ passing through the edge $e$ obey $f_i \notin \cF_i$. Given a partition $\cF\cup \bcF$ of the set of
internal faces of $\cG$, we shall use the assumption 
that this set decomposes as $\cF=\{f_1,f_2,f_3\}\cup 
(\cup_j  {\widetilde \cF_j} ) $, ${\widetilde \cF_j}=\cF\cap \cF_j$, and the complementary set of faces $\bcF=\cup_k {\overline {\widetilde \cF_k}}$,
with $ {\overline {\widetilde \cF_k}}= \bcF \cap \cF_k$.

Since we chose all $f_i \in \cF$, we have
$\cF/e = \{f_1/e, f_2/e, f_3/e\} \cup (\cup_j {\widetilde \cF_j})$
and $\bcF/e = \cup_k {\overline {\widetilde \cF_k}}$. 
Using Theorem~\ref{theo:tensorrecurrence} and after
some algebras, we obtain
\bea
\label{eq:tensorterminalCa}
&&\cU^{\epsilon, {\bar \epsilon}}_{\cG; \, (\{f_1, f_2, f_3\} \cup 
(\cup_j {\widetilde \cF_j}), \,  \cup_k {\overline {\widetilde \cF_k}})}
= 
t_e^3 \,
\cU^{\epsilon, {\bar \epsilon}}_{\cG_1; \, ({\widetilde \cF_1}, \, \{f_1/e\} \cup {\overline {\widetilde \cF_1}} )}
\cU^{\epsilon, {\bar \epsilon}}_{\cG_2; \, ({\widetilde \cF_2}, \, \{f_2/e\} \cup {\overline {\widetilde \cF_2}})}
\cU^{\epsilon, {\bar \epsilon}}_{\cG_3; \, ({\widetilde \cF_3}, \, \{f_3/e\} \cup {\overline {\widetilde \cF_3}})}
\cr\cr
&&+
t_e^2
\Big (
\cU^{\epsilon, {\bar \epsilon}}_{\cG_1; \, ({\widetilde \cF_1}, \, \{f_1/e\} \cup {\overline {\widetilde \cF_1}})}
\cU^{\epsilon, {\bar \epsilon}}_{\cG_2; \, ({\widetilde \cF_2}, \, \{f_2/e\} \cup {\overline {\widetilde \cF_2}})}
\cU^{\epsilon, {\bar \epsilon}}_{\cG_3; \, ({\widetilde \cF_3} \cup \{f_3/e\}, \, {\overline {\widetilde \cF_3}})}
+ (1\to2\to 3)
\Big )
\cr\cr
&&+ 
t_e
\Big (
\cU^{\epsilon, {\bar \epsilon}}_{\cG_1; \, ({\widetilde \cF_1}, \, \{f_1/e\} \cup  {\overline {\widetilde \cF_1}})}
\cU^{\epsilon, {\bar \epsilon}}_{\cG_2; \, ({\widetilde \cF_2} \cup \{f_2/e\}, \, {\overline {\widetilde \cF_2}})}
\cU^{\epsilon, {\bar \epsilon}}_{\cG_3; \, ({\widetilde \cF_3} \cup \{f_3/e\}, \, {\overline {\widetilde \cF_3}})}
+ (1\to2\to 3) \Big) 
\cr\cr
&& + 
\cU^{\epsilon, {\bar \epsilon}}_{\cG_1; \, ({\widetilde \cF_1} \cup \{f_1/e\}, \,  {\overline {\widetilde \cF_1}})}
\cU^{\epsilon, {\bar \epsilon}}_{\cG_2; \, ({\widetilde \cF_2} \cup \{f_2/e\}, \, {\overline {\widetilde \cF_2}})}
\cU^{\epsilon, {\bar \epsilon}}_{\cG_3; \, ({\widetilde \cF_3} \cup \{f_3/e\}, \, {\overline {\widetilde \cF_3}})}\,,
\label{eq:terminalformC}
\eea
where  $(1\to2\to 3)$ simply refers to a permutation
over the three labels which make the contribution
symmetric in 1, 2, and 3. 

Other choices of the parities of the $f_i$'s can also be made.  
The calculation becomes a little bit involved
but the idea and techniques used above remain   the same. 

Assuming again that $f_i \in \cF$ and $\cF_1 = \cF_2 = \cF_3 = \emptyset$,  further noting that $\{f_1/e\}=\{f_2/e\}=\{f_3/e\}=o$ are all bare vertices, 
and using our conventions \eqref{eq:convention2}, we have from  \eqref{eq:terminalformC}
\beq
\cU^{\epsilon, {\bar \epsilon}}_{\cG; \, (\{f_1, f_2, f_3\}, \, \emptyset)} 
=
\begin{cases}
t_e^3\,, \qquad {\rm for \; \epsilon = \od}\,,
\\
1\,,  \;\qquad {\rm for \; \epsilon = \ev}\,.
\end{cases}
\eeq
These are the values of the polynomial $\cU^{\epsilon, {\bar \epsilon}}$
for the simple tensor graph made with 
one vertex (with two half-edges) and one edge.

\subsection{Illustrations}
\label{subsect:illust}
We provide examples in order to check the recurrence
relations using the polynomial of the second kind 
$\cU^{\eps,  \beps}_{\cG; \; (\cF,\, \bcF )}$ on some particular
nontrivial cases. 

\medskip
\noindent{\bf Matrix case.}
\begin{figure}[h]
\centering
\begin{minipage}[t]{0.8\textwidth}
\centering
\def\svgwidth{0.5\columnwidth}
\tiny{
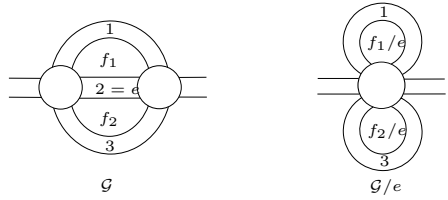
}
\caption{\label{fig:untwisted_self_loop9}\small  A
ribbon graph $\cG$ with an ribbon edge $e$ which passes through two different internal faces $f_1$ and $f_2$.  $\cG/e$  includes $f_1/e$ and $f_2/e$.}
\end{minipage}
\end{figure}
Consider the ribbon 
graph $\cG$  given in Fig.\ref{fig:untwisted_self_loop9}.
We distribute its closed faces as  $\cF = \{f_1\}$ and $\bcF = \{f_2\}$.
From direct evaluation, using \eqref{eq:defcu}, we obtain
\beq
\cU^{\od, \ev}_{\cG; \,\left(\{f_1\}, \, \{f_2\} \right)}
= (t_1 + t_2 ) (1 + t_2 t_3)\,.
\label{eq:ex1direct}
\eeq
Now, we compute the same polynomial using the recurrence
relation. Pick the edge $2$ which is shared by the internal faces $f_1 \ne f_2 $. 
We use ($iii.b$) in Theorem~\ref{theo:gencdel}, noting also \eqref{eq:conventionext}, to write
\bea
\cU^{\od, \ev}_{\cG; \,( \cF, \,\bcF )} 
&&=
\cU^{\od,\ev}_{\cG/e; \,(\cF/e,\, \bcF/e )} 
+
t_e 
\left(
\cU^{\od,\ev}_{\cG/e; \, \left(\cF/e \; \cup \; \{f_2/e\}, \, \bcF \; \setminus \; \{f_2\} \right)}
+
\cU^{\od,\ev}_{\cG/e; \, \left(\cF \; \setminus \{f_1\}, \, \bcF /e \; \cup\{f_1/e\} \right)}
\right)
\crcr
&& + 
\; t_e^2 \;
\cU^{\od, \ev}_{\cG/e; \, \left( \cF \setminus \{f_1\} \; \cup \{f_2 /e \} , \, \bcF\setminus \{f_2 \} \cup \{f_1/e\} \right)} 
\crcr
&& =
 t_1 + t_2   ( t_1 t_3 + 1) + t_2^2 t_3 \,.
\label{fffin}
\eea
Expanding \eqref{eq:ex1direct}, one gets of course \eqref{fffin}.

\medskip

\begin{figure}[h]
\centering
\begin{minipage}[t]{0.8\textwidth}
\centering
\def\svgwidth{0.55\columnwidth}
\tiny{
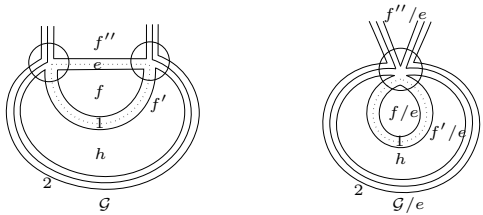
}
\caption{\small \label{fig:tensorex1}  A rank 3 colored tensor graph $\cG$ with internal faces $f$, $f'$ (dotted), and $h$. In $\cG/e$, $f/e$ and $f'/e$ (dotted) pass through the edge $1$.
}
\end{minipage}
\end{figure}

\noindent{\bf Rank 3 colored tensor models.}
Consider the rank 3 graph in Fig.{\ref{fig:tensorex1}}.
We pick the edge $e$ which is shared by two internal faces $f \ne f'$,
 and  $f,f' \in \cF$. We also choose that the remaining
closed face $h \in \cF$. 
Therefore, $\cF = \{f, f', h\}$ and $\bcF = \emptyset$
form a partition of internal faces of $\cG$. 
From direct computation, i.e. using \eqref{eq:defcu}, we obtain
\beq
\cU^{\od, \ev}_{\cG; \, (\cF, \, \bcF)}
=
(t_e + t_1)^2 (t_1 + t_2)\,.
\label{eq:tensorex1}
\eeq
Using the recurrence relations given in 
Theorem~\ref{theo:tensorrecurrence}, one has
\bea
&& 
\cU^{\od, \ev}_{\cG; \, (\cF, \, \bcF)}
 =
t_e^2 \,
\cU^{\od,\ev}_{\cG/e; \, (\cF/e \setminus \{f/e,f'/e\},\, \bcF \cup \{f/e, f'/e\} )} + 
\cU^{\od, \ev}_{\cG/e; \, ( \cF/e \setminus \{f/e, f'/e\} \cup \{f/e, f'/e\}, \bcF)} 
\crcr
&&  + 
t_e  
\left(
\cU^{\od,\ev}_{\cG/e; \, (\cF/e \setminus \{f/e, f'/e\} \cup \{f'/e\}, \, \bcF \cup \{f/e\})}
+
\cU^{\od,\ev}_{\cG/e; \, (\cF/e \setminus \{f/e, f'/e\} \cup \{f/e\}, \, \bcF /e \cup\{f'/e\})}
\right)\\
&& =
t_e^2 
\cU^{\od,\ev}_{\cG/e; \, (\{h\},\, \{f/e, f'/e\} )} 
+
\cU^{\od, \ev}_{\cG/e; \, (\{f/e, f'/e, h\}, \emptyset)} 
\cr\cr
&& 
+ 
t_e 
\left(
\cU^{\od,\ev}_{\cG/e; \, (\{f'/e, h\}, \, \{f/e\})}
+
\cU^{\od,\ev}_{\cG/e; \, (\{f/e, h\}, \, \{f'/e\})}
\right)\nonumber
\eea
and
\bea
&&
\cU^{\od,\ev}_{\cG/e; \, (\{h\},\, \{f/e, f'/e\} )}
 =
t_1 + t_2\,, \qquad \;\;\;\;\,\,
\cU^{\od, \ev}_{\cG/e; \, (\{f/e, f'/e, h\}, \emptyset)}
= t^2_1(t_1+t_2)\,, \crcr
&& 
\cU^{\od,\ev}_{\cG/e; \, (\{f'/e, h\}, \, \{f/e\})}
=
t_1 (t_1+t_2) \,,\qquad 
\cU^{\od,\ev}_{\cG/e; \, (\{f/e, h\}, \, \{f'/e\})}
=
t_1(t_1+t_2)\,.
\eea
Thus, the last equations are consistent with \eqref{eq:tensorex1}.

\section{Conclusion}
\label{concl}

The parametric representation of tensor models over 
the Abelian group $U(1)^{D}$ with a kinetic
term linear in momenta has been investigated in this work. 
We have first introduced a dimensional regularization scheme  
and perform the ensuing renormalization procedure
on amplitudes of specific tensor models. 
An important fact revealed by this work 
is that these well-known procedures can be made compatible
with the Feynman amplitudes depending on
stranded graph structures. We have also shown that the amplitudes
define analytic functions in the complex $D$, for $\Re(D)$
small enough. These graph amplitudes $A_{\cG}$ can be extended in
meromorphic functions in the strip $0<\Re(D)< \bdel +\varepsilon_{\cG}$,
where $\bdel$ is a given dimension of the group 
in the model considered and $\varepsilon_{\cG}$ is a small positive
quantity depending on the graph $\cG$. Due to the presence
of another independent parameter in this class of
models, namely the theory rank $d$, it also seems possible 
to define a new ``rank regularization'' procedure of the amplitudes
by complexifying the parameter $d$. This deserves to
be fully investigated.

 In a second part, we have thoroughly investigated and extended the Symanzik
polynomials yielded by the parametric representation
of generic Abelian models. The ordinary contraction/deletion
rules satisfied by Symanzik polynomials
are now clearly broken by the stranded graph structure. 
We have introduced an abstract class of polynomials which depends both on the graph 
$\cG$ but also on a peculiar decomposition of its set $\cF_{\inter;\cG}$ of faces. 
Then, we prove that these new polynomials satisfy (only) contraction rules. 
We have also provided some terminal form recurrence rules and several illustrations. 
Let us  emphasize that the fact that one might incorporate
more information in graph polynomials which 
depend not only on the graph but also on the sets of its constituents opens an avenue of new investigations. To be clearer, the
Tutte polynomial $T_{\cG}$ is defined by a state sum over the set $\cP(\cG)$ of spanning subgraphs of $\cG$. Using insights of the
present work, the question is whether or not $T_{\cG}$ could have been identified as a function of $\cG$ and $\cP(\cG)$ itself. If the answer to this
question is positive then it will prove that the Tutte polynomial 
can be read differently. All of its consequences and its 
ramification in higher dimensions, like the
Bollob\`as-Riordan polynomial, might find
a different  representation which might lead to a richer
interpretation. This must also be investigated elsewhere. 

Finally, the present work has addressed the simplest setting that
one could envisage using tensor models.
 There exists a $\Phi^6$ model defined
with rank 4 tensors over $U(1)^4$ generating 4D simplicial topologies \cite{BenGeloun:2011rc}. 
This model  is endowed with a kinetic term including a quadratic 
dependence in momenta: $\sum_s p_s^2 +\mu$.  Finding a complete parametric
representation of its amplitudes will be a true challenge. 
The present work might be helpful for understanding a way 
to perform a dimensional regularization for this model and 
for studying the polynomials which will arise from such a representation.
This will be addressed in the forthcoming work. 

\section*{Acknowledgements}
Discussions with Vincent Rivasseau and
Thomas Krajewski are gratefully acknowledged. The authors are thankful to Perimeter Institute for Theoretical Physics, Waterloo, Canada, for having initially fostered this collaboration.

\section*{ Appendix}
\label{app}

\appendix

\renewcommand{\theequation}{\Alph{section}.\arabic{equation}}
\setcounter{equation}{0}

\section{Proof of Proposition \ref{theo:cdel}}
\label{app:theocdel}

In this section we provide the proof of Proposition \ref{theo:cdel}. 
Consider $\cG$ a ribbon graph with sets $\cF_{\inter;\cG}$
and $\cF_{\ext;\cG}$ of internal and external faces, respectively, and
$e$ an edge  of $\cG$.

\begin{enumerate}
\item[(i)]  The polynomial $U^{\od/\ev}_{\cG}$ \eqref{eq:defu} only takes into consideration internal faces. If $e$ only belongs to external faces, then a contraction of $e$ will not affect  $U^{\od/\ev}_{\cG}$. This proves \eqref{ext2}.  The points about the deletion of $e$ and
the creation or not of a new internal face are also direct
by definition.  

\item[(ii)] Let us assume now that $e \in f$, $f \in \cF_{\inter;\cG}$
and $e$ belongs to another external face $f'$. 
Then, we decompose $U^{\od/\ev}_{\cG}$ using Lemma \ref{lem:facontr} 
as follows 
\bea
U^{\od/\ev}_{\cG} &=& (t_e \, A^{\ev/\od}_{f/e} + A^{\od/\ev}_{f/e}) \, \prod_{ {\rm f} \in \cF_{\inter;\cG} ,\, {\rm f} \ne f } A^{\od/\ev}_{{\rm f}} \crcr
&=&t_e \, A^{\ev/\od}_{f/e} \, U^{\od/\ev}_{\cG -e} +U^{\od/\ev}_{\cG/e} \,,
\eea
where we used the fact that the set of internal faces of $\cG/e$
is given by $\{f/e \}\cup \cF_{\inter;\cG} \setminus \{f\}$ 
and, after removing $e$ in $\cG$, the face $f$ merges 
to the external face $f'$. As a result, 
the set of internal faces of $\cG-e$ coincides with $\cF_{\inter;\cG} \setminus \{f\}$. Finally, one observes that either cutting or deleting $e$
has the same effect on the set of internal faces of $\cG\vee e$
and $\cG -e$ (these both loose $f$). This achieves the proof of \eqref{extint}.

\item[(iii)] Consider that $e \in f$ and $e\in f'$, $f\ne f'$ 
and both internal. Still by Lemma \ref{lem:facontr}, 
we expand $U_\cG$ as
\bea
U^{\od/\ev}_{\cG} 
&=& (t_e \, A^{\ev/\od}_{f/e} + A^{\od/\ev}_{f/e})
(t_e \, A^{\ev/\od}_{f'/e} + A^{\od/\ev}_{f'/e}) \, \prod_{ {\rm f} \in \cF_{\inter;\cG} ,\, {\rm f} \ne f,f' } A^{\od/\ev}_{{\rm f}}
\crcr
&=& 
t^2_e \, A^{\ev/\od}_{f/e}A^{\ev/\od}_{f'/e} \prod_{ {\rm f} \in \cF_{\inter;\cG} ,\, {\rm f} \ne f,f' } A^{\od/\ev}_{{\rm f}}
\crcr
&&  +t_e [  A^{\od/\ev}_{f/e} A^{\ev/\od}_{f'/e} + (f \leftrightarrow f') ]\prod_{ {\rm f} \in \cF_{\inter;\cG} ,\, {\rm f} \ne f,f' } A^{\od/\ev}_{{\rm f}}
+ \; 
U^{\od/\ev}_{\cG/e} 
\crcr
&=&
t_e^2  \,A^{\ev/\od}_{f/e}  A^{\ev/\od}_{f^\prime/e}  \, U^{\od/\ev}_{\cG \vee e} 
+  
t_e  \,U^{\od}_{\cG-e}  
+  
U^{\od/\ev}_{\cG/e}\, ,
\eea
where, clearly, by cutting $e$ in $\cG$, one looses $f$ and $f'$
so that $\cF_{\inter;\cG \vee e} = \cF_{\inter;\cG}\setminus \{f,f'\}$,
 and where $\cF_{\inter;\cG/e} = \{f/e,f'/e\} \cup \cF_{\inter;\cG}\setminus \{f,f'\}$. The middle term is more
subtle. 
The removal of $e$ merges $f$ and $f'$ into a unique internal face. The complete odd face polynomial for this new face is given by summing over  odd subsets in $f/e \cup f'/e$. To get an odd subset, one must take an odd part from one and an even part from the other. In the end the new face polynomial exactly corresponds to $[  A^{\od}_{f/e} A^{\ev}_{f'/e} + (f \leftrightarrow f') ]$. This achieves \eqref{fnf}.

\item[(iv)] We have $e^2 \in f$, $f \in \cF_{\inter;\cG}$. 

\subitem(a) Let us assume that the deletion of $e$ gives rise to two distinct internal faces $f_1$ and $f_2$. Lemma \ref{lem:facontr} helps us to write 
\bea\label{aa}
U^{\od/\ev}_{\cG} 
&=& (2t_e \, A^{\ev/\od}_{f/e} + (t_e^2+1)A^{\od/\ev}_{f/e} )
\prod_{ {\rm f} \in \cF_{\inter;\cG} ,\, {\rm f} \ne f } A^{\od/\ev}_{{\rm f}}
\\
&=& 
\left(1 + t_e^2\right)  \; U^{\od/\ev}_{\cG/e} 
+
2 \; t_e 
\Big(
  A^{\od/\ev}_{f_1} \; A^{\od}_{f_2}
+
 A^{\ev/\od}_{f_1} \; A^{\ev}_{f_2}
\Big)\;
\prod_{ {\rm f} \in \cF_{\inter;\cG} ,\, {\rm f} \ne f } A^{\od/\ev}_{{\rm f}}
 \cr
\cr
 \left\{ \begin{array}{c} 
U_\cG^{\od}\\
U_{\cG}^{\ev}\end{array}
\right. &=&
 \left\{ \begin{array}{c}
\left(1 + t_e^2\right)  \; U^{\od}_{\cG/e}
+
2  t_e \, U^{\od}_{\cG-e} 
+
2 t_e \, A^{\ev}_{f_1} A^{\ev}_{f_2} \, U^{\od}_{\cG  \vee e}
 \\
\left(1 + t_e^2\right)  \; U^{\ev}_{\cG/e}
+
2  t_e( A^{\ev}_{f_1} \; A^{\od}_{f_2}
+
 A^{\od}_{f_1} \; A^{\ev}_{f_2}) \, U^{\ev}_{\cG  \vee e} 
\end{array} \right. \,.
\nonumber
\eea
The fact that we have $U^{\od/\ev}_{\cG/e}$ goes by the same argument as before.
We have split $A^{\ev/\od}_{f/e}$ into two types of contributions which come from the face polynomials associated with $f_1$ and $f_2$. The set of internal faces of $\cG-e$ are readily obtained from $\cF_{\inter;\cG} \setminus \{f\} \cup \{f_1, f_2\}$ whereas  $\cF_{\inter;\cG} \setminus \{f\} $ coincides again with the set of faces of $\cG \vee e$.
We get \eqref{e2ff}.
\subitem(b) Finally, we consider that the removal of $e$ generates one internal face $f_{12}$. The first line of \eqref{aa} remains the same. We identify $f/e$ with $f_{12}$, and the rest follows:
\bea
U^{\od/\ev}_{\cG} 
=(1 + t_e^2)  \, U^{\od/\ev}_{\cG/e} + 2 t_e \, A^{\ev/\od}_{f_{12}} \, 
U^{\od/\ev}_{\cG \vee  e} \,.
\eea

\end{enumerate}

\section{Proof of Theorem \ref{theo:gencdel}}
\label{app:theogencdel}

In this section, we give the proof of Theorem \ref {theo:gencdel}.
Let $\cG$ be a ribbon graph with half-ribbons, $\cF_{\inter}$ being
its set of  internal faces. Let $\cF$ and $\bcF$ be subsets
of $\cF_{\inter}$ as stated in the theorem. 
In the following, the 
face polynomial expansions are always performed using Lemma \ref {lem:facontr}.

\item[($0$)] External faces under contraction remain
external and do not affect $\cU^{\eps,\beps}$. 
This is also why $\cF/e=\cF$ and $\bcF/e= \bcF$.

\item[($i$)] Consider $e$ which belongs to an external face and an internal face denoted by $f \in \cF \subset \cF^{\inter}$. We have
\bea
\cU^{\epsilon, \bar\epsilon}_{\cG;\, (\cF, \, \bcF)}
& = & 
(t_e \, A^{\bar{\epsilon}}_{f/e} + A^{\epsilon}_{f/e} ) 
\Big(\prod_{\shortstack{ $_{{\rm f} \in \cF}$ \\ $_{{\rm f} \ne f}$ }}  A^{\epsilon}_{\rm f}\Big) 
\Big(\prod_{{\rm f} \in \bcF}  A^{\bar{\epsilon}}_{\rm f}\Big) \crcr
& = &
t_e \,
\cU^{\epsilon,\bar\epsilon}_{\cG/e;\, (\cF\setminus\{f\},\,\bcF\cup \{f/e\})}
+ \,
\cU^{\epsilon,\bar\epsilon}_{\cG/e;\, (\cF \setminus \{f\} \cup \{f/e\},\,\bcF)}  \,.
\eea
One notices that $(\cF \setminus \{f\}) \cup \{f/e\}$ coincides with $\cF/e$ which is the subset of faces corresponding to $\cF$ in the graph $\cG/e$. 
We have also $\cF \setminus \{f\} = (\cF/e) \setminus \{f/e\}$
and $\bcF/e = \bcF$ as $e$ does not belong  to any 
faces in $\bcF$.  We get \eqref{fexfin}. 

\item[($ii$)] If $e^2 \in f$,  $f \in \cF$,   
\bea
\cU^{\epsilon, \bar{\epsilon}}_{\cG;\, (\cF, \, \bcF)}
& = &
A^{\epsilon}_f  \,
\Big(\prod_{\shortstack{ $_{{\rm f} \in \cF}$ \\ $_{{\rm f} \ne f}$ }}  A^{\epsilon}_{\rm f}\Big) 
\Big(\prod_{{\rm f} \in \bcF} A^{\bar{\epsilon}}_{\rm f}\Big) \cr
\cr
& = & 
\left( (t_e^2 + 1) \, A^{\epsilon}_{f/e} + 2 t_e \, A^{\bar \epsilon}_{f/e} \right) \Big(\prod_{\shortstack{ $_{{\rm f} \in \cF}$ \\ $_{{\rm f} \ne f}$ }}  A^{\epsilon}_{\rm f}\Big) 
\Big(\prod_{{\rm f} \in \bcF}  A^{\bar \epsilon}_{\rm f}\Big) \crcr
& = &
( t_e^2 + 1) \, 
\cU^{\epsilon,\bar\epsilon}_{\cG/e;\, (\cF\setminus\{f\} \cup \{f/e\},\,\bcF)}
+ 
2  t_e \,
\cU^{\epsilon,\bar\epsilon}_{\cG/e;\, (\cF \setminus \{f\},\,\bcF \cup \; \{f/e\} )}  \,. 
\eea
Finally, to get \eqref{f2fin}, we apply the same identities as 
in ($i$). 

\item[$(iii.a)$] If $e \in f_i$, $f_1\neq f_2$, with $f_{i} \in \cF$, then 
\bea
\cU^{\epsilon, \bar{\epsilon}}_{\cG ;\, ( \cF,\, \bcF) }
&=&  
A^{\epsilon}_{f_1}  A^{\epsilon}_{f_2} \, 
\Big(\prod_{\shortstack{ $_{{\rm f} \in \cF}$ \\ $_{{\rm f} \ne f_1,f_2}$ }}  A^{\epsilon}_{\rm f}\Big) 
\Big(\prod_{\shortstack{ $_{{\rm f} \in \bcF}$ }} 
A^{\bar \epsilon}_{\rm f}\Big) \crcr
& = &
\left(t_e^2  A^{\bar \epsilon}_{f_1/e}A^{\bar \epsilon}_{f_2/e}  
+   t_e  A^{\epsilon}_{f_1/e}  A^{\bar \epsilon}_{f_2/e}  
+   t_e  A^{\epsilon}_{f_2/e}  A^{\bar \epsilon}_{f_1/e} 
+   A^{\epsilon}_{f_1/e}  A^{\epsilon}_{f_2/e} \right) 
\Big[\prod_{\shortstack{ {$_{{\rm f} \in \cF}$}\\$_{{\rm f} \ne f_1,f_2}$ }}  A^{\epsilon}_{\rm f}\Big]  
\Big[\prod_{\shortstack{ {$_{{\rm f} \in \bcF}$} }} 
 A^{\bar \epsilon}_{\rm f}\Big]
\cr
& = &
t_e^2 \; 
\cU^{\epsilon, {\bar \epsilon}}_{\cG/e ; \,( \cF \setminus \{f_1, f_2\}, \, \bcF \cup \{f_1/e, f_2/e\})} 
\cr
\cr
&& + 
t_e 
\left(
\cU^{\epsilon, {\bar \epsilon}}_{\cG/e ;\, (\cF \setminus \{f_1, f_2\} \cup  \{f_1/e\} , \, \bcF \cup  \{f_2/e\}  )}
+
\cU^{\epsilon, {\bar \epsilon}}_{\cG/e;\, (\cF \setminus \{f_1, f_2\} \cup  \{f_2/e\} , \, \bcF \cup \{f_1/e\} )} 
\right) 
\cr
\cr
&& + \; 
\cU^{\epsilon, {\bar \epsilon}}_{\cG/e ; (\cF \setminus \{f_1, f_2\} \cup \; \{f_1/e, f_2/e\} , \; \bcF) }\;,
\eea
where $(\cF \setminus \{f_1, f_2\} )\cup \; \{f_1/e, f_2/e\}=\cF/e $, 
$(\cF \setminus \{f_1, f_2\}) \cup  \{f_1/e\}= (\cF/e) \setminus \{f_2/e\}$,
$\cF \setminus \{f_1, f_2\}= (\cF/e) \setminus \{f_1/e, f_2/e\}$,
and $\bcF/e =\bcF$. 
One gets \eqref{f1f2in}.

\item[$(iii.b)$] If $e \in f_i$, $f_1 \in \cF$ and $f_2 \in \bcF$
\bea
\cU^{\epsilon, \bar{\epsilon}}_{\cG ;\, ( \cF,\, \bcF)}
& = &
A^{\epsilon}_{f_1} \; A^{\bar \epsilon}_{f_2} \; 
\bigg(\prod_{\shortstack{ $_{{\rm f} \in \cF}$ \\ $_{{\rm f} \ne f_1}$ }} \; A^{\epsilon}_{\rm f}\bigg) \; 
\bigg(\prod_{\shortstack{ $_{{\rm f} \in \bcF}$ \\ $_{{\rm f} \ne f_2}$ }} 
\; A^{\bar \epsilon}_{\rm f}\bigg) \; \cr
& = &
\left(t_e^2  A^{\epsilon}_{f_2/e}A^{\bar \epsilon}_{f_1/e}  
+   t_e  A^{\epsilon}_{f_1/e}  A^{\epsilon}_{f_2/e}  
+   t_e  A^{\bar \epsilon}_{f_1/e}  A^{\bar \epsilon}_{f_2/e} 
+   A^{\epsilon}_{f_1/e}  A^{\bar \epsilon}_{f_2/e} \right) 
\Big[\prod_{\shortstack{ {$_{{\rm f} \in \cF}$}\\{$_{{\rm f} \ne f_1}$} }}  A^{\epsilon}_{\rm f}\Big]
\Big[\prod_{\shortstack{ {$_{{\rm f} \in \bcF}$}\\{$_{{\rm f} \ne f_2}$} }} 
 A^{\bar \epsilon}_{\rm f}\Big] \cr
& = &
t_e^2 \,
\cU^{\epsilon, {\bar \epsilon}}_{\cG/e ; \,
( \cF \setminus\{f_1\}\cup \{f_2/e\}, 
\, \bcF  \setminus \{f_2\} \cup \{f_1/e\})} \; \cr
\cr
&& + t_e
\left(
\cU^{\epsilon, {\bar \epsilon}}_{\cG/e ; \,
(\cF \setminus\{f_1\}  \cup  \{f_1/e, f_2/e\}, 
\, \bcF \setminus \{f_2\})} 
+
\cU^{\epsilon, {\bar \epsilon}}_{\cG/e; \,
(\cF \setminus \{f_1\} ,\, \bcF  \setminus \{f_2\} \cup \{f_1/e, f_2/e\})} 
\right)
\cr
\cr
&& + 
\cU^{\epsilon, {\bar \epsilon}}_{\cG/e ;\, (\cF \setminus  \{f_1\} \cup  \{f_1/e\} , \,\bcF \setminus \{f_2\} \cup  \{f_2/e\}) }\;.
\eea
We conclude to \eqref{f1nf2in} after identifying 
$(\cF \setminus  \{f_1\}) \cup  \{f_1/e\}= \cF/e$ and $(\bcF \setminus  \{f_2\}) \cup  \{f_2/e\}= \bcF/e$, 
$(\cF \setminus\{f_1\} ) \cup  \{f_1/e, f_2/e\} = 
(\cF/e)  \cup \{ f_2/e\}$ and $\bcF \setminus \{f_2\} = 
\, (\bcF/e) \setminus \{f_2/e\}$, 
 and $(\cF \setminus\{f_1\})\cup \{f_2/e\}= ((\cF/e)\setminus\{f_1/e\})
\cup \{f_2/e\}$ and  $(\bcF  \setminus \{f_2\}) \cup \{f_1/e\}
= ((\bcF/e) \cup \{f_1/e\})\setminus \{f_2/e\} $. 
The rest of the equalities are obtained in a similar way.

\end{document}